\documentclass[twocolumn]{aastex701}
\usepackage{graphics,graphicx}
\usepackage{natbib}
\usepackage{xcolor}
\usepackage{wrapfig}
\usepackage{tcolorbox}
\usepackage{ragged2e}
\usepackage{caption}
\usepackage{amsmath}
\usepackage{multirow}
\usepackage{threeparttable}
\usepackage{CJK}

\newcommand{\oiii}{[O\,{\sc iii}]}
\newcommand{\oii}{[O\,{\sc ii}]}

\newcommand{\hb}{H$\beta$}
\newcommand{\ha}{H$\alpha$}
\newcommand{\hg}{H$\gamma$}
\newcommand{\hd}{H$\delta$}
\newcommand{\he}{H$\epsilon$}

\newcommand{\neiii}{[Ne\,{\sc iii}]}
\newcommand{\mgii}{Mg\,{\sc ii}}

\newcommand{\betauv}{$\beta_{\rm UV}$}
\newcommand{\muv}{$M_{\rm UV}$}
\newcommand{\sersic}{S$\rm \acute{e}$rsic}

\newcommand{\Luv}{$\nu L_{\nu , \rm UV}$}
\newcommand{\Lha}{$L_{\rm H\alpha}$}

\begin{document}
\begin{CJK*}{UTF8}{gbsn}

\title{PRISMS GHZ8: Evidence for a Young Burst and Potential Merger Activity at $z=10.23$}

\author[0000-0003-3987-0858]{Ruqiu Lin ({\CJKfamily{gbsn}林如秋})}\email[show]{ruqiulin@umass.edu}
\affiliation{University of Massachusetts Amherst, 710 North Pleasant Street, Amherst, MA 01003-9305, USA}

\author[0000-0002-7051-1100]{Jorge Zavala}\email{jzavala@umass.edu}
\affiliation{University of Massachusetts Amherst, 710 North Pleasant Street, Amherst, MA 01003-9305, USA}

\author[0000-0002-7093-1877]{Javier \'Alvarez-M\'arquez}\email{jalvarez@cab.inta-csic.es}
\affiliation{Centro de Astrobiolog\'ia (CAB), CSIC–INTA, Cra. de Ajalvir Km.~4, 28850- Torrej\'on de Ardoz, Madrid, Spain}

\author[0000-0002-9090-4227]{Luis Colina}\email{colina@cab.inta-csic.es}
\affiliation{Centro de Astrobiolog\'ia (CAB), CSIC–INTA, Cra. de Ajalvir Km.~4, 28850- Torrej\'on de Ardoz, Madrid, Spain}

\author[]{Marco Castellano}\email{marco.castellano@inaf.it}
\affiliation{INAF – Osservatorio Astronomico di Roma, via Frascati 33, 00078, Monteporzio Catone, Italy}

\author[]{Adriano Fontana}\email{adriano.fontana@inaf.it}
\affiliation{INAF – Osservatorio Astronomico di Roma, via Frascati 33, 00078, Monteporzio Catone, Italy}

\author[]{Paola Santini}\email{paola.santini@inaf.it}
\affiliation{INAF – Osservatorio Astronomico di Roma, via Frascati 33, 00078, Monteporzio Catone, Italy}

\author[]{Lorenzo Napolitano}\email{lorenzo.napolitano@inaf.it}
\affiliation{INAF – Osservatorio Astronomico di Roma, via Frascati 33, 00078, Monteporzio Catone, Italy}

\author[]{Sarah Kendrew}\email{skendrew@stsci.edu}
\affiliation{Space Telescope Science Institute (STScI), 3700 San Martin Drive, Baltimore, MD 21218, USA}

\author[0000-0003-2119-277X]{Alejandro Crespo G\'omez}\email{acrespo@stsci.edu}
\affiliation{Space Telescope Science Institute (STScI), 3700 San Martin Drive, Baltimore, MD 21218, USA}

\author[]{Victoria Salazar}\email{vsalazar@umass.edu}
\affiliation{University of Massachusetts Amherst, 710 North Pleasant Street, Amherst, MA 01003-9305, USA}

\author[0000-0002-5258-8761]{Abdurro'uf}\email{fnuabdur@iu.edu}
\affiliation{Department of Astronomy, Indiana University, 727 East Third Street, Bloomington, IN 47405, USA}

\author[0000-0002-6719-380X]{Stefano Carniani}\email{stefano.carniani@sns.it}
\affiliation{Scuola Normale Superiore, Piazza dei Cavalieri 7, I-56126 Pisa, Italy}

\author[]{Tiger Hsiao}\email{tiger.hsiao@utexas.edu}
\affiliation{Department of Astronomy, University of Texas, Austin, TX 78712,  USA}

\author[]{Yuichi Harikane}\email{hari@icrr.u-tokyo.ac.jp}
\affiliation{Institute for Cosmic Ray Research, The University of Tokyo, 5-1-5  Kashiwanoha, Kashiwa, Chiba 277-8582, Japan}

\author[orcid=0000-0001-9626-9642]{Mahmoud Hamed}\email{mhamed@cab.inta-csic.es}
\affiliation{Centro de Astrobiolog\'ia (CAB), CSIC–INTA, Cra. de Ajalvir Km.~4, 28850- Torrej\'on de Ardoz, Madrid, Spain}

\author[]{Takuya Hashimoto}\email{hashimoto.takuya.ga@u.tsukuba.ac.jp}
\affiliation{Division of Physics, Faculty of Pure and Applied Sciences, University of Tsukuba, Tsukuba, Ibaraki 305-8571, Japan}
\affiliation{Tomonaga Center for the History of the Universe (TCHoU), Faculty of Pure and Applied Sciences, University of Tsukuba, Tsukuba, Ibaraki 305-8571, Japan}

\author[]{Akio Inoue}\email{akinoue@aoni.waseda.jp}
\affiliation{Waseda Research Institute for Science and Engineering, Faculty of Science and Engineering, Waseda University, 3-4-1 Okubo, Shinjuku, Tokyo 169-8555, Japan}
\affiliation{Department of Physics, School of Advanced Science and Engineering, Faculty of Science and Engineering, Waseda University, 3-4-1 Okubo, Shinjuku, Tokyo 169-8555, Japan}

\author[orcid=0000-0001-5710-8395]{Danial Langeroodi}\email{danial.langeroodi@nbi.ku.dk}
\affiliation{Kavli Institute for Cosmology, University of Cambridge, Madingley Road, Cambridge CB3 0HA, UK}
\affiliation{Cavendish Laboratory, University of Cambridge, 19 JJ Thomson Avenue, Cambridge CB3 0HE, UK}
\affiliation{DARK, Niels Bohr Institute, University of Copenhagen, Jagtvej 155A, 2200 Copenhagen, Denmark}

\author[]{Rui Marques-Chaves}\email{rui.marquescoelhochaves@unige.ch}
\affiliation{Geneva Observatory, Department of Astronomy, University of Geneva, Chemin Pegasi 51, CH-1290 Versoix, Switzerland}

\author[]{Ikki Mitsuhashi}\email{ikki0913astr@gmail.com}
\affiliation{Department for Astrophysical \& Planetary Science, University of Colorado, Boulder, CO 80309, USA}

\author[]{Yurina Nakazato}\email{ynakazato@flatironinstitute.org}
\affiliation{Center for Computational Astrophysics, Flatiron Institute, 162 5th Avenue, New York, NY 10010, USA}

\author[]{Borja~P\'{e}rez-D\'{\i}az}\email{borja.perezdiaz@inaf.it}
\affiliation{INAF – Osservatorio Astronomico di Roma, via Frascati 33, 00078, Monteporzio Catone, Italy}

\author[0000-0002-5104-8245]{Pierluigi Rinaldi}\email{prinaldi@stsci.edu}
\affiliation{Steward Observatory, University of Arizona, 933 North Cherry Avenue, Tucson, AZ 85721, USA}

\author[]{Carlota Prieto-Jim\'enez}\email{cprieto@cab.inta-csic.es}
\affiliation{Centro de Astrobiolog\'ia (CAB), CSIC–INTA, Cra. de Ajalvir Km.~4, 28850- Torrej\'on de Ardoz, Madrid, Spain}
\affiliation{Departamento de F\'{i}sica de la Tierra y Astrof\'{i}sica, Facultad de Ciencias F\'{i}sicas, Universidad Complutense de Madrid, E-28040, Madrid, Spain}

\author[0009-0005-7278-1541]{Ana Varo-O'Ferrall}\email{varooferrallana@gmail.com}
\affiliation{Centro de Astrobiolog\'ia (CAB), CSIC–INTA, Cra. de Ajalvir Km.~4, 28850- Torrej\'on de Ardoz, Madrid, Spain}
\affiliation{Departamento de F\'isica de la Tierra y Astrof\'isica, Facultad de Ciencias F\'isicas, Universidad Complutense de Madrid, E-28040, Madrid, Spain}

\author[0000-0001-6763-5869]{Fang-Ting Yuan}\email{yuanft@shao.ac.cn}
\affiliation{Shanghai Astronomical Observatory, Chinese Academy of Sciences, 80 Nandan Road, Shanghai 200030, People's Republic of China}

\author[]{Shuairu Zhu}\email{shuairuz@gmail.com}
\affiliation{Shanghai Astronomical Observatory, Chinese Academy of Sciences, 80 Nandan Road, Shanghai 200030, People's Republic of China}

\author[0000-0002-9634-2923]{Zhen-Ya Zheng}\email{zhengzy@shao.ac.cn}
\affiliation{Shanghai Astronomical Observatory, Chinese Academy of Sciences, 80 Nandan Road, Shanghai 200030, People's Republic of China}

\begin{abstract}
UV-bright galaxies in the early Universe provide important constraints on the rapid build-up of stellar mass and the sources responsible for cosmic reionization. However, the physical processes that drive their high UV luminosities remain unclear. 
Here we report JWST/MIRI LRS spectroscopy of GHZ8, a UV-bright galaxy at $z=10.23$ with a magnification $\mu=1.23$. The spectrum shows a $\sim 5\sigma$ detection for the \hb--\oiii\ complex and $\sim 4\sigma$ for \ha\ line, with fluxes of $3.2\pm0.9\times10^{-18}$ and $1.7\pm0.4\times10^{-18}$ erg s$^{-1}$ cm$^{-2}$, respectively. Based on JWST/NIRCam imaging, we find that GHZ8 consists of two components separated only by 0\farcs16 ($\approx600$\,pc): a bright, extended main component and a faint, compact southeastern clump, hereafter referred to as GHZ8-main and GHZ8-SE. Through morphological decomposition, we find demagnified \muv $=-19.91\pm0.10$ and $-18.03\pm0.15$ mag for GHZ8-main and GHZ8-SE, with blue UV slopes of \betauv $=-2.53\pm0.22$ and $-3.45\pm0.49$, respectively, indicating that both components host young stellar populations. In particular, the extremely blue UV continuum of GHZ8-SE requires an extremely young, ISM-poor stellar population with a high ionizing photon escape fraction, $f_{\rm esc}= 0.6-1.0$. 
Combining the UV and \ha\ luminosities with the results from the spectro-photometric SED fitting, we constrain the star-formation activity of GHZ8-main, and find either a recent, short-lived burst of star formation that is now in its final phases, or continuous star formation over a relatively short period ($<50\,{\rm Myr}$).
GHZ8 therefore provides observational evidence for bursty star formation in a UV-bright galaxy.
Globally, GHZ8 may be undergoing a merger-induced starburst, although deeper imaging or spatially resolved kinematic information is required for a more definitive interpretation.

\end{abstract}

\keywords{\uat{High-redshift galaxies}{734} --- \uat{Early universe}{435} --- \uat{Reionization}{1383} --- \uat{Galaxy mergers}{608} --- \uat{Ultracompact dwarf galaxies}{1734}}


o\section{Introduction}\label{sec:intro}


JWST deep-field and lensing-field imaging surveys have identified a large number of galaxy candidates at $z>10$ \citep{Finkelstein2023ApJ...946L..13F, Harikane2023ApJS..265....5H, Perez-Gonzalez2023ApJ...951L...1P, Robertson2023NatAs...7..611R, Donnan2023MNRAS.520.4554D, Castellano2023ApJ...948L..14C, Austin2023ApJ...952L...7A, Leung2023ApJ...954L..46L, Perez-Gonzalez2023ApJ...951L...1P, Finkelstein2024ApJ...969L...2F, Chemerynska2024MNRAS.531.2615C, Castellano2025A&A...704A.158C, Perez-Gonzalez2025ApJ...991..179P}, many of which have been subsequently confirmed by spectroscopic follow-up observations \citep{Curtis-Lake2023NatAs...7..622C, Fujimoto2023ApJ...949L..25F, Arrabal_Haro2023ApJ...951L..22A, Finkelstein2024ApJ...969L...2F, Hsiao2024ApJ...973...81H, Napolitano2025A&A...693A..50N,Zavala2025NatAs...9..155Z, Alvarez-Marquez2025A&A...695A.250A, Alvarez_Marquez2026arXiv260202323A, Naidu2026OJAp....956033N, Witstok2025Natur.639..897W,Tang2026ApJ..1001...38T, Carniani2025A&A...696A..87C, Hsiao2024ApJ...973....8H}. A subset of these galaxies shows very high UV luminosities, leading to an apparent excess at the bright end of the UV luminosity function \citep[e.g.,][]{Finkelstein2024ApJ...969L...2F, Casey2024ApJ...965...98C,Castellano2024,Harikane2025ApJ...980..138H, Chemerynska2024MNRAS.531.2615C}. This exceeds the expectations of pre-JWST models and poses a challenge to galaxy-formation theory \citep{Boylan-Kolchin2023}. Within the standard framework of hierarchical structure formation, producing a large population of UV-bright galaxies at such early cosmic times is not straightforward, and may involve the earlier dark-matter halo assembly, higher star-formation efficiency, and the lack of feedback \citep{Dekel2023MNRAS.523.3201D, Finkelstein2024ApJ...969L...2F, Harikane2023ApJS..265....5H, Li2024A&A...690A.108L}. 

Several physical mechanisms have been proposed to explain the population of UV-bright galaxies at $z>10$. One possibility is a bursty star-formation history (SFH), in which intense star formation over a short timescale temporarily boosts the UV luminosity and makes galaxies easier to detect \citep{Mirocha2023MNRAS.519..843M, Shen2023MNRAS.525.3254S, Sun2023ApJ...955L..35S, Mason2023MNRAS.521..497M, Sun2023MNRAS.526.2665S, Pallottini2023A&A...677L...4P, Casey2024ApJ...965...98C, Langeroodi2024arXiv240413045L}. Young, metal-poor stellar populations with less dust can also produce brighter UV continua and higher ionizing photon production efficiencies \citep{Ferrara2023MNRAS.522.3986F, Ferrara2024A&A...684A.207F, Ferrara2025A&A...694A.215F, Harikane2023ApJS..265....5H, Donnan2024MNRAS.533.3222D,Burgarella2026arXiv260509829B}. Binary stellar evolution, or a top-heavy IMF, can further enhance the hard ionizing spectrum and UV output \citep[e.g.,][]{Harikane2023ApJS..265....5H, Yung2024MNRAS.527.5929Y, Hutter2025A&A...694A.254H, Jeong2025ApJ...980...10J, Schaerer2025A&A...693A.271S}. These mechanisms are not mutually exclusive, and multiple processes may operate simultaneously within the same system.

Relying only on rest-frame UV broadband photometry makes it difficult to disentangle the degeneracies among stellar mass, age, dust attenuation, metallicity, nebular emission, and the escape fraction of ionizing photons. 
Multiwavelength observations combining imaging and spectroscopy are therefore critical for understanding the physical nature of UV-bright galaxies at $z>10$.

The PRImordial galaxy Survey with MIRI Spectroscopy at $z\sim10$ (PRISMS; program ID 8051; PIs: J. {\'A}lvarez-M{\'a}rquez \& L. Colina; \citealt{Alvarez_Marquez2026arXiv260202323A}) program obtains MIRI low-resolution spectroscopy (LRS) for a sample of UV-bright and -intermediate galaxies at $z\sim10$, with the aim of further constraining the physical properties of early galaxies. The target of this work, GHZ8, is part of the PRISMS program. It is a galaxy at $z=10.231$ in the Abell 2744 field with a magnification amplititude of $\mu=1.23$ (\citealt{Castellano2023ApJ...948L..14C,Napolitano2025A&A...693A..50N}). This source has rich JWST/NIRCam photometry and NIRSpec/prism spectroscopy, and its UV luminosity (\muv $=-20.32\pm0.07$, \citealt{Napolitano2025A&A...693A..50N}) lies in the intermediate-to-bright range among galaxies at $z>10$. It therefore provides a representative case study of the UV-bright galaxy population in the early Universe.

In this paper, we report JWST/MIRI LRS rest-frame optical spectroscopy of GHZ8 and combine it with existing NIRCam imaging and NIRSpec/prism spectroscopy for a joint analysis. In Section~\ref{sec:observation}, we describe the observations and data reduction. In Section~\ref{sec:linemeasurement}, we measure the \ha, \hb+\oiii, and rest-frame UV emission lines, and use line ratios and the Balmer decrement to constrain its properties. Section~\ref{sec:morpmeasurement} presents the image decomposition and the clumpy morphology of GHZ8, consisting of GHZ8-main and GHZ8-SE. In Section~\ref{sec:sedfitting}, we describe the configuration of the decomposed SED fitting. Sections \ref{sec:result_se} and \ref{sec:result_main} present the fitting results for GHZ8-SE and GHZ8-main, respectively. In Section~\ref{sec:discussion}, we discuss the physical origin of the clumpy structure of GHZ8 and its implications for ionizing photon escape. Finally, we summarize our conclusions in Section~\ref{sec:conclusion}. 
Throughout this paper, we adopted a flat $\Lambda$CDM cosmology with parameters of $H_0$ = 67.66 $\rm km\, s^{-1}\, Mpc^{-1}$ and $\Omega_m$ = 0.30966 \citep{Planck2020A&A...641A...6P}.

\section{Observation and Data Reduction}\label{sec:observation}
\subsection{MIRI/LRS}
GHZ8 was observed with PRISMS program \citep{Alvarez_Marquez2026arXiv260202323A}. The observations were obtained on 2025 October 31 UT using the MIRI LRS fixed-slit mode with the MIRIM\_SLIT aperture, MIRIMAGE detector, and P750L disperser. The data were taken with the FASTR1 readout pattern, 8 integrations, and 149 groups per integration, using a mosaic of two points plus a two-point along-slit nod patter that give four equidistant position in along the slit. The total effective exposure time of the combined Stage 3 spectrum is 26618.18 s.

The data were reduced following a procedure similar to that described in \citet{Alvarez_Marquez2026arXiv260202323A}. Briefly, we used the JWST calibration pipeline version 1.20.2, with CRDS version 13.0.6 and the reference context \texttt{jwst\_1464.pmap}. We also applied customized calibration steps, including wavelength masking, residual background subtraction, and sigma clipping. The final product used in this work is the Stage 3 extracted one-dimensional spectrum. Following \citet{Alvarez_Marquez2026arXiv260202323A}, we adopted the extracted one-dimensional background spectrum as the flux uncertainty spectrum.

The MIRI/LRS spectrum shows a marginal ($\sim1.6\sigma$) continuum detection over 4.8-5.2\,$\mu$m, where it overlaps with the NIRSpec/prism spectrum. The continuum flux level is offset from that of the prism spectrum. Using the corrected prism spectrum (Section~\ref{sec:data_nirspec}) as the reference, we rescale the LRS spectrum based on the integrated flux over the overlapping wavelength range. We obtain a correction factor of 1.19, which is applied to the full LRS spectrum. The corrected LRS spectrum, along with the NIRSpec and NIRCam data, is presented in Figure~\ref{fig:spec}. This spectrum is used for subsequent measurements of the rest-frame optical emission lines and in our spectro-photometric SED fitting analysis. We test the impact of this scaling factor in our results by alternatively using a spectrum without any correction factor (as in \citealt{Alvarez_Marquez2026arXiv260202323A}), and find that this correction leads to changes of approximately 10\% in the inferred physical parameters, including stellar mass and SFR. These changes are small compared with the measurement uncertainties and thus do not significantly affect our subsequent analysis and discussion. 

\subsection{NIRSpec/prism}\label{sec:data_nirspec}
GHZ8 was observed with NIRSpec/prism under GO program \#3073 (PI M. Castellano). Its prism spectrum is reduced with similar procedure as \cite{Castellano2026} and is first reported in \cite{Napolitano2025A&A...693A..50N}. Due to an offset between the prism spectral flux and the NIRCam photometry (Section~\ref{sec:data_nircam}), in practice, the spectrum is usually rescaled to match the photometry in order to ensure consistency between the spectroscopic and photometric data. We use the \texttt{integrate\_spectrum\_filter} function in \texttt{msaexp} to compute the synthetic photometry from the prism spectrum (i.e., spectrophotometric fluxes), and compare it with the observed fluxes in the five detected NIRCam bands: F150W, F200W, F277W, F356W, and F444W. We adopt the broadband photometry from the catalog released by Ultradeep NIRSpec and NIRCam Observations before the Epoch of Reionization (UNCOVER) program\footnote{https://jwst-uncover.github.io/\#} \citep{Suess2024ApJ...976..101S, Bezanson2024ApJ...974...92B}, and fit the ratio between the photometric fluxes and the spectrophotometric fluxes with a second-order polynomial. This gives correction factors ranging from 0.6 to 0.9. This polynomial correction is then applied to the NIRSpec spectrum. The corrected spectrum is used for the subsequent emission-line measurements and SED-fitting analysis.

\subsection{NIRCam Images}\label{sec:data_nircam}
We also use the imaging data retrieved from UNCOVER \citep{Suess2024ApJ...976..101S, Bezanson2024ApJ...974...92B}. Specifically, we used the mosaics release \texttt{abell2744clu-grizli-v7.2}, including \texttt{\_sci.fits} and \texttt{\_wht.fits} images in bands F090W, F115W, F150W, F200W, F277W, F365W, and F444W. These mosaics include data from programs GO-2561 (UNCOVER), ERS-1324 (GLASS), and DD-2767. 

\begin{figure*}
    \centering
    \includegraphics[width=1\linewidth]{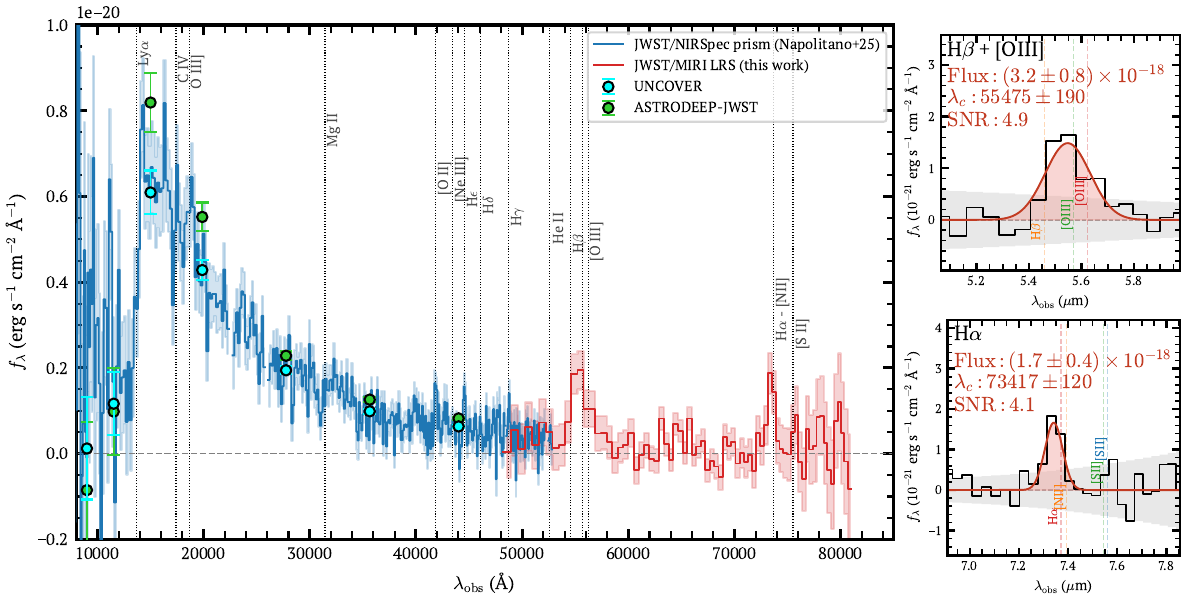}
    \caption{\textit{Left panel}: Rest-frame UV-to-optical SED of GHZ8. The blue and red curves show the JWST NIRSpec/prism spectrum from \cite{Napolitano2025A&A...693A..50N} and the MIRI/LRS spectrum, respectively. The cyan and green points show the broadband photometry from the UNCOVER and ASTRODEEP-JWST catalogs \citep{Suess2024ApJ...976..101S, Merlin2024A&A...691A.240M}. 
    \textit{Right panels}: Detections and best-fit models of the \hb--\oiii\ complex (top) and the H$\alpha$ emission line (bottom). The black and red curves are the observed data and fitting models, respectively. The gray shadow indicates the 1$\sigma$ error. Fluxes are in the unit of $\rm erg\,s^{-1}\,cm^{-2}$ and not corrected for the lensing magnification.
}
    \label{fig:spec}
\end{figure*}

\begin{figure*}
    \centering
    \includegraphics[width=1\linewidth]{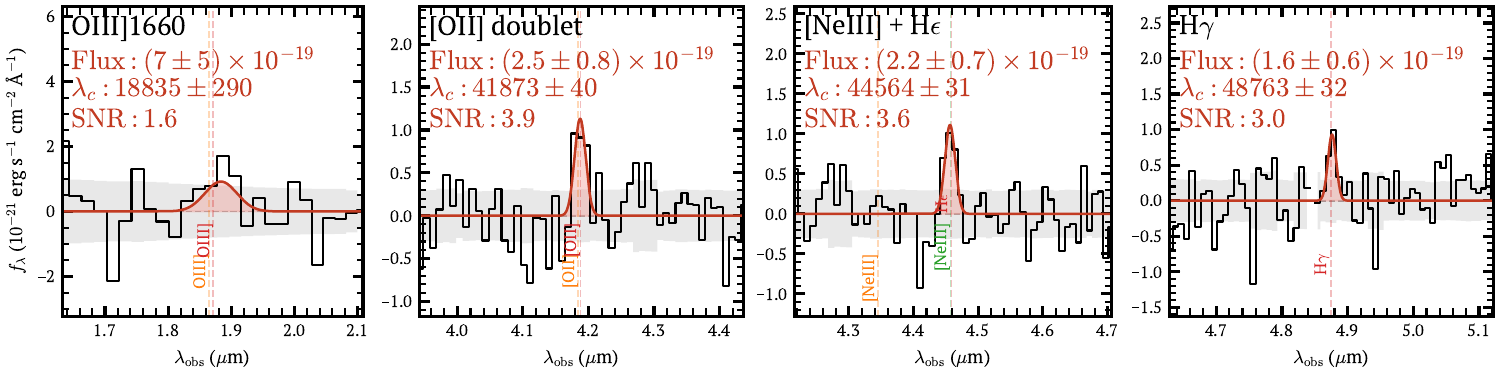}
    \caption{Rest-frame UV emission-line detections and best-fit models. The black and red curves show the continuum-subtracted spectra and fitted models, respectively. Gray shaded regions indicate the $1\sigma$ uncertainties. Fluxes are in the unit of $\rm erg\,s^{-1}\,cm^{-2}$ and not corrected for the lensing magnification.
    }
    \label{fig:uvlines}
\end{figure*}

\section{Emission lines and line-ratio diagnostics}\label{sec:linemeasurement}

\subsection{Line Measurement}
We detect the \oii $\lambda\lambda 3727,3730$ doublet, the \neiii$\lambda 3969$--\he\ blend, and \hg\ in the prism spectrum, consistent with \cite{Napolitano2025A&A...693A..50N}. From the LRS spectrum, we detect the unresolved \hb-\oiii\ complex at $5\sigma$ and the \ha\ line at $4\sigma$. For each emission line, we fit the local continuum with a first-order polynomial and subtract it before measuring the line fluxes. We then fit the emission lines with single Gaussian profile using a Bayesian approach, and estimate the parameter uncertainties through \texttt{emcee} \citep{Foreman-Mackey2013PASP..125..306F}. The emission-lines properties are presented in Figures~\ref{fig:spec} and ~\ref{fig:uvlines}, and summarized in Table~\ref{tab:ghz8_emission_line_table}. 

For the prism spectrum, the \oii\ doublet and the \neiii--\he\ complex cannot be spectrally resolved. To avoid making assumptions about the intrinsic line profiles, we fit each blended feature with a single Gaussian profile.

For the LRS spectrum,  we fit \ha\ and \hb--\oiii\ complex with a single Gaussian profile to measure its integrated flux. 
The redshift measured from the \ha\ line center is $10.18\pm0.02$, which is 0.05 lower than that measured from the prism spectrum. Since this difference does not affect the line flux measurements or our following analysis, we do not apply any additional correction throughout the paper.

We also examine the possibility of deblending the \hb--\oiii\ complex. For a given value of E(B-V), we use the best-fit \ha\ profile convolved with the LRS line-spread function,  and assume Case B recombination at $T=10^4$ K and $n_e=10^3\,\rm cm^{-3}$,  to constrain the \hb\ model. For the \oiii\ doublet, we fix the velocity difference, fix the flux ratio to \oiii$\lambda5008$/\oiii$\lambda4960 = 2.98$ \citep{Storey2000MNRAS.312..813S}, and force the line widths to be the same. However, the goodness of fit changes only weakly for different assumed values of E(B-V), with Bayesian information criterion (BIC) difference of $\rm \Delta BIC \leq 1$. This indicates that the current spectral resolution and signal-to-noise ratio (S/N) are insufficient to deblend the \hb--\oiii\ complex (see Appendix~\ref{app:hboiii_deblending} for details). 

\begin{table}[t]
\centering
\caption{Emission-line measurement and line ratios}
\label{tab:ghz8_emission_line_table}
\setlength{\tabcolsep}{0.5mm}
\begin{tabular}{lcc}
\hline\hline
\multirow{2}{*}{Name} & Flux & $EW_0$ \\
 & ($10^{-18}\,\mathrm{erg\,s^{-1}\,cm^{-2}}$) & (\AA) \\
 (1) & (2) & (3) \\
\hline
O\,{\sc iii}]$\lambda\lambda$1660,1666 & $0.70 \pm 0.53$ & $13.9 \pm 8.9$ \\
Mg\,{\sc ii} & $-0.57 \pm 0.26$ & $-29 \pm 11$ \\
\oii $\lambda\lambda$3726,3728 & $0.246 \pm 0.075$ & $32 \pm 16$ \\
\neiii $\lambda\lambda$3868 & $< 0.16$ & -- \\
\neiii $\lambda\lambda$3970 + \he & $0.223 \pm 0.065$ & $38 \pm 23$ \\
\hd & $< 0.13$ & -- \\
\hg & $0.157 \pm 0.060$ & $35 \pm 30$ \\
\hb\ + \oiii$\lambda\lambda$4959,5007 & $3.17 \pm 0.83$ & -- \\
\ha & $1.68 \pm 0.41$ & -- \\
\hline \hline
\multicolumn{3}{c}{ line-ratio lower limits (90\% confident level)} \\
\hline
log O32 & \multicolumn{2}{l}{$>0.78$} \\
log R3 & \multicolumn{2}{l}{$>0.53$} \\
log R2 & \multicolumn{2}{l}{$>-0.39$} \\
log R23 & \multicolumn{2}{l}{$>0.59$} \\
log Ne3O2 & \multicolumn{2}{l}{$>-0.53$} \\
\hline
\end{tabular}
\begin{tablenotes}
\item Note. \textit{Top}: Column (1): line name. Column (2): integrated line flux. Column (3): rest-frame equivalent width. For detections, flux uncertainties are quoted at the first significant digit of the 1$\sigma$ error. The EW$_0$ entries for \ha\ and \hb\ + \oiii$\lambda\lambda$4959,5007 are set to ``--'' due to the low S/N of continuum. \textit{Bottom}: definition of line ratios:
\item O32 $\equiv\ $ \oiii$\lambda \lambda 4960,5008$/\oii$\lambda \lambda 3727,3730$
\item R3 $\equiv\ $ \oiii$\lambda 5008$/\hb
\item R2 $\equiv\ $ \oii$\lambda \lambda 3727,3730$/\hb
\item R23 $\equiv\ $ (\oiii$\lambda \lambda 5008$+\oii$\lambda \lambda 3727,3730$)/\hb
\item Ne3O2 $\equiv\ $ \neiii$\lambda 3870$/\oii$\lambda \lambda 3727,3730$

\end{tablenotes}
\end{table}

\subsection{Balmer Decrement} \label{sec:balmder_de}
Combining the measurement of prism and LRS spectra, we can derive the Balmer decrement, of which the \ha/\hg\ ratio is $10.6_{-7.1}^{+18.3}$. We adopt a theoretical value of \ha/\hg = 6.10, calculated using \texttt{PyNeb} under Case B recombination with $T_{\rm e}=10^4\,{\rm K}$ and $n_{\rm e}=10^3\,{\rm cm^{-3}}$ \citep{Osterbrock2006agna.book.....O}. Using the \citep{Cardelli1989ApJ...345..245C} extinction curve for nebular, \cite{Calzetti2000ApJ...533..682C} attenuation law for stellar continuum, and $\rm E(B-V)_{star}/E(B-V)_{nebular}=0.44$, the corresponding emission-line color excess and dust attenuation are $\rm E(B-V)_{nebular}=0.37_{-0.10}^{+0.73}$ mag and $A_{V\rm ,\,star} = 0.65_{-0.17}^{+1.30}$ mag. Despite with large uncertainty, this dust attenuation is consistent with the inferred value derived from SED fitting (see Table~\ref{tab:ghz8_sedfit_params}) and comparable to high-$z$ UV-bright galaxies \citep{Carniani2025A&A...696A..87C,Kokorev2025,Zavala2025NatAs...9..155Z}.

\subsection{Line ratio}

Although \hb\ and \oiii\ cannot be deblended directly, the detection of \ha\ allows us to place an upper limit on the \hb\ flux by assuming zero dust attenuation and Case B recombination. We can then infer a lower limit on the total \oiii\ doublet flux. Given the theoretical \oiii $\lambda5008$/$\lambda4960$ ratio of 2.98, we obtain a lower limit on the \oiii $\lambda5008$ flux. Based on the MCMC samples from the line fitting, we therefore calculate 90\% lower limits on different line ratios, whose definitions and values are listed in Table~\ref{tab:ghz8_emission_line_table}.

\begin{figure*}
    \centering
    \includegraphics[width=1\linewidth]{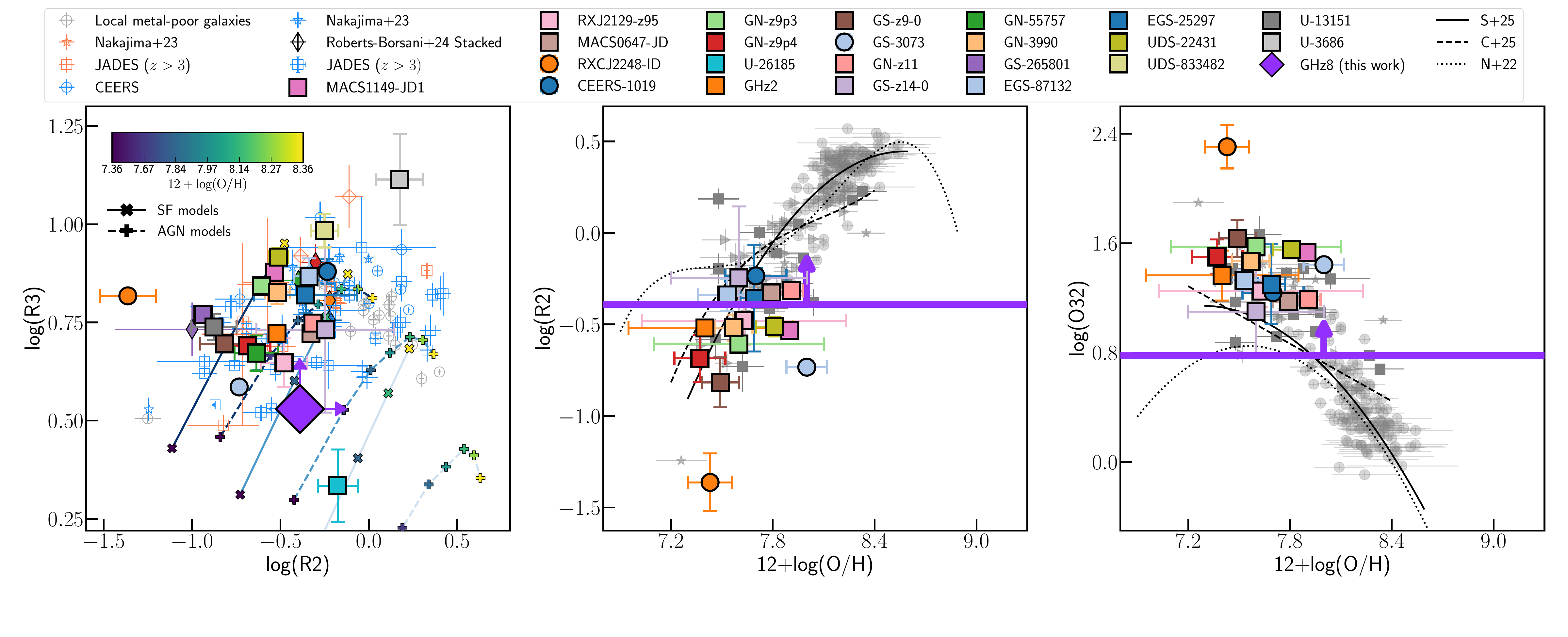}
    \caption{
    \textit{Left panel}: $\log R3$ versus $\log R2$ diagram. The purple diamond shows the lower limits on the line ratios of GHZ8. Other colored and gray markers show literature high-$z$ and N-rich samples compiled by \cite{Alvarez_Marquez2026arXiv260202323A} \footnote{References include \cite{Stiavelli2023ApJ...957L..18S, Williams2023Sci...380..416W, Hsiao2023ApJ...949L..34H, Boyett2024NatAs...8..657B, Schaerer2024A&A...687L..11S, Castellano2024, Zavala2025NatAs...9..155Z, Curti2025A&A...697A..89C, Bunker2023A&A...677A..88B, Alvarez_Marquez2025A&A...695A.250A, Helton2026ApJ..1004L..30H, Donnan2025ApJ...993..224D, Pollock2026A&A...708A.203P, Larson2023ApJ...953L..29L, Zamora2025arXiv251209022Z, Ji2024MNRAS.535..881J, Topping2024MNRAS.529.4087T, Cameron2023A&A...677A.115C, Bunker2024A&A...690A.288B, Curti2024A&A...684A..75C, Sanders2024ApJ...962...24S, Nakajima2023ApJS..269...33N, Roberts-Borsani2024ApJ...976..193R, Izotov2006A&A...448..955I, Calabro2023A&A...679A..80C}.} (see Section~4.3 in \cite{Alvarez_Marquez2026arXiv260202323A} for details). The colored solid and dashed curves show the AGN and star-forming galaxy model line ratios from \citet{Calabro2023A&A...679A..80C}, respectively. The curve colors indicate metallicity, while the color intensity from light to dark corresponds to increasing ionization parameter, with $\log U=[-3,-2.5,-2]$.
    \textit{Center panel}: $\log R2$ versus $12+\log({\rm O/H})$ diagram. The purple line marks GHZ8's lower limit. The solid, dashed, and dotted curves show the best-fit calibrations from \citet{Sanders2026ApJ..1003..228S}, \citet{Chakraborty2025ApJ...985...24C}, and \citet{Nakajima2023ApJS..269...33N}, respectively. Gray markers show local metal-poor galaxies, the galaxy sample from \citet{Nakajima2023ApJS..269...33N}, and JADES galaxies \citep{Cameron2023A&A...677A.115C, Bunker2024A&A...690A.288B, Curti2024A&A...684A..75C}.
    \textit{Right panel}: $\log O32$ versus $12+\log({\rm O/H})$ diagram. Markers and curves are the same as in the center panel.
    }
    \label{fig:line_ratio}
\end{figure*}

Figure~\ref{fig:line_ratio} compares GHZ8 with literature samples. The current lower limit on $\log \rm R3$ does not require an extreme ionization parameter (contrary to sources such as GHZ2;  \citealt{Castellano2026}); a moderate ionization parameter is sufficient. For reference, we also use the R2--$12+\log({\rm O/H})$ and O32--$12+\log({\rm O/H})$ relations to constrain the gas-phase metallicity. Inferred from relation of \citet{Sanders2026ApJ..1003..228S} and \citet{Chakraborty2025ApJ...985...24C}, the lower limit from $\log \rm R2$ gives approximately $12+\log({\rm O/H})\gtrsim7.22$, while the upper limit from $\log \rm O32$ gives $12+\log({\rm O/H})\lesssim7.93$. This corresponds to a metallicity range of $\sim0.03$--$0.2\,Z_\odot$. This range is similar to the gas and stellar metallicities inferred from SED fitting (Table~\ref{tab:ghz8_sedfit_params} and also see Section~\ref{sec:sedfitting}), suggesting that the metallicities adopted in the SED models are within a reasonable range.

\section{Morphology Characteristics} \label{sec:morpmeasurement}

\subsection{Morphology modeling and uncertainty determination} \label{sec:morphology_mea}
We perform simultaneous multi-band morphological fitting of the NIRCam images using \texttt{GALFITM}\footnote{https://www.nottingham.ac.uk/astronomy/megamorph/} \citep{Vika2013MNRAS.435..623V, Peng2002AJ....124..266P}. Since GHZ8 is not detected in F090W and F115W, we fit only the remaining five NIRCam bands. We first create $3\arcsec \times 3\arcsec$ cutouts centered on GHZ8, and reproject all images onto the WCS of the F356W image. For F150W and F200W, which have a pixel scale smaller by a factor of two than the longer-wavelength bands, we first resample the images to a common scale of $0.04\arcsec$ pixel$^{-1}$.

We construct the sigma images from the weight maps and use them to mask bad pixels. Other sources in the field of view are masked using the segmentation map. Since the images are in units of 10 nJy, we set \texttt{EXPTIME}$ = 1$ and adopt a magnitude zero point of \texttt{ZP}$=28.9$, following the flux convention used by \texttt{GALFITM}. This ensures a consistent conversion between the fitted fluxes and AB magnitudes. For each band, we estimate the sky background from nearby source-free regions. The sky level is included in \texttt{GALFITM} as an independent component but with a fixed value for each band. We checked that the total flux of two components is consistent with the published photometry in the UNCOVER catalog.

We use pre-made simulated PSF models generated with \texttt{STPSF} \footnote{https://www.stsci.edu/jwst/science-planning/proposal-planning-toolbox/psf-simulation-tool} \citep{Perrin2014SPIE.9143E..3XP}. As for the science images, the PSFs in F150W and F200W are resampled to match the pixel scale of the longer-wavelength bands.

We adopt the model-based photometry from \texttt{GALFITM}. However, since \texttt{GALFITM} does not account for the uncertainties associated with background subtraction and correlated background noise, the photometric uncertainties are likely underestimated. To take the correlated noise into account, we first create a segmentation map to mask all detected sources within the stamp. We then randomly place 500 blank apertures in the source-free regions to science images and calculate the background rms, denoted as $\sigma_{\rm blank}$. To quantify the uncertainty introduced by background subtraction, we repeatedly compute the background fluxes with annuli of different sizes centered on the source and then estimate the standard deviation of these measurements, denoted as $\sigma_{\rm bg}$. The final photometric uncertainty is then computed as
\begin{equation}
\sigma_{\rm final} = \sqrt{\sigma_{\rm GALFITM}^2 + \sigma_{\rm blank}^2 + \sigma_{\rm bg}^2}.
\end{equation}
The decomposed photometric uncertainties reported in Table \ref{tab:ghz8_galfitm_vertical} correspond to $\sigma_{\rm final}$.

\begin{table}
\centering
\caption{Morphology parameters and decomposed SEDs}
\label{tab:ghz8_galfitm_vertical}
\setlength{\tabcolsep}{1mm}
\begin{tabular}{lcc}
\hline \hline
Parameter & GHZ8-main & GHZ8-SE \\
\hline
R.A. (deg) & 3.451417 & 3.451436 \\
Dec. (deg) & -30.321794 & -30.321836 \\
$M_{\rm UV}$ (mag)$^a$ & -19.91 $\pm$ 0.10 & -18.03 $\pm$ 0.15 \\
$\beta_{\rm UV}$ & -2.53 $\pm$ 0.22 & -3.45 $\pm$ 0.49 \\
Profile & Sersic & point source \\
$n_{\rm Sersic}$ & 1.71 $\pm$ 0.08 & -- \\
$R_{\rm e,F150W}$(pc)$^b$ & $311\pm16\,(0.082\pm0.004\,\arcsec)$ & $\leq 96\,(0.026\,\arcsec)$ \\
$R_{\rm e,F200W}$(pc)$^b$ & $355\pm16\,(0.094\pm0.004\,\arcsec)$ & $\leq 127\,(0.034\,\arcsec)$ \\
$R_{\rm e,F277W}$(pc)$^b$ & $427\pm18\,(0.113\pm0.005\,\arcsec)$ & $\leq 168\,(0.044\,\arcsec)$ \\
$R_{\rm e,F356W}$(pc)$^b$ & $498\pm21\,(0.132\pm0.005\,\arcsec)$ & $\leq 220\,(0.058\,\arcsec)$ \\
$R_{\rm e,F444W}$(pc)$^b$ & $574\pm25\,(0.152\pm0.007\,\arcsec)$ & $\leq 274\,(0.073\,\arcsec)$ \\
$m_{\rm F150W}$ (mag) & 27.43 $\pm$ 0.10 & 29.30 $\pm$ 0.15 \\
$m_{\rm F200W}$ (mag) & 27.10 $\pm$ 0.06 & 29.31 $\pm$ 0.11 \\
$m_{\rm F277W}$ (mag) & 27.29 $\pm$ 0.05 & 29.84 $\pm$ 0.14 \\
$m_{\rm F356W}$ (mag) & 27.40 $\pm$ 0.04 & 30.11 $\pm$ 0.16 \\
$m_{\rm F444W}$ (mag) & 27.31 $\pm$ 0.05 & 30.56 $\pm$ 0.25 \\
\hline
\multicolumn{3}{c}{projected distance between GHZ8-main and GHZ8-SE} \\ 
\hline
$\delta D ^b$ & \multicolumn{2}{c}{0.16\arcsec (601\, pc)} \\ 
\hline
\end{tabular}
\begin{tablenotes}
    \item Note. $^a\ M_{\rm UV}$ is corrected for magnification based on $M_{\rm UV, int} = M_{\rm UV, obs} + 2.5\, \log(\mu)$, where $\mu=1.23$.  $^b$ Length scales given in pc are corrected for lensing magnification by dividing by $\sqrt{\mu}$, while those given in arcseconds are uncorrected.
\end{tablenotes}
\end{table}

\begin{figure*}
    \centering
    \includegraphics[width=1\linewidth]{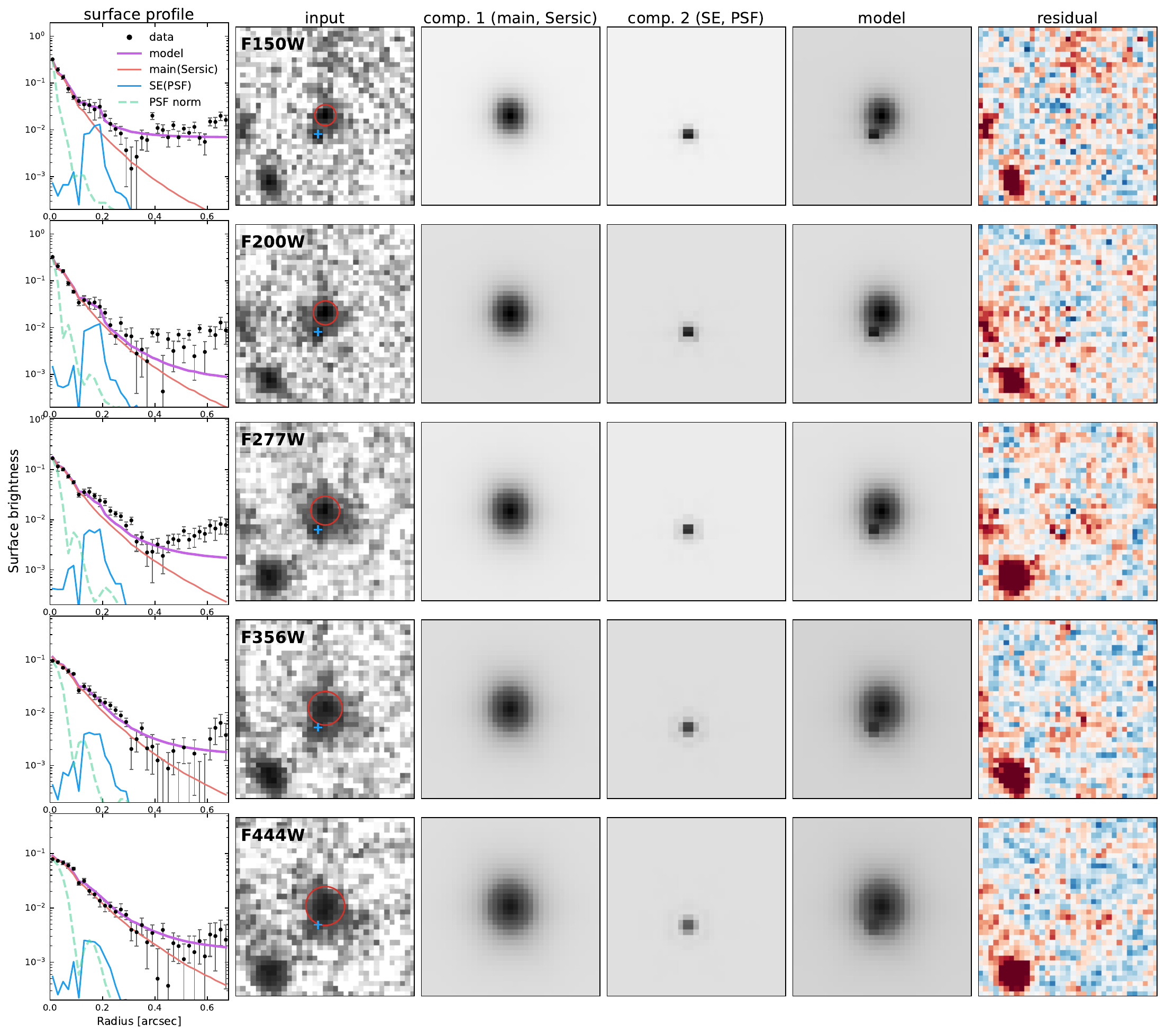}
    \caption{Morphological decomposition of GHZ8 into the main and southeastern (SE) components. From left to right, the columns show the surface-brightness profile, observed image, main-component model, SE-component model, total model, and residual image. In the observed images, the main and SE components are indicated by the red circle and cross, respectively. Different rows correspond to different NIRCam bands.
}
    \label{fig:galfitm_result}
\end{figure*}

\begin{figure*}
    \centering
    \includegraphics[width=0.95\linewidth]{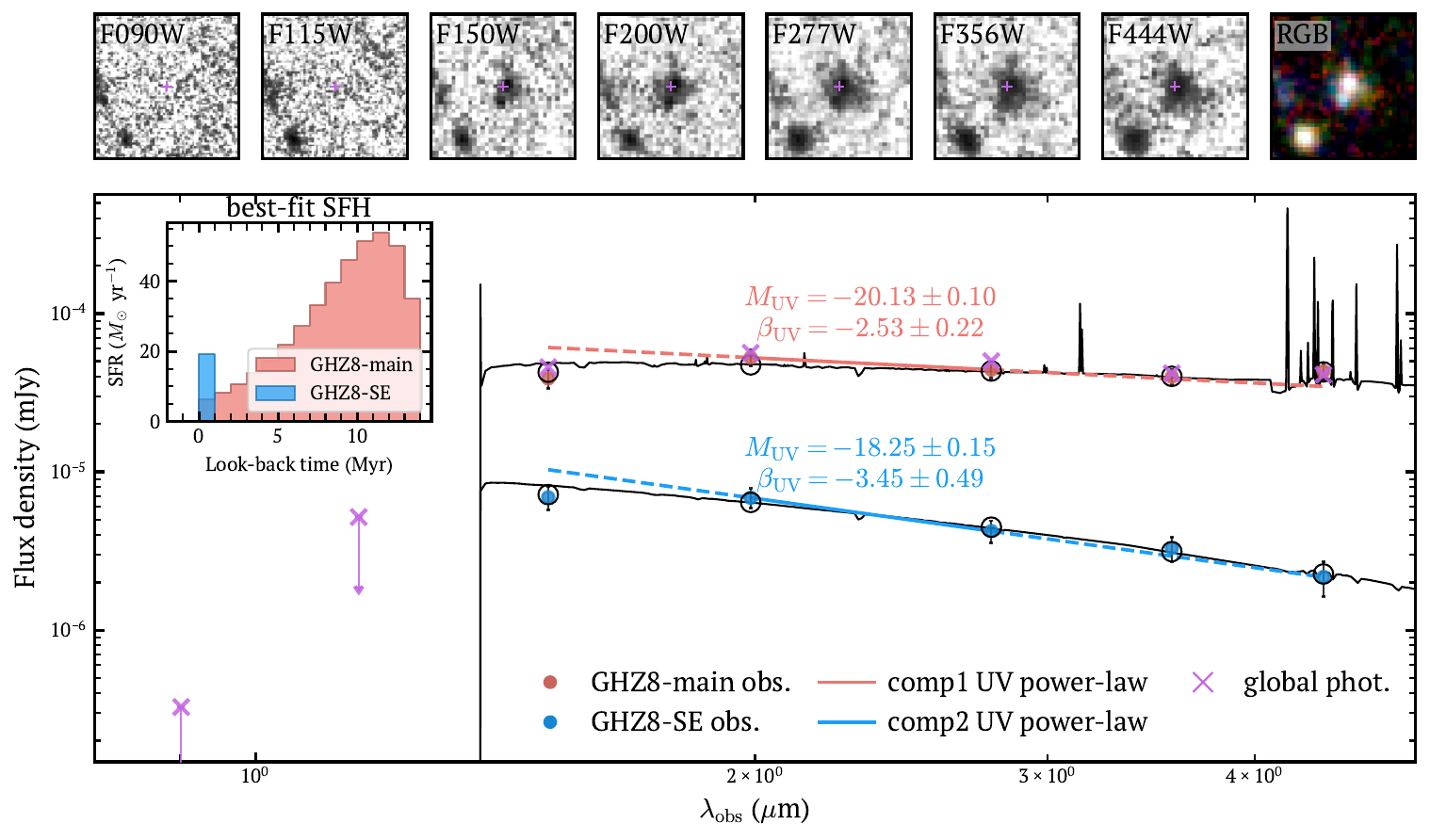}
    \caption{\textit{Top panel}: Images of GHZ8, with the purple cross marking its position in the public catalog. The RGB image is F200W-F277W-F444W color image. The cutout size is 1.4\arcsec $\times$ 1.4\arcsec.
    \textit{Bottom panel}: Decomposed SEDs of the GHZ8-main and SE components and their best-fit BC03 models. The red and blue points show the \texttt{GALFITM} decomposed photometry for the main and SE components, respectively. The red and blue curves show the power-law continua derived from their UV slopes. The UV slopes are measured from the F200W and F277W fluxes, so the line segment between these two bands is solid, while the extrapolated parts are dashed. The upper-left inset shows the best-fit SFH. 
}
    \label{fig:decomposed_sed}
\end{figure*}

\subsection{Clumpy Morphology: GHZ8-main and GHZ8-SE}
Figure~\ref{fig:galfitm_result} presents the morphological decomposition of GHZ8. The galaxy shows a morphology consisting of two components: a main component (\textit{GHZ8-main}) and a southeastern (\textit{GHZ8-SE}) component. The projected separation between the two components is $0.16\arcsec$, corresponding to a demagnified distance of $\sim 600 \rm \, pc$. The results of the multi-band morphological decomposition are listed in Table~\ref{tab:ghz8_galfitm_vertical}. 

To rule out the possibility that the fainter GHZ8-SE component is a foreground system at a very different redshift, we perform a photometric redshift fitting individually for each component. We first derived the detection upper limits in the F090W and F115W at their corresponding positions. For GHZ8-main we use the F150W-band profile derived from \texttt{GALFITM} and for GHZ8-SE we use the profile of F150W-band PSF. The uncertainties are measured using the same method as described above. 
We then run EA{\sc z}Y \citep{Brammer2008ApJ...686.1503B} with decomposed photometry and upper limits in the seven bands. This gives $z_{\rm phot, main}=10.42_{-0.26}^{+0.26}$ for GHZ8-main and $z_{\rm phot, SE}=10.04_{-0.37}^{+0.5}$ for GHZ8-SE. These results are consistent, within the uncertainties, with the inferred spectroscopic redshift, indicating that the two components are most likely at similar redshifts. 

We also estimate the probability of finding an unrelated galaxy within an angular radius $\theta=0\farcs16$. That can be estimated as $P \simeq \Sigma(M_{\rm UV}<-18,z\sim10)\,\pi\theta^2$, where $\Sigma$ is the surface density of galaxies at comparable redshifts and luminosities. 
The corresponding search area is only $\pi(0\farcs16)^2 \simeq 2.2\times10^{-5}\,{\rm arcmin}^2$. 
We consider the redshift difference between GHZ8-SE and GHZ8-main is $\pm0.38$ estimated by EA{\sc z}Y, corresponding to $\sim 14$ pMpc at $z=10.231$. Thus, the surface density of $z\sim10$ galaxies with $M_{\rm UV}\lesssim -18$ is $\rm 2.8 \times 10^{-4}\, arcmin^{-2}$ \citep{Harikane2025ApJ...980..138H}. Even if there is evidence showing A2744 field is an overdensity region, the surface number density is unlikely to be enhanced by more than an order of magnitude \citep{Castellano2023ApJ...948L..14C}, and a larger redshift difference up to 0.57 does not dramatically increase the surface density as well. Therefore, the chance-alignment probability would remain of order $10^{-8}-10^{-9}$. This resulting the probability of a random projection within $0\farcs16$ is extreme low and is unlikely to explain the observed configuration.
Given their small angular separation and the extremely low probability of a chance projection, the two components are almost certainly physically associated.

Based on the spectroscopic redshift,  we then estimate their \muv\ based on the F150W band (corresponding to rest-frame $\sim 1340\,$\AA) and the UV continuum slope \betauv, defined as $f_\lambda \propto \lambda^\beta$ and calculated using F200W and F277W photometry.
Although \betauv\ varies slightly depending on its definition, we have checked that this does not affect our results (see Appendix~\ref{app:betadefinition} for details).  

GHZ8-main is bright and spatially extended, with demagnified \muv $=-19.91\pm0.10$ mag and effective radius of $r_e\sim 310-580\, \rm pc$. It is well described by a \sersic\ profile, with a \sersic\ index of $n=1.71\pm0.08$ and an axis ratio of $q=0.82\pm0.02$. Its light profile is relatively symmetric.
In contrast, GHZ8-SE is fainter (with demagnified $M_{\rm UV}=-18.03\pm0.15$ mag) with a very compact and unresolved morphology.
We constraint its radius upper limit as half of PSF full width of half maximum (FWHM$_{\rm PSF}$/2) in the F150W band \citep{Mowla2024Natur.636..332M}, which gives $r_e \lesssim 100\, \rm pc$. It exhibits a bluer UV continuum slope of \betauv $=-3.45\pm 0.49$, compared to GHZ8-main of \betauv $=-2.53\pm0.22$. The extreme properties of GHZ8-SE are discussed in detail in Section~\ref{sec:result_se}.

\begin{table*}
\centering
\small
\caption{GHZ8-main, GHZ8-SE, and global parameters inferred from SED fitting.}
\label{tab:ghz8_sedfit_params}
\setlength{\tabcolsep}{2pt}
\begin{tabular}{llcccccc}
\hline \hline
\multirow{2}{*}{Parameter} & \multirow{2}{*}{Unit} & \multicolumn{2}{c}{GHZ8-main} & \multicolumn{2}{c}{GHZ8-SE} & \multicolumn{2}{c}{Global} \\
 & & BC03 & BPASS & BC03 & BPASS & BC03 & BPASS \\
\hline

Age & Myr & $44 \pm 29$ & $43 \pm 30$ & $6 \pm 5$ & $4 \pm 3$ & $35 \pm 4$ & $39 \pm 28$ \\
$\log M_* \, ^\dagger$ & $\rm M_\odot$ & $8.6 \pm 0.2$ & $8.6 \pm 0.3$ & $7.1 \pm 0.2$ & $6.9 \pm 0.2$ & $8.6 \pm 0.1$ & $8.6 \pm 0.3$ \\
$\log \Sigma_{M_*} \, ^\dagger$ & $\rm M_\odot\,{\rm pc}^{-2}$ & $2.8 \pm 0.2$ & $2.8 \pm 0.3$ & $\gtrsim 2.3$ & $\gtrsim 2.1$ & -- & -- \\
$\rm E(B-V)_{\rm nebular}$ & -- & $0.23 \pm 0.10$ & $0.15 \pm 0.08$ & $0.03 \pm 0.05$ & $0.03 \pm 0.05$ & $0.20 \pm 0.02$ & $0.10 \pm 0.08$ \\
$Z_{\rm gas}$ & $Z_\odot$ & $0.06 \pm 0.08$ & $0.03 \pm 0.06$ & $0.06 \pm 0.08$ & $0.06 \pm 0.08$ & $0.02 \pm 0.02$ & $0.04 \pm 0.06$ \\
$Z_{\rm *}$ & $Z_\odot$ & $0.10 \pm 0.09$ & $0.10 \pm 0.08$ & $0.06 \pm 0.08$ & $0.08 \pm 0.07$ & $0.02 \pm 0.01$ & $0.10 \pm 0.09$ \\
$\log U$ & -- & $-2.1 \pm 0.5$ & $-2.1 \pm 0.5$ & $-2.2 \pm 0.6$ & $-2.2 \pm 0.6$ & $-2.8 \pm 0.3$ & $-2.1 \pm 0.5$ \\
$f_{\rm esc}$ & -- & $0^a$ & $0^a$ & $0.8 \pm 0.2$ & $0.9 \pm 0.1$ & $0^a$ & $0^a$ \\
$\log Q_H \, ^\dagger$ & Hz$\rm \,erg^{-1}$ & $54.3 \pm 0.2$ & $54.2 \pm 0.2$ & $53.7 \pm 0.2$ & $53.7 \pm 0.2$ & $54.3 \pm 0.1$ & $54.2 \pm 0.2$ \\
SFR$^{\dagger \, b} $ & $M_\odot\,{\rm yr}^{-1}$ & $8.2 \pm 3.6$ & $5.7 \pm 3.0$ & $5.2 \pm 4.8$ & $4.9 \pm 3.6$ & $6.6 \pm 1.2$ & $4.3 \pm 2.4$ \\
$\Sigma_{\rm SFR} \,^\dagger$ & $M_\odot\,{\rm yr}^{-1}\,{\rm kpc}^{-2}$ & $13.4 \pm 5.9$ & $9.3 \pm 5.0$ & $\gtrsim 89.1$ & $\gtrsim 85.5$ & -- & -- \\
$\log {\rm sSFR} \,^\dagger$ & $\rm Gyr^{-1}$ & $1.3 \pm 0.3$ & $1.1 \pm 0.3$ & $2.7 \pm 0.4$ & $2.8 \pm 0.4$ & $1.2 \pm 0.1$ & $1.1 \pm 0.4$ \\
$f_{\rm H\beta+[OIII]}$ & $10^{-5}\,{\rm mJy}$ & $8.4 \pm 2.2$ & $8.3 \pm 1.8$ & $0.3 \pm 0.2$ & $0.3 \pm 0.2$ & $5.9 \pm 0.9$ & $8.0 \pm 1.9$ \\
$F_{\rm H\alpha}$ & $10^{-18}\,{\rm erg\,s^{-1}\,cm^{-2}}$ & $1.50 \pm 0.35$ & $1.44 \pm 0.34$ & $0.05 \pm 0.05$ & $0.04 \pm 0.05$ & $1.41 \pm 0.16$ & $1.36 \pm 0.32$ \\
$F_{\rm H\gamma}$ & $10^{-19}\,{\rm erg\,s^{-1}\,cm^{-2}}$ & $1.81 \pm 0.39$ & $1.95 \pm 0.43$ & $0.08 \pm 0.09$ & $0.07 \pm 0.08$ & $1.78 \pm 0.19$ & $2.01 \pm 0.46$ \\
$\beta_{\rm UV}$ & -- & $-2.28 \pm 0.17$ & $-2.30 \pm 0.17$ & $-2.97 \pm 0.17$ & $-2.95 \pm 0.19$ & $-2.38 \pm 0.03$ & $-2.41 \pm 0.17$ \\
$\beta_{\rm UV, C94}$ & -- & $-2.24$ & $-2.23$ & $-3.06$ & $-3.11$ & $-2.35$ & $-2.42$ \\
$M_{\rm UV} \,^\dagger$ & mag & $-20.13 \pm 0.07$ & $-20.11 \pm 0.06$ & $-18.01 \pm 0.07$ & $-17.99 \pm 0.08$ & $-20.29 \pm 0.05$ & $-20.25 \pm 0.06$ \\
\hline
\end{tabular}
\begin{tablenotes}
\item Notes: The listed parameters are Bayesian estimates from CIGALE with BC03 and BPASS (binary) models. The UV slope $\beta_{\rm UV}$ is measured from the model photometry in F200W and F277W, assuming $f_\lambda \propto \lambda^\beta$, following the same way as for the observed photometry. $\beta_{\rm UV, C94}$ is measured from the best-fit model using the UV slope definition of \cite{Calzetti1994ApJ...429..582C}. $M_{\rm UV}$ is estimated from the best-fit model spectrum over rest-frame 1300--1800 \AA. The SFR surface density is computed as $\Sigma_{\rm SFR} = {\rm SFR}/(2\pi R_e^2)$ using the GHZ8-main $R_e$ value in F150W band measured by \texttt{GALFITM}; the GHZ8-SE entry is reported as an lower limit using PSF $R_e$ in F150W band. Missing quantities are shown as --. $^\dagger$ parameters are corrected for magnification. $^a$ $f_{\rm esc}$ is fixed to zero. $^b$ Instantaneous SFR.
\end{tablenotes}
\end{table*}

\section{Decomposed SED fitting} \label{sec:sedfitting}
Based on the decomposed photometry from \texttt{GALFITM}, we separately perform SED fittings for GHZ8-main and GHZ8-SE, using Code Investigating GALaxy Emission (\texttt{CIGALE}\footnote{https://cigale.lam.fr} \citealt{Boquien2019A&A...622A.103B, Burgarella2025A&A...699A.336B, Burgarella2026arXiv260509829B}).

\subsection{CIGALE configuration for GHZ8-SE}\label{sec:sedfitting_se}
GHZ8-SE's extremely blue UV slope places strong constraints on the stellar age and nebular properties. 
Using the \texttt{CIGALE} (\texttt{2025.v1} version), we adopt a delayed-$\tau$ SFH model, with the age and $\tau_{\rm main}$ of the main stellar population allowed to vary between 1 and 20 Myr. We use the BC03 stellar population models \citep{Bruzual2003MNRAS.344.1000B} with metallicities of 0.0001, 0.0004, and 0.004.
We also perform a separate fit using BPASS models \citep{Stanway2018MNRAS.479...75S}, to test the impact of including models with the binary stellar evolution. For the BPASS models, we adopt an IMF with a slope of $\alpha=-1.30$ over the mass range $\rm 0.1 - 0.5\,M_\odot$ and $\alpha=-2.35$ over $\rm 0.5-300\,M_\odot$, with an upper stellar mass limit of $\rm 300\,M_\odot$. The stellar metallicity is allowed to take values of 0.00001, 0.0001, 0.001, 0.002, 0.003, and 0.004.

Nebular emission is included, with the ionization parameter $\log U$ ranging from -3.0 to -1.5 and electron density of $n_e = 100$ and $10^3 \, \rm cm^{-3}$. 
For dust attenuation, we adopt the \cite{Calzetti2000ApJ...533..682C} attenuation law for stellar emission and the Milky Way extinction curve \citep{Cardelli1989ApJ...345..245C} for nebular emission, with E(B-V)$_{\rm nebular}$ allowed to vary from $10^{-4}$ to 0.50. 

Finally, we allow the escape fraction to vary from 0 to 1. This is motivated by the extremely blue UV slope of GHZ8-SE, which suggests that the stellar population is very young and that the nebular continuum contribution is weak, also naturally implying a high escape fraction \citep{Zackrisson2013ApJ...777...39Z}. We also verify that, when the escape fraction is fixed to zero, the reduced $\chi^2$ is larger than 2 for both the BC03 and BPASS SSP models, significantly worse than the fits in which the escape fraction is allowed to vary (see Appendix~\ref{app:frac_se} for details).

\subsection{CIGALE configuration for GHZ8-main}\label{sec:sedfitting_main}
In the case of  GHZ8-main, we perform a spectro-photometric SED fitting including the deblended NIRCam photometry (\S\ref{sec:morphology_mea}) and the emission lines from the NIRSpec and MIRI spectra (\S\ref{sec:linemeasurement}) assuming the lines are mainly contributed by GHZ8-main. This assumption is based on two considerations: first, as mentioned above, GHZ8-SE shows a very blue UV continuum suggestive of a scenario with a high ionizing photon escape fraction suppressing nebular emission \citep{Zackrisson2013ApJ...777...39Z}; second, 
GHZ8-SE is around 5--20 times fainter than GHZ8-main, and thus, if the two components have similar emission-line equivalent widths, then the expected line flux contribution from GHZ8-SE would be smaller than the current measurement uncertainties (even if the escape fraction is low).   

The spectro-photometric SED fitting is thus preferred since it  improves the constraints on the SFR and SFH. During this procedure, the age of the main stellar population allowed to vary between 10 and 100 Myr, as GHZ8-main has redder UV color than GHZ8-SE. We also allow a recent burst with $\rm age_{\rm burst} \in$ [1, 3, 5] Myr and \texttt{$\tau_{\rm burst}$} $\in$ [1, 3, 5] Myr. We adopt the same model grid of other parameters as for GHZ8-SE, except that the escape fraction of GHZ8-main is fixed to zero. This choice is  motivated by the fact that the strong observed nebular emission and typical dust attenuation is observed in GHZ8-main, indicating that most of its ionizing photons are absorbed locally by the surrounding gas rather than escaping.

\begin{figure*}
    \centering
    \includegraphics[width=0.49\linewidth]{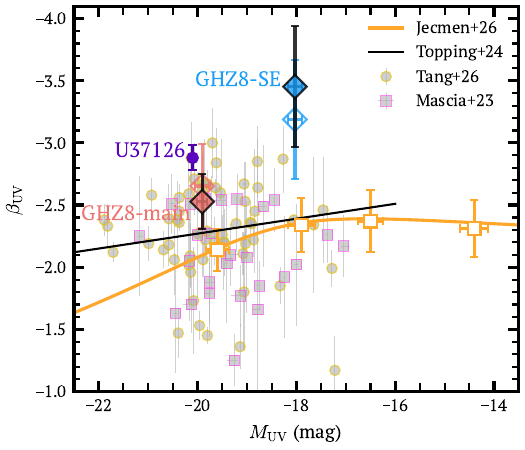}
    \includegraphics[width=0.49\linewidth]{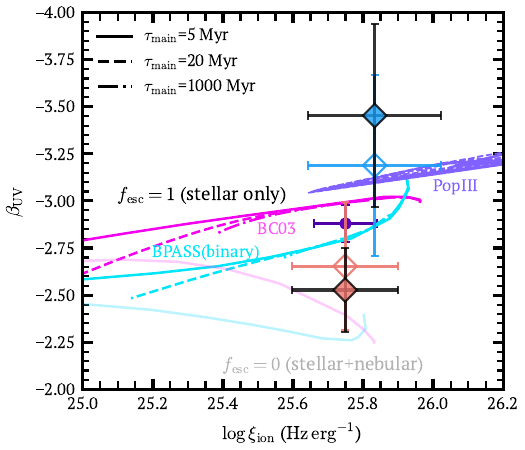}
    \caption{\textit{Left panel}: UV continuum slope versus $M_{\rm UV}$ diagram. Red and blue diamonds denote the main and SE components, respectively. Filled diamonds show the values measured from the decomposed SEDs (see Section~\ref{sec:result_se}), while open diamonds show those measured directly from the images (see Appendix~\ref{app:colormap}). Literature spectroscopic samples presented in \cite{Tang2026ApJ..1001...38T} and \cite{Mascia2023A&A...672A.155M} are also shown for comparison. Black line shows the relation of \cite{Topping2024MNRAS.529.4087T} for photometric sample in $8.5<z_{\rm phot}<11$ bin. Orange curve and orange squares indicate the photometric sample at $z_{\rm pho}\gtrsim 6$ reported in \cite{Jecmen2026ApJ..1001..234J}. 
    \textit{Right panel}: UV slope versus ionizing photon production efficiency. Pink, cyan, and purple tracks indicate the BC03, BPASS binary, and Pop III models with $\log Z_*/Z_\odot = -2.3$, respectively. The model tracks cover stellar ages of 1--100 Myr and assume delayed-$\tau$ SFHs with $\tau_{\rm main} = 5$, 20, and 1000 Myr. The $\tau_{\rm main}$ of 5 and 20 Myr cases represent bursty SFHs and $\tau_{\rm main}$ of 1000 Myr approximate a constant SFH. Light pink and light cyan tracks show the corresponding models including nebular emission.
}
    \label{fig:Muv_slope}
\end{figure*}

\section{GHZ8-SE: Extremely young, lack of ISM, compact pure stellar scenario}\label{sec:result_se}

The most striking property of GHZ8-SE is its extremely blue UV continuum. Based on the photometry obtained from the morphological decomposition, GHZ8-SE has an demagnified \muv $=-18.21\pm0.15$ mag and a UV slope of \betauv $=-3.45\pm0.49$. Although GHZ8-SE is about 2 mag fainter than the main component, its UV continuum is significantly bluer. As seen in Figure~\ref{fig:Muv_slope}, even accounting for the measurement uncertainty, GHZ8-SE lies among the bluest known high-$z$ galaxies in the \muv--\betauv\ diagram. It is comparable or even bluer than the extremely blue galaxies with \betauv$<-3$ at $z\gtrsim 9$ \citep{Topping2022ApJ...941..153T,Topping2024MNRAS.529.4087T, Yanagisawa2025ApJ...988...86Y, Jecmen2026ApJ..1001..234J, Cullen2025MNRAS.540.2176C}.

Such an extremely blue UV slope places strong constraints on the physical properties of GHZ8-SE (see \citealt{Inoue2011MNRAS.415.2920I} Figure 11). First, GHZ8-SE is very young. For both the BC03 and BPASS SSP models, the best-fit SED of GHZ8-SE points to an extremely young stellar population. As seen in the inset panel of Figure~\ref{fig:decomposed_sed}, the best-fit SFH indicates a dominant stellar age of $\sim 1\,{\rm Myr}$, while the MCMC posterior is mainly distributed over $1-10\,{\rm Myr}$. 

Second, GHZ8-SE may lack of surrounding ISM, as the extremely blue UV slope requires deficiency of nebular continuum emission \citep{Zackrisson2013ApJ...777...39Z}. For comparison with stellar models, we derive $\xi_{\rm ion}$-\betauv\ tracks with BC03, BPASS (binary), and Pop III models in Figure~\ref{fig:Muv_slope} using delayed-$\tau$ SFH with different delayed timescale $\tau_{\rm main}$ (5, 20, 1000 Myr) of main population with a metallicity of $\log Z_*/Z_\odot = -2.3$. Compare to other stellar models, GHZ8-SE lies closer to the stellar-only tracks than to the tracks including nebular emission. A interpretation is that the gas covering fraction around GHZ8-SE is low, or that the gas column density is low at least along the line of sight, allowing a large fraction of ionizing photons to escape \citep{Marques_Chaves2026A&A...711A.301M}, which is also self-consistent with the best-fit escape fraction of $f_{\rm esc}\sim0.6-1.0$ given by the SED fitting (see Section~\ref{sec:lycescape} for more details).
We note that  the UV slope computed from the best-fit model is slightly redder than that measured directly from the decomposed photometry, but it remains extremely blue. This difference is not caused by a different definition of \betauv\, since the model-based \betauv\ is calculated using the same bands as the direct measurement. But it more likely reflects the difficulty of current SED models in reproducing such an extremely blue continuum. 

Third, it has strong ionizing photon production efficiency $\xi_{\rm ion}$, that is also self-consistent with its young age, compact size ($r_e <100\,\rm pc$), high specified SFR ($\rm sSFR \sim 500\,\rm Gyr^{-1}$), and high SFR surface density ($\Sigma_{\rm SFR} \gtrsim 80\, \rm M_\odot\,yr^{-1}\,kpc^{-2}$). Its inferred $\log \xi_{\rm ion} = 25.83\pm0.19$ is comparable to the known ISM-naked UV-bright galaxy at $z\sim10$, U37126, reported in \cite{Marques_Chaves2026A&A...711A.301M}. 
Together with the high ionizing photon escape fraction, these properties indicate that faint galaxies but with blue colors which are undergoing a short-lived and intense episode of star formation, like GHZ8-SE, may play an important role in the reionization era (see Section~\ref{sec:lycescape}).

Despite its possibly high SFR, GHZ8-SE contributes little to the currently observed emission lines. The emission-line fluxes predicted from the best-fit SED model are about 30 times fainter than the observed values and are smaller than the observational uncertainties. 

In summary, GHZ8-SE is a faint, compact, and extremely blue component. Both its directly measured UV slope and the SED-fitting results point to an extremely young, ISM-poor star-forming system with weak nebular continuum emission and a high ionizing photon escape fraction. GHZ8-SE may therefore represent a short-lived, intense, and high-escape-fraction star-forming event at $z>10$. 
Although its UV slope is derived from decomposed photometry and may be affected by PSF matching, background estimation, and the assumed morphological model, direct aperture photometry without decomposition gives \betauv$=-2.97\pm0.32$ (also see Appendix~\ref{app:colormap} for more details). We consider this value an upper limit, as any contamination from GHZ8-main would make the measured UV slope redder than the intrinsic slope of GHZ8-SE. Deeper imaging or spatially resolved spectroscopy will be needed to better constrain its color and physical nature.

\section{GHZ8-main: evidence for a recent SF episode} \label{sec:result_main}
GHZ8-main is the dominant component in both the UV continuum and the observed nebular emission of GHZ8. It is a UV-bright system with a demagnified \muv\ of -19.91 mag, lying close to the medium to bright end of the UV luminosity distribution at $z>10$ (Figure~\ref{fig:Muv_slope} right panel). Its \betauv\ of -2.53 is bluer than the typical UV slope measured for most galaxies at $z\gtrsim 6$ (although this is not as extreme as that of GHZ8-SE). This suggests that GHZ8-main is dominated by a young stellar population and has moderate dust attenuation. This is supported by the SED fitting results, with an inferred stellar age of $\approx40\pm30\,$Myr and $\rm E(B-V)_{nebular}\approx0.23$ (see Table 3). In terms of size, stellar mass, and UV brightness, GHZ8 is similar to a post-starburst galaxy, CEERS2-588, at $z=11$ \citep{Harikane2026arXiv260121833H}.

The comparison between \Luv\ and \Lha\ provides a useful diagnostic of recent SFR variations \citep{Mehta2023ApJ...952..133M, Asada2024MNRAS.52711372A, Carvajal-Bohorquez2025A&A...704A.290C, Mitsuhashi2026arXiv260116284M, Munoz2026MNRAS.547ag415M}. \ha\ emission is produced by recombination in ionized gas and is powered by ionizing photons from the youngest and most massive OB stars. It therefore traces star formation over a short timescale of order of $\sim 10\,$Myr \citep{Kennicutt2012ARA&A..50..531K}. In contrast, the UV continuum is contributed not only by OB stars but also by A stars, and is sensitive to star formation over a longer timescale, typically up to $\sim 100\,$Myr \citep{Kennicutt2012ARA&A..50..531K}. Hence, the relative strength of \Luv\ and \Lha\ can be used to infer whether the recent SFR is declining \citep{Mehta2023ApJ...952..133M}. 

\begin{figure}
    \centering
    \includegraphics[width=1.0\linewidth]{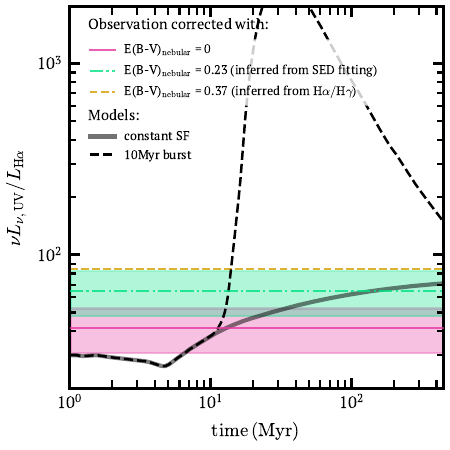}
    \caption{UV-to-H$\alpha$ luminosity ratios of GHZ8 with different dust correction ($\rm E(B-V)_{nebular} =$ [0, 0.23, 0.37], marked in pink, cyan,  and orange, respectively). The dust attenuation is derived from SED fitting and the Balmer decrement. The black solid and dashed curves show the model \Luv / \Lha\ as a function of time after the onset of star formation for constant star formation and 10\,Myr burst, respectively \citep{Mehta2023ApJ...952..133M}. These models have a metallicity of $\log(Z/Z_\odot)=-1$. Shadow regions show the $1\sigma$ errors for the $\rm E(B-V)_{nebular} =$ 0 and 0.23 cases (omitted for the  $\rm E(B-V)_{nebular} =0.37$ case for better visualization).
    }
    \label{fig:uv2ha_lum}
\end{figure}

We thus examine the recent star formation activity using \Luv -to-\Lha\ luminosity ratio. We adopt two dust corrections, one from the SED fitting ($\rm E(B-V)_{nebular} = 0.23$) and the other from the Balmer decrement ($\rm E(B-V)_{nebular} = 0.37$; Section~\ref{sec:balmder_de}). The resulting \Luv / \Lha\ are listed in Table~\ref{tab:ghz8_L_table} and shown in Figure~\ref{fig:uv2ha_lum}. We also include the \Luv / \Lha\ directly derived from SED fittings (showing only the BC03 results, as the predicted fluxes differ negligibly among stellar population models) and the observed value without any dust correction.

\begin{table}[t]
\centering
\caption{Measurements of UV-to-H$\alpha$ luminosity ratio}
\label{tab:ghz8_L_table}
\setlength{\tabcolsep}{20pt}
\begin{tabular}{l|l}
\hline\hline
\multicolumn{1}{c}{Corrected/derived by} & \multicolumn{1}{c}{ $\nu L_{\nu, \rm UV} / L_{\rm H\alpha}$ }  \\
\hline
$\rm E(B-V)_{nebular}=0$ & $ 42\pm11$  \\
$\rm E(B-V)_{nebular, SED}=0.23$ & $ 65\pm17$ \\
$\rm E(B-V)_{nebular, Balmer}=0.37$ & $ 85\pm51$ \\
SED fitting & $56\pm14$ \\
\hline
\end{tabular}
\begin{tablenotes}
\item Note. The $\nu L_{\rm \nu, UV}$ and $L_{\rm H\alpha}$ values are corrected for attenuation using $E(B-V)_{\rm line}$ inferred from both SED fitting and the Balmer decrement. We adopte the Milky Way extinction law and a E(B-V)$_{\rm stellar}$/E(B-V)$_{\rm line}$ = 0.44.
\end{tablenotes}
\end{table}

Figure~\ref{fig:uv2ha_lum} compares the inferred  \Luv/\Lha\ constraints of GHZ8-main with the predictions for two different SFHs: a constant star formation rate and a short burst of 10\,Myr duration \citep{Mehta2023ApJ...952..133M}. Regardless of the SFH model, the dust-free ($\rm E(B-V)_{nebular} = 0$) scenario implies a very young stellar age of $\sim 10\,$Myr (taking into account the $1\sigma$ uncertainty), in agreement with the best-fit SFH inferred from the SED fitting (see inset plot in Figure 5).

The dust-corrected \Luv-to-\Lha\ ratios are also consistent with a very young stellar population ($<20\,$Myr) when compared  with the model predictions for a 10\,Myr episode of star formation. This suggests that GHZ8 is being observed shortly after the end of the burst or during its declining phase, as independently supported by the best-fit SFH derived from \texttt{CIGALE} and non-parametric SFHs obtain from \texttt{BAGPIPES} as well (see Appendix~\ref{app:globalSEDfitting}). 

When comparing the  dust-corrected \Luv-to-\Lha\ ratios with the model predictions for a continuous SFH, we find that the $\rm E(B-V)_{nebular} = 0.23$ scenario allows for older ages from a few tens of Myrs to a couple of hundreds of Myrs. While the lower range is still consistent with the SED fitting results, the higher range seems to be in tension. Assuming, $\rm E(B-V)_{nebular} = 0.37$ would predict a luminosity ratio that is higher than those predicted by a constant SFH for stellar ages younger than $\sim 400\, \rm Myr$. All this suggests that GHZ8-main may have recently experienced a short-lived burst of star formation that has already ended or in a declining phase.

We note that the above discussion assumes $f_{\rm esc}=0$ for GHZ8-main, such that its \ha\ luminosity is not suppressed by the escape of ionizing photons. Although we do not expect a very high escape fraction for GHZ8-main, even a moderate $f_{\rm esc}$ of 10--30\% would increase the intrinsic \ha\ luminosity and lower the inferred \Luv/\Lha\ ratio. This would  result in even younger ages in  most the scenarios discussed above.

Overall, the current measurements suggest that GHZ8-main has either experienced a recent, short-lived burst followed by declining star formation or is undergoing a phase of constantly active star formation that began within the past few to a few tens of Myrs.

\subsection{Implication on UV-bright Galaxies}
The combined evidence suggests the following physical picture. The high UV luminosity of GHZ8-main indicates that it formed a substantial stellar population within the past several tens of Myr, while the most recent formation of massive stars may be weakened. This scenario of a declining SFH is independently supported by the SED fitting results, as seen in the SFH of the best-fit SED shown in Figure~\ref{fig:spec}. This short  episode of star formation was able to form a total stellar mass of $\log(M_\star/M_\odot)\approx8.6-8.8$ in $\sim 50\rm\,Myr$.

This result has implications for the origin of UV-bright galaxies at $z>10$. JWST observations has revealed a surprisingly large number of UV-bright galaxies in the early Universe, leading to a possible excess or slow evolution at the bright end of the UV luminosity \citep[e.g.,][]{Finkelstein2023ApJ...946L..13F, Harikane2023ApJS..265....5H, Perez-Gonzalez2023ApJ...951L...1P, Donnan2023MNRAS.520.4554D}. Bursty SFHs provide a natural way to temporarily enhance the UV luminosity of early galaxies and increase their probability of entering UV-selected samples \citep[e.g.,][]{Sun2023ApJ...955L..35S, Mason2023MNRAS.521..497M, Mirocha2023MNRAS.519..843M,Pallottini2023A&A...677L...4P,Santini2026A&A...710A.249S}. GHZ8-main offers a direct example of this scenario that, it is UV bright and blue, but both the \Luv/\Lha\ ratio and the SED-inferred SFHs suggest that its SFR may be declining. The weak CIII]$\lambda\lambda 1907, 1909 \AA$ line is also consistent with this scenario \citep{Roberts-Borsani2026MNRAS.548ag701R}.

This supports the idea that bursty SFHs can boost the UV luminosities of early galaxies and increase their probability of being selected in UV-bright samples. Other factors, including the young stellar age, compact star formation, and possibly efficient star formation in an early-forming halo, may also contribute to the observed UV brightness \citep{Finkelstein2024ApJ...969L...2F}. Additionally, merger/interaction may also trigger the burst in GHZ8 \citep{Harikane2025ApJ...980..138H, Bik2026arXiv260422460B}. The relative importance of these effects cannot be determined from a single object. Thus, GHZ8-main demonstrates that burstiness is present and may be important, but statistical samples are needed to assess whether it is a dominant driver of the UV-bright galaxy population at $z>10$.

\begin{figure*}
    \centering
    \includegraphics[width=1\linewidth]{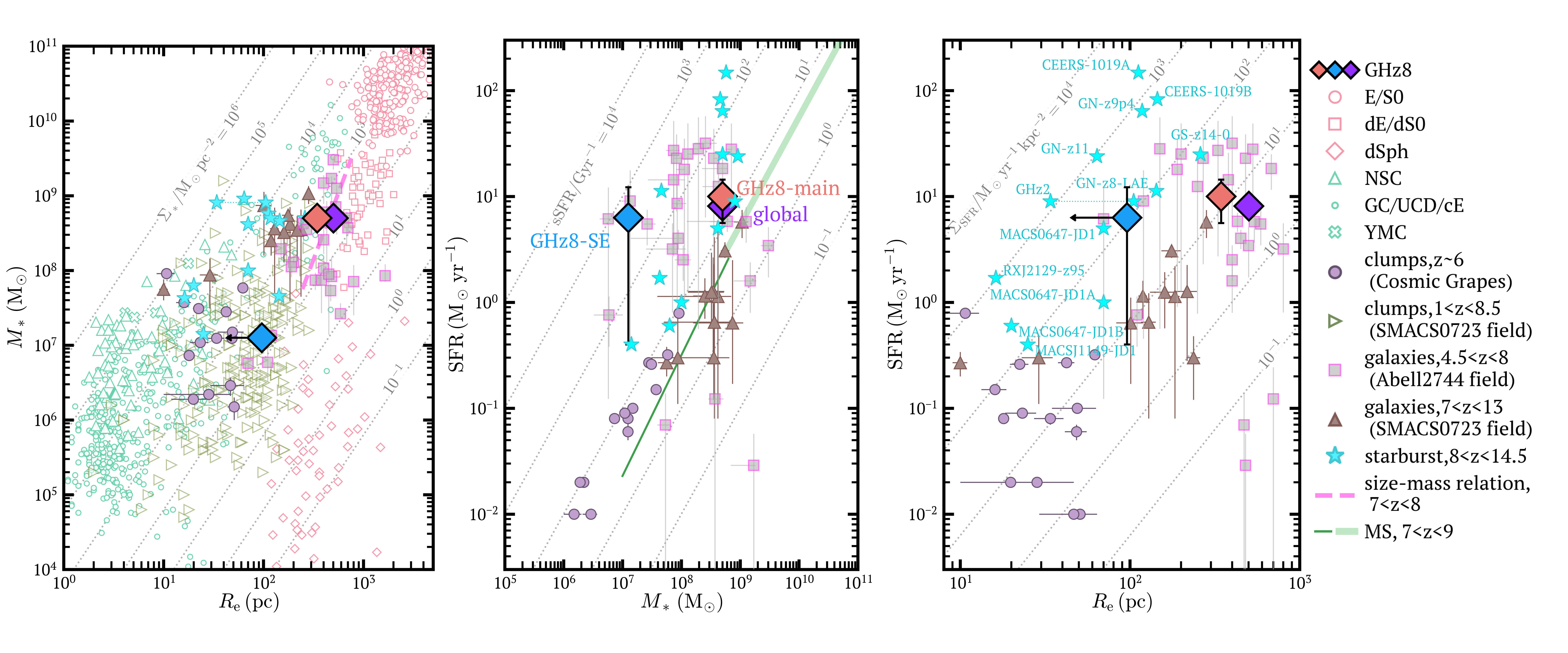}
    \caption{\textit{Left}: stellar mass vs. effective radius. Red and blue diamonds indicate GHZ8-main and SE, respectively. The comparison samples of elliptical/S0 galaxies (E/S0), dwarf elliptical/S0 galaxies (dE/dS0), dwarf spheroids (dSph), nuclear star clusters (NSC), Global cluster/ultra-compact dwarfs/compact elliptical galaxies (GC/UCD/cE), and young massive star clusters (YMC) are compiled in \cite{Norris2014MNRAS.443.1151N}. We also show clump measurements of \textit{Cosmic Grapes} at $z=6$ \citep{Fujimoto2025NatAs...9.1553F} and galaxy clumps in SMACA0723 field at $1<z<8.5$ \citep{Claeyssens2023MNRAS.520.2180C}. For comparison with high-$z$ samples, we also show galaxy measurements in the Abell 2744 field at $4.5<z<8$ \citep{Mascia2023A&A...672A.155M} and in the SMACA0723 field at $7<z<13$ \citep{Morishita2023ApJ...946L..35M}, and starburst galaxies at $8<z<14.5$ collected in literature \citep{Alvarez_Marquez2025A&A...695A.250A} are include as well. The pink dashed line indicates the size-mass relation for galaxies at $7<z<8$ in \cite{Miller2025ApJ...988..196M}. 
    \textit{Center}: SFR vs. stellar mass. SFRs of GHZ8 are adopted the instantaneous SFRs derived from SED fitting with BC03 stellar population (see Section~\ref{sec:sedfitting}). The dark green and light green lines show the main sequence at $7<z<9$ for low-mass regimes and all galaxies, respectively \citep{Merida2026A&A...707A...5M}.
    \textit{Right}: SFR vs. effective radius. }
    \label{fig:sfr_mass_re}
\end{figure*}

\section{Discussion} \label{sec:discussion}

\subsection{The Nature of GHZ8: A Merger?}\label{sec:merger}
The two-component morphology of GHZ8 raises the question of whether the system is a merger and an internally formed star-forming clump (i.e., in situ star formation). 
The merger or interaction scenario provides the more natural explanation for the current data. 

Figure~\ref{fig:sfr_mass_re} compares the sizes, masses, and SFRs of GHZ8 with samples in the literature. 
GHZ8 is globally more extended than other SFGs at the similar redshift ($z>8$; \citealt{Alvarez_Marquez2025A&A...695A.250A, Morishita2023ApJ...946L..35M}), whereas it has a size comparable to lower-redshift ($z<8$) galaxies and aligns with the size-mass relation at $7<z<8$ \citep{Miller2025ApJ...988..196M}. Such a difference between GHZ8 and typical galaxies at similar redshifts suggests that external processes, such as mergers, have played a role in its formation.

GHZ8-SE is compact, with a size that falls in the range of among high-$z$ starburst galaxies, low-$z$ dwarf galaxies, and star-forming clumps. Its stack rest-UV image (F150W, F200W, and F277W) gives a $\rm Gini = 0.55$ and $\rm M_{20}=-1.53$, which meet with the merger selection criteria of $\rm Gini + 0.14\,M_{20}>0.33$ \citep{Calabro2026arXiv260218068C}.
It shows a extremely high sSFR and $\Sigma_{\rm SFR}$ (also see Section~\ref{sec:result_se}), that are significantly higher than high-$z$ clumps \citep{Fujimoto2025NatAs...9.1553F} but closer to high-$z$ galaxies \citep{Alvarez_Marquez2025A&A...695A.250A, Morishita2023ApJ...946L..35M}.
Meanwhile, its extremely blue UV slope and high inferred escape fraction indicate stellar and nebular properties that are markedly different from those of GHZ8-main. This supports the merger-driven clump scenario \citep{Calabro2019A&A...632A..98C, Nakazato2024ApJ...975..238N}.

The mass ratio is roughly $10^{-2}$, consistent with a minor merger or accretion event. The small projected separation ($\sim 600\, \rm pc$) further suggests that the two components are already in a close interaction phase. This interpretation can naturally explain the presence of two stellar populations with very different UV slopes, sizes, and inferred escape fractions. It may also be connected to the bursty SFH inferred for GHZ8-main and GHZ8-SE, since galaxy interactions can trigger or enhance short-lived star formation episodes \citep{Barnes1991ApJ...370L..65B, Kaviraj2014MNRAS.440.2944K, Renaud2014MNRAS.442L..33R}.

An alternative possibility is that GHZ8-SE formed \textit{in situ} as a massive star-forming clump within GHZ8-main. Clumpy morphologies are common among high-redshift galaxies, and giant star-forming clumps are expected in gas-rich, turbulent systems where disk instabilities or local gravitational collapse can operate efficiently \citep[][]{Noguchi1999ApJ...514...77N}. In this picture, GHZ8-SE would represent an extreme star-forming complex formed within GHZ8-main. However, several aspects of GHZ8-SE make this interpretation less straightforward. 
Theoretically, giant star-forming clumps can form rapidly through gravitational instabilities in gas-rich disks \citep{Pawlik2011ApJ...731...54P,Mandelker2014MNRAS.443.3675M, Mandelker2017MNRAS.464..635M}. However, disk fragmentation often produces multiple clumps distributed across the disk, resulting in a relatively symmetric clumpy structure \cite[e.g., Cosmic Grapes;][]{Fujimoto2023ApJ...949L..25F}. This differs from the observed morphology of GHZ8, which consists of two distinct components with very different physical properties.
Thus, although \textit{in situ} clump formation cannot be ruled out, GHZ8-SE would have to be an unusually star-forming complex in the very early Universe, possibly among the highest-$z$ giant star-forming clumps known to date \citep{de_la_Vega2026ApJ..1005..169D, Adamo2024Natur.632..513A, Mowla2024Natur.636..332M}. This will further raises up an important question of how such clumps form in the early galaxies.

Taken together, the available evidence favors the interpretation that GHZ8 is a minor merger or interacting system, with GHZ8-SE being a low-mass companion. This scenario naturally accounts for the small projected separation, the dwarf-galaxy-like mass scale of GHZ8-SE, and the distinct stellar population properties of the two components. It also provides a plausible physical context for the bursty star formation activity inferred in GHZ8. 
A similar structure has also been observed in MACS1149-JD and MACS0647–JD at $z>9$, consisting of an extended main component and a companion \citep{Alvarez-Marquez2024A&A...686A..85A, Hsiao2024ApJ...973....8H}. In MACS1149-JD1, the velocity difference between the two components, measured from emission lines, may also support a merger scenario \citep{Alvarez-Marquez2024A&A...686A..85A}.

Nevertheless, this interpretation remains tentative. GHZ8-SE does not yet have an independent spectroscopic redshift, and the current data do not provide spatially resolved kinematic information. Its size and stellar mass also depend on the morphological decomposition and SED modeling. Deeper imaging could reveal whether diffuse tidal features or a bridge connects the two components, while higher-resolution spectroscopy would be needed to determine whether GHZ8-main and GHZ8-SE share the same systemic redshift and velocity structure. Such observations will be essential for distinguishing between a minor merger and an extreme \textit{in situ} star-forming clump.

\subsection{Implication for LyC Escape}\label{sec:lycescape}
The extremely blue UV continuum of GHZ8-SE suggests that it has a high ionizing photon (i.e., Lyman continuum photon, LyC photon) escape fraction. This motivates a discussion of both its possible contribution to the ionizing photon budget and the mechanisms that allows LyC photons to escape from GHZ8. 

GHZ8-SE appears to be an efficient ionizing source. Its young stellar population, compact size, blue UV slope, and high inferred $f_{\rm esc}$ point to a system with both strong ionizing photon production and weak nebular reprocessing. In this respect, GHZ8-SE may be analog to U37126, a blue $z=10.3$ galaxy reported to have a nearly unity LyC escape fraction \citep{Marques_Chaves2026A&A...711A.301M}. Although it is fainter and less massive than U37126 (Figure~\ref{fig:Muv_slope}), its escaped ionizing photon production rate $\dot{N}_{\rm LyC, esc}$ ($\equiv f_{\rm esc}\,\xi_{\rm ion}\, L_{\rm UV}\equiv f_{\rm esc}\, Q_{\rm H}$) is $\sim 4.9\pm2.0 \times 10^{53}\, \rm Hz\,erg^{-1}$ given the SED-fitting-inferred $f_{\rm esc}=0.8$. Considering the survey area of 37 arcmin$^2$ \citep{Castellano2023ApJ...948L..14C} and a redshift range of $\Delta z=0.5$, this single source in this field contributes $8.4\times10^{48}\, \rm photon\,s^{-1}\,cMpc^{-3}$. This is even comparable to the escaped photon rate of the $10^8-10^9\,\rm M_\odot$ galaxy population at $z\sim 10$ in simulation \citep{Yeh2023MNRAS.520.2757Y}. Such object may represent a short-lived important phase in which a small fraction of galaxies contributes a large fraction of the escaping ionizing photons, which could represent an important ionizing population in the reionization era.

The escape of ionizing photons from GHZ8-SE can be understood in terms of two related mechanisms. The first is a density-bounded nebulae scenario \citep{Zackrisson2013ApJ...777...39Z, Nakajima2014MNRAS.442..900N}. In this case, the source is sufficiently young, compact, and efficient at producing ionizing photons that it can ionize much of the surrounding gas. The second is a ionization-bounded scenario, in which ionizing photons escape through low-column-density sightlines \citep{Zackrisson2013ApJ...777...39Z, Nakajima2014MNRAS.442..900N}, originated from inhomogeneous ISM distribution. The latter interpretation is particularly relevant for GHZ8-SE, because its weak nebular contribution to continuum and high $f_{\rm esc}$ suggest that at least along line of sight the gas column density is low. 

The origin of such channels remains uncertain. Stellar and supernovae feedback can clear out gas and open low-density channels \citep[e.g.,][]{Ferrara2024A&A...684A.207F, Carr2025arXiv251021197C}. Another possible mechanism is rapid gas consumption through highly efficient star formation. 
A feedback-free starburst scenario with highly efficient star formation can also consume most of surrounding gas \citep{Dekel2023MNRAS.523.3201D}, leaving a low gas fraction and low-column-density sightlines.
At the same time, galaxy interactions or mergers can disturb the gas distribution, strip gas, and create escape paths, which are also seen in lower-$z$ LyC leakers \citep{Yuan2024ApJ...975...53Y, Gupta2024ApJ...973..169G, Zhu2025ApJ...982L..58Z, Ejdetjarn2026MNRAS.546ag211E}. In GHZ8, all mechanisms may be relevant. GHZ8-SE has an extremely high sSFR of hundreds $\rm Gyr^{-1}$, placing it well above the critical value of $\rm sSFR^\star = 25\, Gyr^{-1}$ proposed for radiation-driven outflows \citep{Ferrara2024A&A...684A.207F}. The tentative outflow signature (see Appendix~\ref{app:outflow} for details) would support this feedback-driven origin, but gas stripping or disturbed gas kinematics driven by the potential ongoing merger (see Section~\ref{sec:merger}) could produce similar observational features. Spatially resolved kinematics will therefore be required to distinguish between these possibilities.

In general, GHZ8-SE may represent a density-dominant LyC escape case, in which ionizing photons leak through low-density channels produced by feedback, interaction, or both. The case also infers that LyC escape in the reionization era may not be controlled by a single mechanism, which is also discussed by low-$z$ LyC leaker studies \citep[e.g.,][]{Yuan2026arXiv260301472Y}. Instead, the dominant escape pathway may vary from source to source, especially at very high redshift, where galaxies are more compact, burstier, and more frequently interacting than their low-$z$ counterparts. A statistical sample with deep rest-frame UV-optical imaging and spectroscopic data will be essential for determining whether GHZ8-SE is rare event.

\section{Conclusion} \label{sec:conclusion}

In this work, we report JWST/MIRI LRS rest-frame optical emission-line measurements of GHZ8, a UV-bright galaxy at $z=10.231$ in the Abell 2744 field. We also investigate its morphology and physical properties. Our main results are summarized as follows:

\begin{itemize}
    \item After self-calibrating the NIRSprec/prism and MIRI/LRS spectra with available photometry, we detect H$\alpha$ and the \hb+\oiii\ complex, with fluxes of $1.7\pm0.4\times10^{-18}\,{\rm erg\,s^{-1}\,cm^{-2}}$ ($\sim 4\sigma$) and $3.2\pm0.8 \times 10^{-18}\,{\rm erg\,s^{-1}\,cm^{-2}}$ ($\sim 5\sigma$), respectively. Combining these measurements with rest-frame UV emission lines from the prism spectrum, we derive $\rm E(B-V)_{nebular}=0.37_{-0.73}^{+0.10}$ from the \ha/\hg\ Balmer decrement. Although the uncertainty is large, this value is consistent with the attenuation inferred from SED fitting, $\rm E(B-V)_{nebular}=0.23\pm0.10$.

    \item GHZ8 shows two components in the rest-frame UV images: a bright and relatively extended main component (GHZ8-main), and a fainter, compact, unresolved southeastern component (GHZ8-SE). Through image decomposition, we find that GHZ8-main has demagnified $M_{\rm UV}=-19.91\pm0.10$ and $\beta=-2.53\pm0.22$, while GHZ8-SE has demagnified $M_{\rm UV}=-18.03\pm0.15$ and $\beta=-3.45\pm0.49$. These results indicate that both components are young and actively forming stars. In particular, GHZ8-SE exhibits an extremely blue UV slope.

    \item The decomposed SED fitting shows that GHZ8-SE is extremely young, with an age younger than $10\,{\rm Myr}$, a stellar mass of $\sim10^7\,M_\odot$, little dust attenuation, low metallicity, and active star formation. It has extreme sSFR and $\Sigma_{\rm SFR}$, comparable to those of starburst galaxies at similar redshifts. Its extremely blue \betauv\ requires a very young stellar population with little nebular contribution, implying an exceptionally high ionizing photon escape fraction of $\sim 0.6-0.9$. 
    
    GHZ8-main is also young, but older than GHZ8-SE, and is closer to normal main-sequence galaxies at this redshift. The best-fit SED suggests that GHZ8-main recently experienced a burst of star formation and is currently observed during its declining phase.
    The  UV-to-H$\alpha$ luminosity ratio independently supports this picture, suggesting that bursty star formation may also be important for explaining the abundance of UV-bright galaxies at very high redshift.

    \item The clumpy morphology of GHZ8 may originate from a minor merger. By comparing the sizes, stellar masses, and star-formation properties of GHZ8-main and GHZ8-SE with different classes of objects, we find that GHZ8-SE is similar to high-redshift starburst galaxies. Given the distinct physical properties of the two components, a merger scenario provides a more natural explanation, which may also relate to the recent burst of star formation in these systems. However, other explanations like \textit{in situ} star formation could not still be ruled out. Spatially resolved kinematic information is required to draw a solid conclusion.
    
    \item Since GHZ8-SE requires a low gas column density at least along our line of sight, the data favor a density-dominated LyC escape scenario. However, the physical origin of the low-density channels remains unclear. Both merger-driven gas redistribution,  stellar and supernovae feedback (i.e. outflows), and highly efficient star formation may play important roles. The outflow scenario might be supported by a tentative absorption \mgii\ feature in the NIRSpec spectrum, but deeper observations with higher spectral resolution would be needed for confirmation.
\end{itemize}

Although this work focuses on a single source, GHZ8 may provide a representative case study of UV-bright galaxies at $z>10$. Its UV luminosity is moderate among galaxies at this redshift, and it shows weak emission lines and an extended morphology. In this work, we reveal its clumpy structure and bursty star-formation activity, providing insight into the star formation, merging process, and possible reionization contribution of UV-bright galaxies in the early Universe.

\begin{acknowledgments}
We would like to thank Vhang Mehta for kindly sharing the $\nu L_{\rm \nu, UV}$-$L_{\rm H\alpha}$ tracks. We thank Denis Burgarella, Sinclaire Manning, Carlos Garcia Diaz, and Caitlin Casey for the constructive discussions and comments.
J.A.-M., C.P.-J., A.V.-O., M.H. acknowledge support by grant PID2024-158856NA-I00 from the Spanish Ministry of Science and Innovation/State Agency of Research MCIN/AEI/10.13039/501100011033 and by “ERDF A way of making Europe”. 
J.A.-M., L.C., C.P.-J., A.V.-O. acknowledge support by grant CSIC/BILATERALES2025/BIJSP25022.
L.C. acknowledges support by grants PIB2021-127718NB-I00 and PID2025-169195NB-I00 from the Spanish Ministry of Science and Innovation/State Agency of Research MCIN/AEI/10.13039/501100011033 and by “ERDF A way of making Europe”
M.C. acknowledges INAF GO Grant 2024 ``Revealing the nature of bright galaxies at cosmic dawn with deep JWST spectroscopy".
P.S. acknowledges financial support from INAF RF2024 Large Grant "UNDUST: UNveiling the Dawn of the Universe with JWST".
A.C.G. acknowledges support from NASA through STScI award JWST-GO-8051.
Y.N. acknowledges Flatiron Research Fellowship. The Flatiron Institute is a division of the Simons Foundation.
T.H. was supported by the Leading Initiative for Excellent Young Researchers, MEXT, Japan (HJH02007), and by JSPS KAKENHI Grant Numbers 22H01258, 23K22529, and 25K00020. T.H. was also supported by the ALMA Japan Research Grant of NAOJ ALMA Project, NAOJ-ALMA-2025-28A.
A.I. acknowledges JSPS KAKENHI Grant No. 26H02069.
D.L. acknowledges support from the Carlsberg Foundation, grant CF25-0662 (DL).
A.V.-O. is supported by the European Research Council (ERC) under the European Union’s Horizon 2020 research and innovation programme (DistantDust, Grant agreement No. 101117541).

\end{acknowledgments}





%
\facilities{JWST(MIRI, NIRCam, NIRSpec) }


\software{astropy~\citep{2013A&A...558A..33A,2018AJ....156..123A,2022ApJ...935..167A},  
            emcee~\citep{Foreman-Mackey2013PASP..125..306F}, 
            GALFITM~\citep{Vika2013MNRAS.435..623V, Peng2002AJ....124..266P}, CIGALE~\citep{Boquien2019A&A...622A.103B,Burgarella2025A&A...699A.336B, Burgarella2026arXiv260509829B}, photutils~\citep{bradley_2026_19636730}, 
            msaexp~\citep{Brammer2022zndo...7299500B}, 
            BAGPIPES~\citep{Carnall2018MNRAS.480.4379C, Carnall2019MNRAS.490..417C}, PyNeb~\citep{Luridiana2015A&A...573A..42L, Morisset2020Atoms...8...66M, Mendoza2023Atoms..11...63M}
          }


\newpage
\appendix

\section{\hb--\oiii\ deblending} \label{app:hboiii_deblending}
To test whether \hb\ and \oiii\ can be deblended, we construct models under different assumptions for the attenuation. For a given attenuation, we adopt the Milky Way extinction curve and assume Case B recombination to predict the \hb\ model from the \ha\ flux and line width, while accounting for the spectral resolution. The widths of the \oiii\ lines are also determined from the \ha\ line width and the spectral resolution. The \oiii\ flux is left as a free parameter, and the \oiii $\lambda5008/\lambda4960$ flux ratio is fixed to the laboratory value of 2.98 \citep{Storey2000MNRAS.312..813S}. We then compute BIC as $ {\rm BIC} = \kappa \ln(n) - \ln(L_{\rm max})$ for each model, where $\kappa$ is the dimension of free parameters, $n$ is the number of data points, and $L_{\rm max}$ is the maximum likelihood.

\begin{figure}[h]
    \centering
    \includegraphics[width=0.4\linewidth]{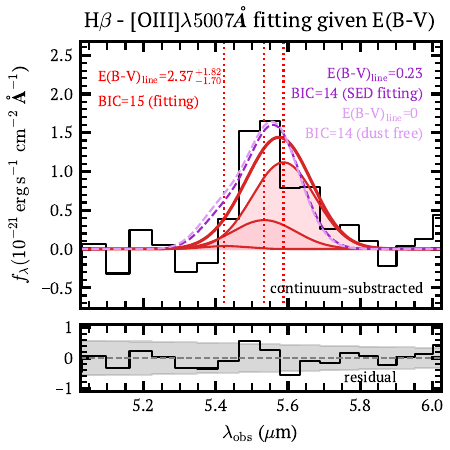}
    \caption{\hb--\oiii\ deblending models and fits for given values of E(B-V). The \hb\ flux is inferred from \ha\ for each assumed value of E(B-V). The light purple and purple curves show the models assuming no dust correction and the E(B-V) value derived from SED fitting, respectively. The red curve shows the fit in which E(B-V) is treated as a free parameter. The red shaded components represent \hb\ and the \oiii\ doublet.
}
    \label{fig:hbeta_oiii_model}
\end{figure}

Figure~\ref{fig:hbeta_oiii_model} shows the fitting results for three assumed values of $\rm E(B-V)_{nebular}=0$, 0.23, and 2.37, corresponding to no dust attenuation, the attenuation inferred from the SED fitting, and the case where E(B-V) is treated as a free parameter, respectively. However, the BIC values do not change significantly among these three cases. This indicates that the current \hb--\oiii\ complex does not favor any of these models. Therefore, \hb\ and \oiii\ cannot be robustly deblended using the \ha\ constraint with the current data, mainly because of the limited spectral resolution and sensitivity.

\begin{figure}
    \centering
    \includegraphics[width=0.4\linewidth]{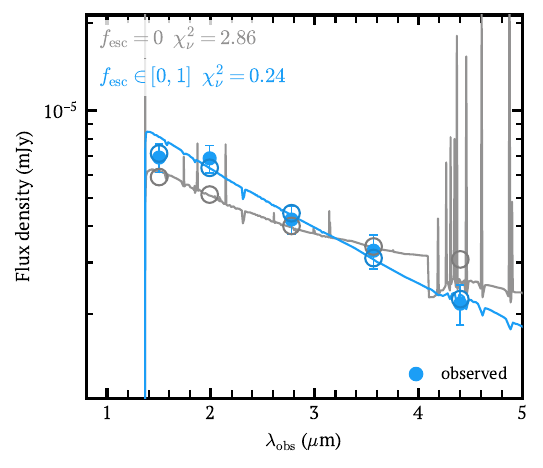}
    \caption{Comparison between SED fits with non-zero $f_{\rm esc}$ (blue) and zero $f_{\rm esc}$ (gray), based on the BC03 stellar population models. The non-zero $f_{\rm esc}$ model  produces a significant better fit. The reduced $\chi^2$ values are shown in the upper-left corner.  
    }
    \label{fig:SEsedComparison}
\end{figure}

\section{GHZ8-SE SED fitting given $f_{\rm esc} = 0$} \label{app:frac_se}
We here test whether the SED fitting of GHZ8-SE prefers a non-zero escape fraction. To do this, we compare two fits in which $f_{\rm esc}$ is fixed to zero and allowed to vary freely. Both fits use the BC03 stellar population models, and the only difference between them is the treatment of $f_{\rm esc}$. As shown in Figure~\ref{fig:SEsedComparison}, the fit with non-zero $f_{\rm esc}$ is significantly better than the fit with $f_{\rm esc}=0$. This is expected because, when $f_{\rm esc}=0$, the nebular continuum contribution is stronger and the UV continuum becomes redder, making it difficult to reproduce the extremely blue UV slope of GHZ8-SE.

We have also verified that the same conclusion is obtained when binary stellar populations are included. This indicates that the high $f_{\rm esc}$ inferred for GHZ8-SE is unlikely to be caused by model degeneracy, but instead reflects the physical requirement imposed by its extremely blue UV continuum.

\section{Integrated SED Fitting and Global Properties} \label{app:globalSEDfitting}
We here perform integrated SED fitting for GHZ8. We run multiple fits in order to compare the results obtained with the BC03 stellar population models and those obtained with BPASS, which includes binary stellar evolution. The resulting parameters are listed in Table~\ref{tab:ghz8_sedfit_params}. As a comparison, we also use \texttt{BAGPIPES} with a non-parametric SFH to examine how the inferred SFH depends on the assumed SFH parameterization. 

For the integrated SED fitting, we use the \texttt{CIGALE-spec} version, which allows us to jointly fit the Prism spectrum and the NIRCam broadband photometry. Since \texttt{CIGALE-spec} cannot simultaneously include the LRS spectrum and the LRS continuum has a low S/N ratio, we instead include the measured \ha\ integrated flux and the \hb--\oiii\ flux density as additional constraints. The \hb--\oiii\ flux density is measured by defining a top-hat filter at the corresponding observed wavelength. These additional constraints help to better constrain the recent star formation rate and dust attenuation.

\begin{figure}
    \centering
    \includegraphics[width=0.4\linewidth]{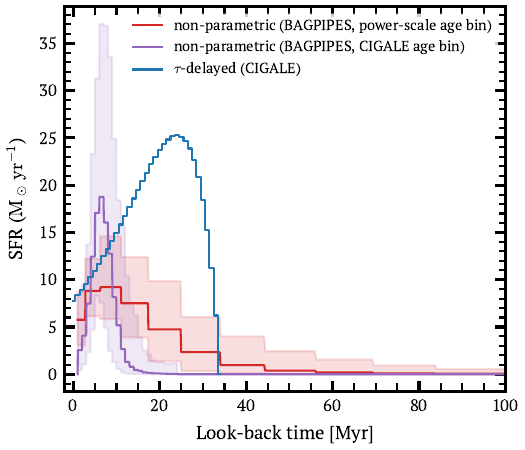}
    \caption{Comparison of SFHs between non-parametric (red and purple curves) and $\tau$-delayed (blue curve) models for GHZ8 (including both components).  }
    \label{fig:sfh_comparison}
\end{figure}

For comparison, we also perform SED fitting including binary stellar populations using the \texttt{CIGALE-2025.v1} version, which incorporates the BPASS models. Since this version cannot jointly fit spectra, we measure the emission lines from the prism spectrum and include them as additional constraints in the fitting. For isolated emission lines, such as \hg\ and \hd\ , we use their integrated line fluxes. For blended features, such as \neiii--\he\ , we measure the flux density using a top-hat filter.

To assess how different delayed-SFH assumptions affect the fitting results, we use \texttt{BAGPIPES} to examine the inferred SFH shape based on the NIRCam photometry and the prism spectrum. We adopt a continuous non-parametric SFH model and test two different age-binning schemes: (a) 20 power-law-spaced bins over 0-200 Myr, which provide finer sampling at young ages and (b) bins defined to match the age grid of the best-fit \texttt{CIGALE} model, with a bin width of 1 Myr.

Figure~\ref{fig:sfh_comparison} shows the comparison of SFHs between non-parametric and $\tau$-delayed models. All of them show a recent burst and undergoes a declining-SFR phase, although the peaks of burst vary from model to model.

\section{\betauv\ measured directly from images and \betauv\ Map} \label{app:colormap}

\begin{figure*}
    \centering
    \includegraphics[width=1.0\linewidth]{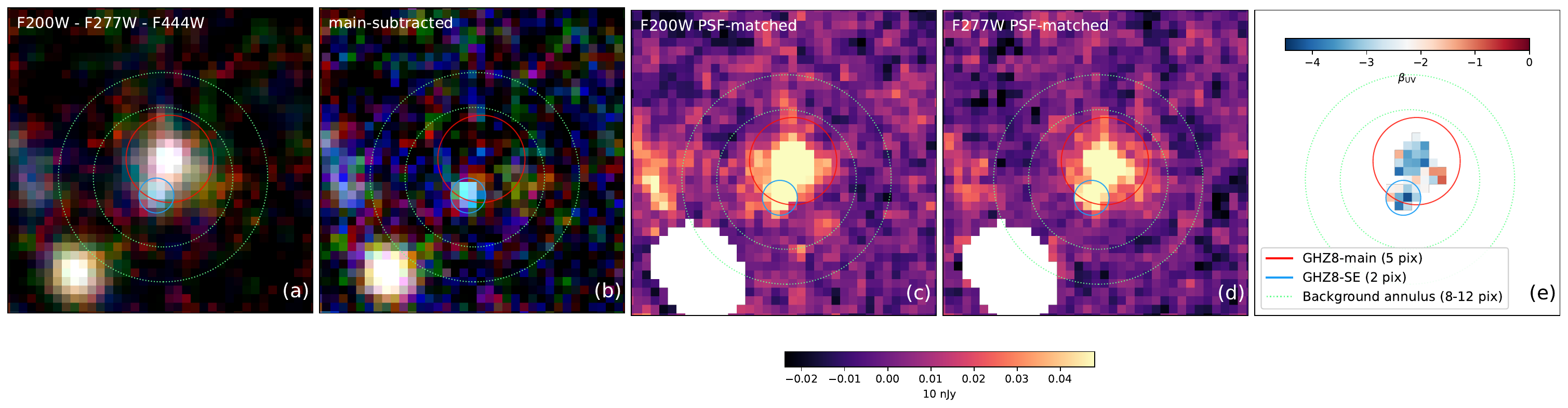}
    \caption{(a) F200W--F277W--F444W pseudo-color image of GHZ8. The red and blue circles are centered on the main and SE components, respectively. The green annulus marks the background region, centered on the global centroid. 
    (b) Pseudo-color image after subtraction of the main component, using the main-component model derived from the morphological decomposition. 
    (c) PSF-matched F200W image. 
    (d) PSF-matched F277W image. 
    (e) $\beta_{\rm UV}$ map.
    }
    \label{fig:colormap}
\end{figure*}

Figure~\ref{fig:colormap} shows the F200W--F277W--F444W pseudo-color image (panel a) and the pseudo-color image after subtracting the GHZ8-main component (panel b). GHZ8-SE is clearly more compact and bluer than GHZ8-main. Since the \texttt{GALFITM} decomposed photometry is based on model fluxes integrated to infinity, we additionally measure the UV slopes of GHZ8-main and GHZ8-SE directly from the images using fixed apertures. This provides an independent check of any possible bias in the UV slopes derived from the \texttt{GALFITM} decomposed photometry.

For GHZ8-SE, we first subtract the main component. We then estimate and subtract the local background using an annulus with an inner radius of 8 pixels and an outer radius of 12 pixels, shown as the green annulus in Figure~\ref{fig:colormap}. We perform aperture photometry in F200W and F277W using a circular aperture with a radius of 2 pixels, which encloses most of the light from GHZ8-SE. The UV slope is then calculated from the aperture fluxes in these two bands. To estimate the uncertainty in \betauv, we include both the background rms and the uncertainty introduced by the background subtraction,
\begin{equation}
\sigma_\beta = \sqrt{\sigma_{\rm \beta ,blank}^2+\sigma_{\rm \beta, bg}^2}.
\end{equation}
Here, $\sigma_{\rm blank}$ and $\sigma_{\rm bg}$ are estimated in the same way as described in Section~\ref{sec:morphology_mea}. For GHZ8-main, we follow the same procedure after subtracting the GHZ8-SE component, but use a larger aperture with a radius of 8 pixels.

This aperture-based measurement gives \betauv $=-3.19\pm0.48$ for GHZ8-SE and \betauv $=-2.65\pm0.34$ for GHZ8-main. These results confirm that both components of GHZ8 have blue UV continua. In particular, GHZ8-SE remains extremely blue, although the uncertainty is relatively large. For comparison, direct aperture photometry \textit{without} component decomposition gives \betauv$=-2.97\pm0.32$ and $-2.70\pm0.30$ for GHZ8-SE and GHZ8-main, respectively.

In addition, we construct a \betauv\ map from the F200W and F277W images. Because the PSFs of the two bands are slightly different, we first perform PSF matching to avoid flux differences caused by resolution mismatch. The PSF-matched images are shown in panels c and d. We then subtract the background in the same way as above. Since the F200W and F277W images are aligned on the same pixel grid, we calculate \betauv\ pixel by pixel. For visualization, we only keep pixels with flux S/N $>3$ both in two images. The resulting \betauv\ map shows again that GHZ8-SE is bluer than GHZ8-main in the UV. GHZ8-main has a blue central region, while a redder clump is seen toward the southwest. This may be related to the dust distribution or stellar population variations within GHZ8-main. However, given the limited spatial resolution of the current data, we do not discuss this feature further.

\section{Impact of different definitions on \betauv} \label{app:betadefinition}
We measure the UV slope using the F200W and F277W bands, corresponding to rest-frame pivot wavelengths of $\sim1900$ and $2500\,{\rm \AA}$, respectively. We exclude F150W because it covers the DLA absorption feature, which would bias the measured UV slope toward redder values. To examine how the choice of wavelength coverage affects the UV slope measurement, we use the best-fit SED models to compare different methods of measuring $\beta$.

In Table~\ref{tab:ghz8_sedfit_params}, we present the UV slopes measured from F200W and F277W (denoted as \betauv) and those measured by \texttt{CIGALE} following the definition of \citet{Calzetti1994ApJ...429..582C} (denoted as $\beta_{\rm UV,C94}$). For the global system, the two measurements show little difference. For GHZ8-main, $\beta_{\rm UV}$ is slightly bluer than $\beta_{\rm UV,C94}$, mainly because rest-frame FUV emission lines contribute to a slight excess in the F200W flux. In contrast, for GHZ8-SE, $\beta_{\rm UV}$ is somewhat redder than $\beta_{\rm UV,C94}$, but both measurements indicate an extremely blue UV continuum.

Another point worth noting is that the UV slopes derived from the best-fit SED models are slightly redder than those measured directly from the decomposed photometry. This mainly reflects the difficulty of current SED models in reproducing the extremely blue UV continuum of GHZ8-SE. Nevertheless, the best-fit models remain consistent with the measurements within their uncertainties. Deeper imaging and improved separation of GHZ8-SE from GHZ8-main will be needed to better constrain its extreme UV slope.

\begin{figure*}
    \centering
    \includegraphics[width=0.325\linewidth]{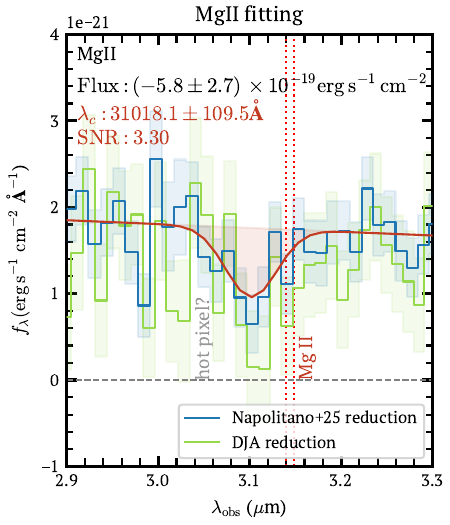}
    \includegraphics[width=0.645\linewidth]{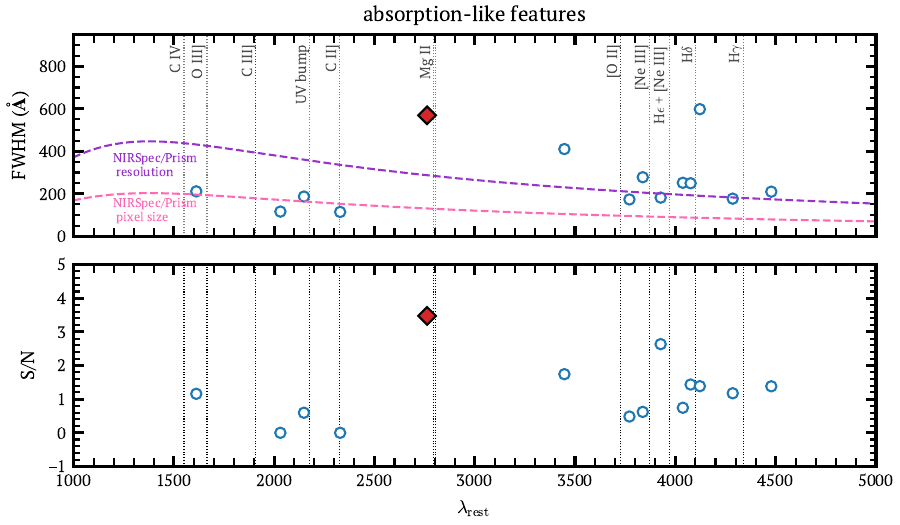}
    \caption{\textit{Left panel}: Tentative \mgii\ absorption feature. The blue and green curves represent reductions by \cite{Napolitano2025A&A...693A..50N} and DJA. The red curve shows the best-fit model. The red dotted lines indicate the expected \mgii\ position based on the redshift estimated with emission lines. \textit{Right panel}: FWHM (top) and S/N (bottom) of absorption-like features in the prism spectrum. The tentative \mgii\ detection is highlighted with a red diamond.
    }
    \label{fig:absorptionfeature}
\end{figure*}

\section{The tentative evidence of \mgii\ outflow} \label{app:outflow}
We tentatively detect a $\sim 3\sigma$ \mgii\ absorption at rest-frame $2761.8 \pm 9.7\rm \, \AA$, with a flux of $-5.8\pm2.7 \times 10^{-19} \rm \, erg\,s^{-1}\,cm^{-2}$, a velocity offset of $-4092 \pm 1059 \, \rm km\,s^{-1}$, and a equivalent width of $\rm EW_0 = -29 \pm 11\, \rm \AA$. Figure~\ref{fig:absorptionfeature} shows the \mgii\ fitting, based on the prism spectrum reduced by \cite{Castellano2026} and the DAWN JWST data Archive (DJA) \footnote{https://dawn-cph.github.io/dja/index.html} \citep{de_Graaff2025A&A...697A.189D, Heintz2024Sci...384..890H}. This feature is both presented in these two reduction, suggesting that it is not induced by the data reduction. 

To compare the tentative \mgii\ absorption with other absorption-like features, we use the \texttt{find\_peaks} function in the \texttt{scipy} Python package to search for absorption-like features in the rest-frame UV spectrum. We require each feature to have a minimum prominence of $1\sigma$, to extend over at least 3 pixels, and to be separated from neighboring features by at least 2 pixels. We then fit each feature with a single Gaussian profile. This procedure identifies 15 absorption-like features, including the tentative \mgii\ absorption. Among them, the tentative \mgii\ feature has the largest FWHM and the highest S/N. In other words, this feature is the most significant among the absorption-like features identified in the prism spectrum. Still, caution is required. Known instrumental and reduction-related effects in NIRSpec spectra can produce spurious positive or negative features in the extracted 1-D spectra, often on the scale of individual pixels \footnote{https://jwst-docs.stsci.edu/known-issues/nirspec-known-issues?utm\_source=chatgpt.com\#NIRSpecKnownIssues-replaceSpuriousfeaturesinextracted1-Dspectra}. Therefore, higher-resolution and higher-sensitivity medium-resolution spectroscopy is needed to solidly confirm the presence of this feature.


\bibliography{ref}{}

@ARTICLE{Harikane2026arXiv260121833H,
       author = {{Harikane}, Yuichi and {Perez-Gonzalez}, Pablo G. and {Alvarez-Marquez}, Javier and {Ouchi}, Masami and {Nakazato}, Yurina and {Ono}, Yoshiaki and {Nakajima}, Kimihiko and {Umeda}, Hiroya and {Isobe}, Yuki and {Xu}, Yi and {Zhang}, Yechi},
        title = "{A UV-Luminous Galaxy at z=11 with Surprisingly Weak Star Formation Activity}",
      journal = {arXiv e-prints},
         year = 2026,
        month = jan,
          eid = {arXiv:2601.21833},
        pages = {arXiv:2601.21833},
          doi = {10.48550/arXiv.2601.21833},
archivePrefix = {arXiv},
       eprint = {2601.21833},
 primaryClass = {astro-ph.GA},
       adsurl = {https://ui.adsabs.harvard.edu/abs/2026arXiv260121833H}
}

@ARTICLE{Bezanson2024ApJ...974...92B,
       author = {{Bezanson}, Rachel and {Labbe}, Ivo and {Whitaker}, Katherine E. and {Leja}, Joel and {Price}, Sedona H. and {Franx}, Marijn and {Brammer}, Gabriel and {Marchesini}, Danilo and {Zitrin}, Adi and {Wang}, Bingjie and {Weaver}, John R. and {Furtak}, Lukas J. and {Atek}, Hakim and {Coe}, Dan and {Cutler}, Sam E. and {Dayal}, Pratika and {van Dokkum}, Pieter and {Feldmann}, Robert and {F{\"o}rster Schreiber}, Natascha M. and {Fujimoto}, Seiji and {Geha}, Marla and {Glazebrook}, Karl and {de Graaff}, Anna and {Greene}, Jenny E. and {Juneau}, St{\'e}phanie and {Kassin}, Susan and {Kriek}, Mariska and {Khullar}, Gourav and {Maseda}, Michael and {Mowla}, Lamiya A. and {Muzzin}, Adam and {Nanayakkara}, Themiya and {Nelson}, Erica J. and {Oesch}, Pascal A. and {Pacifici}, Camilla and {Pan}, Richard and {Papovich}, Casey and {Setton}, David J. and {Shapley}, Alice E. and {Smit}, Renske and {Stefanon}, Mauro and {Taylor}, Edward N. and {Williams}, Christina C.},
        title = "{The JWST UNCOVER Treasury Survey: Ultradeep NIRSpec and NIRCam Observations before the Epoch of Reionization}",
      journal = {\apj},
         year = 2024,
        month = oct,
       volume = {974},
       number = {1},
          eid = {92},
        pages = {92},
          doi = {10.3847/1538-4357/ad66cf},
archivePrefix = {arXiv},
       eprint = {2212.04026},
 primaryClass = {astro-ph.GA},
       adsurl = {https://ui.adsabs.harvard.edu/abs/2024ApJ...974...92B}
}

@ARTICLE{Alvarez-Marquez2024A&A...686A..85A,
       author = {{{\'A}lvarez-M{\'a}rquez}, J. and {Colina}, L. and {Crespo G{\'o}mez}, A. and {Rinaldi}, P. and {Melinder}, J. and {{\"O}stlin}, G. and {Annunziatella}, M. and {Labiano}, A. and {Bik}, A. and {Bosman}, S. and {Greve}, T.~R. and {Wright}, G. and {Alonso-Herrero}, A. and {Boogaard}, L. and {Azollini}, R. and {Caputi}, K.~I. and {Costantin}, L. and {Eckart}, A. and {Garc{\'\i}a-Mar{\'\i}n}, M. and {Gillman}, S. and {Hjorth}, J. and {Iani}, E. and {Ilbert}, O. and {Jermann}, I. and {Langeroodi}, D. and {Meyer}, R. and {Pei{\ss}ker}, F. and {P{\'e}rez-Gonz{\'a}lez}, P. and {Pye}, J.~P. and {Tikkanen}, T. and {Topinka}, M. and {van der Werf}, P. and {Walter}, F. and {Henning}, Th. and {Ray}, T.},
        title = "{Spatially resolved H{\ensuremath{\alpha}} and ionizing photon production efficiency in the lensed galaxy MACS1149-JD1 at a redshift of 9.11}",
      journal = {\aap},
         year = 2024,
        month = jun,
       volume = {686},
          eid = {A85},
        pages = {A85},
          doi = {10.1051/0004-6361/202347946},
archivePrefix = {arXiv},
       eprint = {2309.06319},
 primaryClass = {astro-ph.GA},
       adsurl = {https://ui.adsabs.harvard.edu/abs/2024A&A...686A..85A}
}

@ARTICLE{Bik2026arXiv260422460B,
       author = {{Bik}, Arjan and {{\'A}lvarez-M{\'a}rquez}, Javier and {Crespo G{\'o}mez}, Alejandro and {Colina}, Luis and {P{\'e}rez-Gonz{\'a}lez}, Pablo G. and {{\"O}stlin}, G{\"o}ran and {Blanco Prieto}, Carmen and {Melinder}, Jens and {Langeroodi}, Danial and {Wright}, Gillian and {Algera}, Hiddo S.~B. and {Boogaard}, Leindert A. and {Caputi}, Karina and {Gillman}, Steven and {Greve}, Thomas and {Hjorth}, Jens and {Iani}, Edoardo and {Kendrew}, Sarah and {Labiano Ortega}, Alvaro and {Perna}, Michele and {Prieto Jimenez}, Carlota and {Pye}, John and {Rinaldi}, Pierluigi and {van der Werf}, Paul and {Walter}, Fabian and {Pei{\ss}ker}, Florian and {Eckart}, Andreas and {Henning}, Thomas and {G{\"u}del}, Manuel},
        title = "{Spatially resolved metallicity and ionization in the merging system Gz9p3 at z=9.3}",
      journal = {arXiv e-prints},
         year = 2026,
        month = apr,
          eid = {arXiv:2604.22460},
        pages = {arXiv:2604.22460},
          doi = {10.48550/arXiv.2604.22460},
archivePrefix = {arXiv},
       eprint = {2604.22460},
 primaryClass = {astro-ph.GA},
       adsurl = {https://ui.adsabs.harvard.edu/abs/2026arXiv260422460B}
}

@ARTICLE{Roberts-Borsani2026MNRAS.548ag701R,
       author = {{Roberts-Borsani}, Guido and {Oesch}, Pascal A. and {Ellis}, Richard and {Weibel}, Andrea and {Giovinazzo}, Emma and {Bouwens}, Rychard and {Dayal}, Pratika and {Fontana}, Adriano and {Heintz}, Kasper E. and {Matthee}, Jorryt and {Meyer}, Romain A. and {Pentericci}, Laura and {Shapley}, Alice and {Tacchella}, Sandro and {Treu}, Tommaso and {Walter}, Fabian and {Atek}, Hakim and {Bose}, Sownak and {Castellano}, Marco and {Fudamoto}, Yoshinobu and {Morishita}, Takahiro and {Naidu}, Rohan P. and {Sanders}, Ryan L. and {van der Wel}, Arjen},
        title = "{JWST spectroscopic insights into the diversity of galaxies in the first 500 Myr: short-lived snapshots along a common evolutionary pathway}",
      journal = {\mnras},
         year = 2026,
        month = may,
       volume = {548},
       number = {3},
          eid = {stag701},
        pages = {stag701},
          doi = {10.1093/mnras/stag701},
archivePrefix = {arXiv},
       eprint = {2508.21708},
 primaryClass = {astro-ph.GA},
       adsurl = {https://ui.adsabs.harvard.edu/abs/2026MNRAS.548ag701R}
}

@ARTICLE{Pawlik2011ApJ...731...54P,
       author = {{Pawlik}, Andreas H. and {Milosavljevi{\'c}}, Milo{\v{s}} and {Bromm}, Volker},
        title = "{The First Galaxies: Assembly of Disks and Prospects for Direct Detection}",
      journal = {\apj},
         year = 2011,
        month = apr,
       volume = {731},
       number = {1},
          eid = {54},
        pages = {54},
          doi = {10.1088/0004-637X/731/1/54},
archivePrefix = {arXiv},
       eprint = {1011.0438},
 primaryClass = {astro-ph.CO},
       adsurl = {https://ui.adsabs.harvard.edu/abs/2011ApJ...731...54P}
}

@ARTICLE{Munoz2026MNRAS.547ag415M,
       author = {{Mu{\~n}oz}, Julian B. and {Chisholm}, John and {Sun}, Guochao and {Samuel}, Jenna and {Mirocha}, Jordan and {Bregou}, Emily and {Venditti}, Alessandra and {Qezlou}, Mahdi and {Simmonds}, Charlotte and {Endsley}, Ryan},
        title = "{Relatively fast and reasonably furious: evidence for increased burstiness in smaller haloes at cosmic dawn}",
      journal = {\mnras},
         year = 2026,
        month = apr,
       volume = {547},
       number = {4},
          eid = {stag415},
        pages = {stag415},
          doi = {10.1093/mnras/stag415},
archivePrefix = {arXiv},
       eprint = {2601.07912},
 primaryClass = {astro-ph.GA},
       adsurl = {https://ui.adsabs.harvard.edu/abs/2026MNRAS.547ag415M}
}

@ARTICLE{Langeroodi2024arXiv240413045L,
       author = {{Langeroodi}, Danial and {Hjorth}, Jens},
        title = "{NIRSpec View of the Appearance and Evolution of Balmer Breaks and the Transition from Bursty to Smooth Star Formation Histories from Deep Within the Epoch of Reionization to Cosmic Noon}",
      journal = {arXiv e-prints},
         year = 2024,
        month = apr,
          eid = {arXiv:2404.13045},
        pages = {arXiv:2404.13045},
          doi = {10.48550/arXiv.2404.13045},
archivePrefix = {arXiv},
       eprint = {2404.13045},
 primaryClass = {astro-ph.GA},
       adsurl = {https://ui.adsabs.harvard.edu/abs/2024arXiv240413045L}
}

@ARTICLE{Stanway2018MNRAS.479...75S,
       author = {{Stanway}, E.~R. and {Eldridge}, J.~J.},
        title = "{Re-evaluating old stellar populations}",
      journal = {\mnras},
         year = 2018,
        month = sep,
       volume = {479},
       number = {1},
        pages = {75-93},
          doi = {10.1093/mnras/sty1353},
archivePrefix = {arXiv},
       eprint = {1805.08784},
 primaryClass = {astro-ph.GA},
       adsurl = {https://ui.adsabs.harvard.edu/abs/2018MNRAS.479...75S}
}

@ARTICLE{Bruzual2003MNRAS.344.1000B,
       author = {{Bruzual}, G. and {Charlot}, S.},
        title = "{Stellar population synthesis at the resolution of 2003}",
      journal = {\mnras},
         year = 2003,
        month = oct,
       volume = {344},
       number = {4},
        pages = {1000-1028},
          doi = {10.1046/j.1365-8711.2003.06897.x},
archivePrefix = {arXiv},
       eprint = {astro-ph/0309134},
 primaryClass = {astro-ph},
       adsurl = {https://ui.adsabs.harvard.edu/abs/2003MNRAS.344.1000B}
}

@BOOK{Osterbrock2006agna.book.....O,
       author = {{Osterbrock}, Donald E. and {Ferland}, Gary J.},
        title = "{Astrophysics of gaseous nebulae and active galactic nuclei}",
         year = 2006,
       adsurl = {https://ui.adsabs.harvard.edu/abs/2006agna.book.....O}
}

@ARTICLE{Mendoza2023Atoms..11...63M,
       author = {{Mendoza}, Claudio and {M{\'e}ndez-Delgado}, Jos{\'e} E. and {Bautista}, Manuel and {Garc{\'\i}a-Rojas}, Jorge and {Morisset}, Christophe},
        title = "{Atomic Data Assessment with PyNeb: Radiative and Electron Impact Excitation Rates for [Fe II] and [Fe III]}",
      journal = {Atoms},
         year = 2023,
        month = apr,
       volume = {11},
       number = {4},
          eid = {63},
        pages = {63},
          doi = {10.3390/atoms11040063},
archivePrefix = {arXiv},
       eprint = {2304.01298},
 primaryClass = {astro-ph.GA},
       adsurl = {https://ui.adsabs.harvard.edu/abs/2023Atoms..11...63M}
}

@ARTICLE{Morisset2020Atoms...8...66M,
       author = {{Morisset}, Christophe and {Luridiana}, Valentina and {Garc{\'\i}a-Rojas}, Jorge and {G{\'o}mez-Llanos}, Ver{\'o}nica and {Bautista}, Manuel and {Mendoza}, Claudio},
        title = "{Atomic Data Assessment with PyNeb}",
      journal = {Atoms},
         year = 2020,
        month = oct,
       volume = {8},
       number = {4},
          eid = {66},
        pages = {66},
          doi = {10.3390/atoms8040066},
archivePrefix = {arXiv},
       eprint = {2009.10586},
 primaryClass = {astro-ph.GA},
       adsurl = {https://ui.adsabs.harvard.edu/abs/2020Atoms...8...66M}
}

@ARTICLE{Luridiana2015A&A...573A..42L,
       author = {{Luridiana}, V. and {Morisset}, C. and {Shaw}, R.~A.},
        title = "{PyNeb: a new tool for analyzing emission lines. I. Code description and validation of results}",
      journal = {\aap},
         year = 2015,
        month = jan,
       volume = {573},
          eid = {A42},
        pages = {A42},
          doi = {10.1051/0004-6361/201323152},
archivePrefix = {arXiv},
       eprint = {1410.6662},
 primaryClass = {astro-ph.IM},
       adsurl = {https://ui.adsabs.harvard.edu/abs/2015A&A...573A..42L}
}

@ARTICLE{Hsiao2024ApJ...973....8H,
       author = {{Hsiao}, Tiger Yu-Yang and {Abdurro'uf} and {Coe}, Dan and {Larson}, Rebecca L. and {Jung}, Intae and {Mingozzi}, Matilde and {Dayal}, Pratika and {Kumari}, Nimisha and {Kokorev}, Vasily and {Vikaeus}, Anton and {Brammer}, Gabriel and {Furtak}, Lukas J. and {Adamo}, Angela and {Andrade-Santos}, Felipe and {Antwi-Danso}, Jacqueline and {Brada{\v{c}}}, Maru{\v{s}}a and {Bradley}, Larry D. and {Broadhurst}, Tom and {Carnall}, Adam C. and {Conselice}, Christopher J. and {Diego}, Jose M. and {Donahue}, Megan and {Eldridge}, Jan J. and {Fujimoto}, Seiji and {Henry}, Alaina and {Hernandez}, Svea and {Hutchison}, Taylor A. and {James}, Bethan L. and {Norman}, Colin and {Park}, Hyunbae and {Pirzkal}, Norbert and {Postman}, Marc and {Ricotti}, Massimo and {Rigby}, Jane R. and {Vanzella}, Eros and {Welch}, Brian and {Wilkins}, Stephen M. and {Windhorst}, Rogier A. and {Xu}, Xinfeng and {Zackrisson}, Erik and {Zitrin}, Adi},
        title = "{JWST NIRSpec Spectroscopy of the Triply Lensed z = 10.17 Galaxy MACS0647─JD}",
      journal = {\apj},
         year = 2024,
        month = sep,
       volume = {973},
       number = {1},
          eid = {8},
        pages = {8},
          doi = {10.3847/1538-4357/ad5da8},
archivePrefix = {arXiv},
       eprint = {2305.03042},
 primaryClass = {astro-ph.GA},
       adsurl = {https://ui.adsabs.harvard.edu/abs/2024ApJ...973....8H}
}

@ARTICLE{Ferrara2025A&A...694A.215F,
       author = {{Ferrara}, A. and {Carniani}, S. and {di Mascia}, F. and {Bouwens}, R.~J. and {Oesch}, P. and {Schouws}, S.},
        title = "{ALMA observations of super-early galaxies: Attenuation-free model predictions}",
      journal = {\aap},
         year = 2025,
        month = feb,
       volume = {694},
          eid = {A215},
        pages = {A215},
          doi = {10.1051/0004-6361/202452368},
archivePrefix = {arXiv},
       eprint = {2409.17223},
 primaryClass = {astro-ph.GA},
       adsurl = {https://ui.adsabs.harvard.edu/abs/2025A&A...694A.215F}
}

@ARTICLE{Carniani2025A&A...696A..87C,
       author = {{Carniani}, Stefano and {D'Eugenio}, Francesco and {Ji}, Xihan and {Parlanti}, Eleonora and {Scholtz}, Jan and {Sun}, Fengwu and {Venturi}, Giacomo and {Bakx}, Tom J.~L.~C. and {Curti}, Mirko and {Maiolino}, Roberto and {Tacchella}, Sandro and {Zavala}, Jorge A. and {Hainline}, Kevin and {Witstok}, Joris and {Johnson}, Benjamin D. and {Alberts}, Stacey and {Bunker}, Andrew J. and {Charlot}, St{\'e}phane and {Eisenstein}, Daniel J. and {Helton}, Jakob M. and {Jakobsen}, Peter and {Kumari}, Nimisha and {Robertson}, Brant and {Saxena}, Aayush and {{\"U}bler}, Hannah and {Williams}, Christina C. and {Willmer}, Christopher N.~A. and {Willott}, Chris},
        title = "{The eventful life of a luminous galaxy at z = 14: metal enrichment, feedback, and low gas fraction?}",
      journal = {\aap},
         year = 2025,
        month = apr,
       volume = {696},
          eid = {A87},
        pages = {A87},
          doi = {10.1051/0004-6361/202452451},
archivePrefix = {arXiv},
       eprint = {2409.20533},
 primaryClass = {astro-ph.GA},
       adsurl = {https://ui.adsabs.harvard.edu/abs/2025A&A...696A..87C}
}

@ARTICLE{Witstok2025Natur.639..897W,
       author = {{Witstok}, Joris and {Jakobsen}, Peter and {Maiolino}, Roberto and {Helton}, Jakob M. and {Johnson}, Benjamin D. and {Robertson}, Brant E. and {Tacchella}, Sandro and {Cameron}, Alex J. and {Smit}, Renske and {Bunker}, Andrew J. and {Saxena}, Aayush and {Sun}, Fengwu and {Alberts}, Stacey and {Arribas}, Santiago and {Baker}, William M. and {Bhatawdekar}, Rachana and {Boyett}, Kristan and {Cargile}, Phillip A. and {Carniani}, Stefano and {Charlot}, St{\'e}phane and {Chevallard}, Jacopo and {Curti}, Mirko and {Curtis-Lake}, Emma and {D'Eugenio}, Francesco and {Eisenstein}, Daniel J. and {Hainline}, Kevin N. and {Jones}, Gareth C. and {Kumari}, Nimisha and {Maseda}, Michael V. and {P{\'e}rez-Gonz{\'a}lez}, Pablo G. and {Rinaldi}, Pierluigi and {Scholtz}, Jan and {{\"U}bler}, Hannah and {Williams}, Christina C. and {Willmer}, Christopher N.~A. and {Willott}, Chris and {Zhu}, Yongda},
        title = "{Witnessing the onset of reionization through Lyman-{\ensuremath{\alpha}} emission at redshift 13}",
      journal = {\nat},
         year = 2025,
        month = mar,
       volume = {639},
       number = {8056},
        pages = {897-901},
          doi = {10.1038/s41586-025-08779-5},
archivePrefix = {arXiv},
       eprint = {2408.16608},
 primaryClass = {astro-ph.GA},
       adsurl = {https://ui.adsabs.harvard.edu/abs/2025Natur.639..897W}
}

@ARTICLE{Naidu2026OJAp....956033N,
       author = {{Naidu}, Rohan P. and {Oesch}, Pascal A. and {Brammer}, Gabriel and {Weibel}, Andrea and {Li}, Yijia and {Matthee}, Jorryt and {Chisolm}, John and {Pollock}, Clara L. and {Heintz}, Kasper E. and {Johnson}, Benjamin D. and {Shen}, Xuejian and {Hviding}, Raphael E. and {Leja}, Joel and {Tacchella}, Sandro and {Ganguly}, Arpita and {Witten}, Callum and {Atek}, Hakim and {Belli}, Siro and {Bose}, Sownak and {Bouwens}, Rychard and {Dayal}, Pratika and {Decarli}, Roberto and {de Graaff}, Anna and {Fudamoto}, Yoshinobu and {Giovinazzo}, Emma and {Greene}, Jenny E. and {Illingworth}, Garth and {Inoue}, Akio K. and {Kane}, Sarah G. and {Labbe}, Ivo and {Leonova}, Ecaterina and {Marques-Chaves}, Rui and {Meyer}, Roman A. and {Nelson}, Erica J. and {Roberts-Borsani}, Guido and {Schaerer}, Daniel and {Simcoe}, Robert A. and {Stefanon}, Mauro and {Sugahara}, Yuma and {Toft}, Sune and {van der Wel}, Arjen and {van Dokkum}, Pieter and {Walter}, Fabian and {Watson}, Darrach and {Weaver}, John R. and {Whitaker}, Katherine E.},
        title = "{A Cosmic Miracle: A Remarkably Luminous Galaxy at zspec = 14.44 Confirmed with JWST}",
      journal = {The Open Journal of Astrophysics},
         year = 2026,
        month = jan,
       volume = {9},
        pages = {56033},
          doi = {10.33232/001c.156033},
archivePrefix = {arXiv},
       eprint = {2505.11263},
 primaryClass = {astro-ph.GA},
       adsurl = {https://ui.adsabs.harvard.edu/abs/2026OJAp....956033N}
}

@ARTICLE{Perez-Gonzalez2025ApJ...991..179P,
       author = {{P{\'e}rez-Gonz{\'a}lez}, Pablo G. and {{\"O}stlin}, G{\"o}ran and {Costantin}, Luca and {Melinder}, Jens and {Finkelstein}, Steven L. and {Somerville}, Rachel S. and {Annunziatella}, Marianna and {{\'A}lvarez-M{\'a}rquez}, Javier and {Colina}, Luis and {Dekel}, Avishai and {Ferguson}, Henry C. and {Li}, Zhaozhou and {Yung}, L.~Y. Aaron and {Bagley}, Micaela B. and {Boogaard}, Leindert A. and {Burgarella}, Denis and {Calabr{\`o}}, Antonello and {Caputi}, Karina I. and {Cheng}, Yingjie and {Dickinson}, Mark and {Eckart}, Andreas and {Giavalisco}, Mauro and {Gillman}, Steven and {Greve}, Thomas R. and {Hamed}, Mahmoud and {Hathi}, Nimish P. and {Hjorth}, Jens and {Huertas-Company}, Marc and {Kartaltepe}, Jeyhan S. and {Koekemoer}, Anton M. and {Kokorev}, Vasily and {Labiano}, {\'A}lvaro and {Langeroodi}, Danial and {Leung}, Gene C.~K. and {Natarajan}, Priyamvada and {Papovich}, Casey and {Peissker}, Florian and {Pentericci}, Laura and {Pirzkal}, Nor and {Rinaldi}, Pierluigi and {van der Werf}, Paul and {Walter}, Fabian},
        title = "{The Rise of the Galactic Empire: Ultraviolet Luminosity Functions at z {\ensuremath{\sim}} 17 and z {\ensuremath{\sim}} 25 Estimated with the MIDIS+NGDEEP Ultra-deep JWST/NIRCam Data Set}",
      journal = {\apj},
         year = 2025,
        month = oct,
       volume = {991},
       number = {2},
          eid = {179},
        pages = {179},
          doi = {10.3847/1538-4357/adf8c9},
archivePrefix = {arXiv},
       eprint = {2503.15594},
 primaryClass = {astro-ph.GA},
       adsurl = {https://ui.adsabs.harvard.edu/abs/2025ApJ...991..179P}
}

@ARTICLE{Castellano2025A&A...704A.158C,
       author = {{Castellano}, M. and {Fontana}, A. and {Merlin}, E. and {Santini}, P. and {Napolitano}, L. and {Menci}, N. and {P{\'e}rez-Gonz{\'a}lez}, P.~G. and {Calabr{\`o}}, A. and {Paris}, D. and {Pentericci}, L. and {Zavala}, J.~A. and {Dickinson}, M. and {Finkelstein}, S.~L. and {Treu}, T. and {Amorin}, R.~O. and {Arrabal Haro}, P. and {Bergamini}, P. and {Bisigello}, L. and {Catone}, M. and {Daddi}, E. and {Dayal}, P. and {Dekel}, A. and {Ferrara}, A. and {Fortuni}, F. and {Gandolfi}, G. and {Giavalisco}, M. and {Grillo}, C. and {Guida}, S.~T. and {Hathi}, N.~P. and {Holwerda}, B.~W. and {Koekemoer}, A.~M. and {Kokorev}, V. and {Li}, Z. and {Llerena}, M. and {Lucas}, R.~A. and {Mascia}, S. and {Metha}, B. and {Morishita}, T. and {Nanayakkara}, T. and {Pacucci}, F. and {Roberts-Borsani}, G. and {Rodighiero}, G. and {Rosati}, P. and {Salazar}, V. and {Schneider}, R. and {Somerville}, R.~S. and {Taylor}, A. and {Trenti}, M. and {Trinca}, A. and {Wang}, X. and {Watson}, P.~J. and {Yang}, L. and {Yung}, L.~Y.~A.},
        title = "{Pushing JWST to the extremes: Search and scrutiny of bright galaxy candidates at z ≃ 15─30}",
      journal = {\aap},
         year = 2025,
        month = dec,
       volume = {704},
          eid = {A158},
        pages = {A158},
          doi = {10.1051/0004-6361/202555082},
archivePrefix = {arXiv},
       eprint = {2504.05893},
 primaryClass = {astro-ph.GA},
       adsurl = {https://ui.adsabs.harvard.edu/abs/2025A&A...704A.158C}
}

@ARTICLE{Calzetti1994ApJ...429..582C,
       author = {{Calzetti}, Daniela and {Kinney}, Anne L. and {Storchi-Bergmann}, Thaisa},
        title = "{Dust Extinction of the Stellar Continua in Starburst Galaxies: The Ultraviolet and Optical Extinction Law}",
      journal = {\apj},
         year = 1994,
        month = jul,
       volume = {429},
        pages = {582},
          doi = {10.1086/174346},
       adsurl = {https://ui.adsabs.harvard.edu/abs/1994ApJ...429..582C}
}

@ARTICLE{Brammer2008ApJ...686.1503B,
       author = {{Brammer}, Gabriel B. and {van Dokkum}, Pieter G. and {Coppi}, Paolo},
        title = "{EAZY: A Fast, Public Photometric Redshift Code}",
      journal = {\apj},
         year = 2008,
        month = oct,
       volume = {686},
       number = {2},
        pages = {1503-1513},
          doi = {10.1086/591786},
archivePrefix = {arXiv},
       eprint = {0807.1533},
 primaryClass = {astro-ph},
       adsurl = {https://ui.adsabs.harvard.edu/abs/2008ApJ...686.1503B}
}

@ARTICLE{Alvarez-Marquez2025A&A...695A.250A,
       author = {{{\'A}lvarez-M{\'a}rquez}, J. and {Crespo G{\'o}mez}, A. and {Colina}, L. and {Langeroodi}, D. and {Marques-Chaves}, R. and {Prieto-Jim{\'e}nez}, C. and {Bik}, A. and {Alonso-Herrero}, A. and {Boogaard}, L. and {Costantin}, L. and {Garc{\'\i}a-Mar{\'\i}n}, M. and {Gillman}, S. and {Hjorth}, J. and {Iani}, E. and {Jermann}, I. and {Labiano}, A. and {Melinder}, J. and {Meyer}, R. and {{\"O}stlin}, G. and {P{\'e}rez-Gonz{\'a}lez}, P.~G. and {Rinaldi}, P. and {Walter}, F. and {van der Werf}, P. and {Wright}, G.},
        title = "{Insight into the starburst nature of Galaxy GN-z11 with JWST MIRI spectroscopy}",
      journal = {\aap},
         year = 2025,
        month = mar,
       volume = {695},
          eid = {A250},
        pages = {A250},
          doi = {10.1051/0004-6361/202451731},
archivePrefix = {arXiv},
       eprint = {2412.12826},
 primaryClass = {astro-ph.GA},
       adsurl = {https://ui.adsabs.harvard.edu/abs/2025A&A...695A.250A}
}

@ARTICLE{Hsiao2024ApJ...973...81H,
       author = {{Hsiao}, Tiger Yu-Yang and {{\'A}lvarez-M{\'a}rquez}, Javier and {Coe}, Dan and {Crespo G{\'o}mez}, Alejandro and {Abdurro'uf} and {Dayal}, Pratika and {Larson}, Rebecca L. and {Bik}, Arjan and {Blanco-Prieto}, Carmen and {Colina}, Luis and {P{\'e}rez-Gonz{\'a}lez}, Pablo Guillermo and {Costantin}, Luca and {Prieto-Jim{\'e}nez}, Carlota and {Adamo}, Angela and {Bradley}, Larry D. and {Conselice}, Christopher J. and {Fujimoto}, Seiji and {Furtak}, Lukas J. and {Hutchison}, Taylor A. and {James}, Bethan L. and {Jim{\'e}nez-Teja}, Yolanda and {Jung}, Intae and {Kokorev}, Vasily and {Mingozzi}, Matilde and {Norman}, Colin and {Ricotti}, Massimo and {Rigby}, Jane R. and {Sharon}, Keren and {Vanzella}, Eros and {Welch}, Brian and {Xu}, Xinfeng and {Zackrisson}, Erik and {Zitrin}, Adi},
        title = "{JWST MIRI Detections of H{\ensuremath{\alpha}} and [O III] and a Direct Metallicity Measurement of the z = 10.17 Lensed Galaxy MACS0647{\ensuremath{-}}JD}",
      journal = {\apj},
         year = 2024,
        month = oct,
       volume = {973},
       number = {2},
          eid = {81},
        pages = {81},
          doi = {10.3847/1538-4357/ad6562},
archivePrefix = {arXiv},
       eprint = {2404.16200},
 primaryClass = {astro-ph.GA},
       adsurl = {https://ui.adsabs.harvard.edu/abs/2024ApJ...973...81H}
}

@ARTICLE{Planck2020A&A...641A...6P,
       author = {{Planck Collaboration} and {Aghanim}, N. and {Akrami}, Y. and {Ashdown}, M. and {Aumont}, J. and {Baccigalupi}, C. and {Ballardini}, M. and {Banday}, A.~J. and {Barreiro}, R.~B. and {Bartolo}, N. and {Basak}, S. and {Battye}, R. and {Benabed}, K. and {Bernard}, J.-P. and {Bersanelli}, M. and {Bielewicz}, P. and {Bock}, J.~J. and {Bond}, J.~R. and {Borrill}, J. and {Bouchet}, F.~R. and {Boulanger}, F. and {Bucher}, M. and {Burigana}, C. and {Butler}, R.~C. and {Calabrese}, E. and {Cardoso}, J.-F. and {Carron}, J. and {Challinor}, A. and {Chiang}, H.~C. and {Chluba}, J. and {Colombo}, L.~P.~L. and {Combet}, C. and {Contreras}, D. and {Crill}, B.~P. and {Cuttaia}, F. and {de Bernardis}, P. and {de Zotti}, G. and {Delabrouille}, J. and {Delouis}, J.-M. and {Di Valentino}, E. and {Diego}, J.~M. and {Dor{\'e}}, O. and {Douspis}, M. and {Ducout}, A. and {Dupac}, X. and {Dusini}, S. and {Efstathiou}, G. and {Elsner}, F. and {En{\ss}lin}, T.~A. and {Eriksen}, H.~K. and {Fantaye}, Y. and {Farhang}, M. and {Fergusson}, J. and {Fernandez-Cobos}, R. and {Finelli}, F. and {Forastieri}, F. and {Frailis}, M. and {Fraisse}, A.~A. and {Franceschi}, E. and {Frolov}, A. and {Galeotta}, S. and {Galli}, S. and {Ganga}, K. and {G{\'e}nova-Santos}, R.~T. and {Gerbino}, M. and {Ghosh}, T. and {Gonz{\'a}lez-Nuevo}, J. and {G{\'o}rski}, K.~M. and {Gratton}, S. and {Gruppuso}, A. and {Gudmundsson}, J.~E. and {Hamann}, J. and {Handley}, W. and {Hansen}, F.~K. and {Herranz}, D. and {Hildebrandt}, S.~R. and {Hivon}, E. and {Huang}, Z. and {Jaffe}, A.~H. and {Jones}, W.~C. and {Karakci}, A. and {Keih{\"a}nen}, E. and {Keskitalo}, R. and {Kiiveri}, K. and {Kim}, J. and {Kisner}, T.~S. and {Knox}, L. and {Krachmalnicoff}, N. and {Kunz}, M. and {Kurki-Suonio}, H. and {Lagache}, G. and {Lamarre}, J.-M. and {Lasenby}, A. and {Lattanzi}, M. and {Lawrence}, C.~R. and {Le Jeune}, M. and {Lemos}, P. and {Lesgourgues}, J. and {Levrier}, F. and {Lewis}, A. and {Liguori}, M. and {Lilje}, P.~B. and {Lilley}, M. and {Lindholm}, V. and {L{\'o}pez-Caniego}, M. and {Lubin}, P.~M. and {Ma}, Y.-Z. and {Mac{\'\i}as-P{\'e}rez}, J.~F. and {Maggio}, G. and {Maino}, D. and {Mandolesi}, N. and {Mangilli}, A. and {Marcos-Caballero}, A. and {Maris}, M. and {Martin}, P.~G. and {Martinelli}, M. and {Mart{\'\i}nez-Gonz{\'a}lez}, E. and {Matarrese}, S. and {Mauri}, N. and {McEwen}, J.~D. and {Meinhold}, P.~R. and {Melchiorri}, A. and {Mennella}, A. and {Migliaccio}, M. and {Millea}, M. and {Mitra}, S. and {Miville-Desch{\^e}nes}, M.-A. and {Molinari}, D. and {Montier}, L. and {Morgante}, G. and {Moss}, A. and {Natoli}, P. and {N{\o}rgaard-Nielsen}, H.~U. and {Pagano}, L. and {Paoletti}, D. and {Partridge}, B. and {Patanchon}, G. and {Peiris}, H.~V. and {Perrotta}, F. and {Pettorino}, V. and {Piacentini}, F. and {Polastri}, L. and {Polenta}, G. and {Puget}, J.-L. and {Rachen}, J.~P. and {Reinecke}, M. and {Remazeilles}, M. and {Renzi}, A. and {Rocha}, G. and {Rosset}, C. and {Roudier}, G. and {Rubi{\~n}o-Mart{\'\i}n}, J.~A. and {Ruiz-Granados}, B. and {Salvati}, L. and {Sandri}, M. and {Savelainen}, M. and {Scott}, D. and {Shellard}, E.~P.~S. and {Sirignano}, C. and {Sirri}, G. and {Spencer}, L.~D. and {Sunyaev}, R. and {Suur-Uski}, A.-S. and {Tauber}, J.~A. and {Tavagnacco}, D. and {Tenti}, M. and {Toffolatti}, L. and {Tomasi}, M. and {Trombetti}, T. and {Valenziano}, L. and {Valiviita}, J. and {Van Tent}, B. and {Vibert}, L. and {Vielva}, P. and {Villa}, F. and {Vittorio}, N. and {Wandelt}, B.~D. and {Wehus}, I.~K. and {White}, M. and {White}, S.~D.~M. and {Zacchei}, A. and {Zonca}, A.},
        title = "{Planck 2018 results. VI. Cosmological parameters}",
      journal = {\aap},
         year = 2020,
        month = sep,
       volume = {641},
          eid = {A6},
        pages = {A6},
          doi = {10.1051/0004-6361/201833910},
archivePrefix = {arXiv},
       eprint = {1807.06209},
 primaryClass = {astro-ph.CO},
       adsurl = {https://ui.adsabs.harvard.edu/abs/2020A&A...641A...6P}
}

@ARTICLE{Schaerer2025A&A...693A.271S,
       author = {{Schaerer}, D. and {Guibert}, J. and {Marques-Chaves}, R. and {Martins}, F.},
        title = "{Observable and ionizing properties of star-forming galaxies with very massive stars and different initial mass functions}",
      journal = {\aap},
         year = 2025,
        month = jan,
       volume = {693},
          eid = {A271},
        pages = {A271},
          doi = {10.1051/0004-6361/202451454},
archivePrefix = {arXiv},
       eprint = {2407.12122},
 primaryClass = {astro-ph.GA},
       adsurl = {https://ui.adsabs.harvard.edu/abs/2025A&A...693A.271S}
}

@ARTICLE{Heintz2024Sci...384..890H,
       author = {{Heintz}, Kasper E. and {Watson}, Darach and {Brammer}, Gabriel and {Vejlgaard}, Simone and {Hutter}, Anne and {Strait}, Victoria B. and {Matthee}, Jorryt and {Oesch}, Pascal A. and {Jakobsson}, P{\'a}ll and {Tanvir}, Nial R. and {Laursen}, Peter and {Naidu}, Rohan P. and {Mason}, Charlotte A. and {Killi}, Meghana and {Jung}, Intae and {Hsiao}, Tiger Yu-Yang and {Abdurro'uf} and {Coe}, Dan and {Arrabal Haro}, Pablo and {Finkelstein}, Steven L. and {Toft}, Sune},
        title = "{Strong damped Lyman-{\ensuremath{\alpha}} absorption in young star-forming galaxies at redshifts 9 to 11}",
      journal = {Science},
         year = 2024,
        month = may,
       volume = {384},
       number = {6698},
        pages = {890-894},
          doi = {10.1126/science.adj0343},
archivePrefix = {arXiv},
       eprint = {2306.00647},
 primaryClass = {astro-ph.GA},
       adsurl = {https://ui.adsabs.harvard.edu/abs/2024Sci...384..890H}
}

@ARTICLE{de_Graaff2025A&A...697A.189D,
       author = {{de Graaff}, Anna and {Brammer}, Gabriel and {Weibel}, Andrea and {Lewis}, Zach and {Maseda}, Michael V. and {Oesch}, Pascal A. and {Bezanson}, Rachel and {Boogaard}, Leindert A. and {Cleri}, Nikko J. and {Cooper}, Olivia R. and {Gottumukkala}, Rashmi and {Greene}, Jenny E. and {Hirschmann}, Michaela and {Hviding}, Raphael E. and {Katz}, Harley and {Labb{\'e}}, Ivo and {Leja}, Joel and {Matthee}, Jorryt and {McConachie}, Ian and {Miller}, Tim B. and {Naidu}, Rohan P. and {Price}, Sedona H. and {Rix}, Hans-Walter and {Setton}, David J. and {Suess}, Katherine A. and {Wang}, Bingjie and {Whitaker}, Katherine E. and {Williams}, Christina C.},
        title = "{RUBIES: A complete census of the bright and red distant Universe with JWST/NIRSpec}",
      journal = {\aap},
         year = 2025,
        month = may,
       volume = {697},
          eid = {A189},
        pages = {A189},
          doi = {10.1051/0004-6361/202452186},
archivePrefix = {arXiv},
       eprint = {2409.05948},
 primaryClass = {astro-ph.GA},
       adsurl = {https://ui.adsabs.harvard.edu/abs/2025A&A...697A.189D}
}

@INPROCEEDINGS{Perrin2014SPIE.9143E..3XP,
       author = {{Perrin}, Marshall D. and {Sivaramakrishnan}, Anand and {Lajoie}, Charles-Philippe and {Elliott}, Erin and {Pueyo}, Laurent and {Ravindranath}, Swara and {Albert}, Lo{\"\i}c.},
        title = "{Updated point spread function simulations for JWST with WebbPSF}",
    booktitle = {Space Telescopes and Instrumentation 2014: Optical, Infrared, and Millimeter Wave},
         year = 2014,
       editor = {{Oschmann}, Jr., Jacobus M. and {Clampin}, Mark and {Fazio}, Giovanni G. and {MacEwen}, Howard A.},
       series = {Society of Photo-Optical Instrumentation Engineers (SPIE) Conference Series},
       volume = {9143},
        month = aug,
          eid = {91433X},
        pages = {91433X},
          doi = {10.1117/12.2056689},
       adsurl = {https://ui.adsabs.harvard.edu/abs/2014SPIE.9143E..3XP}
}

@ARTICLE{Foreman-Mackey2013PASP..125..306F,
       author = {{Foreman-Mackey}, Daniel and {Hogg}, David W. and {Lang}, Dustin and {Goodman}, Jonathan},
        title = "{emcee: The MCMC Hammer}",
      journal = {\pasp},
         year = 2013,
        month = mar,
       volume = {125},
       number = {925},
        pages = {306},
          doi = {10.1086/670067},
archivePrefix = {arXiv},
       eprint = {1202.3665},
 primaryClass = {astro-ph.IM},
       adsurl = {https://ui.adsabs.harvard.edu/abs/2013PASP..125..306F}
}

@ARTICLE{Mowla2024Natur.636..332M,
       author = {{Mowla}, Lamiya and {Iyer}, Kartheik and {Asada}, Yoshihisa and {Desprez}, Guillaume and {Tan}, Vivian Yun Yan and {Martis}, Nicholas and {Sarrouh}, Ghassan and {Strait}, Victoria and {Abraham}, Roberto and {Brada{\v{c}}}, Maru{\v{s}}a and {Brammer}, Gabriel and {Muzzin}, Adam and {Pacifici}, Camilla and {Ravindranath}, Swara and {Sawicki}, Marcin and {Willott}, Chris and {Estrada-Carpenter}, Vince and {Jahan}, Nusrath and {Noirot}, Ga{\"e}l and {Matharu}, Jasleen and {Rihtar{\v{s}}i{\v{c}}}, Gregor and {Zabl}, Johannes},
        title = "{Formation of a low-mass galaxy from star clusters in a 600-million-year-old Universe}",
      journal = {\nat},
         year = 2024,
        month = dec,
       volume = {636},
       number = {8042},
        pages = {332-336},
          doi = {10.1038/s41586-024-08293-0},
archivePrefix = {arXiv},
       eprint = {2402.08696},
 primaryClass = {astro-ph.GA},
       adsurl = {https://ui.adsabs.harvard.edu/abs/2024Natur.636..332M}
}

@ARTICLE{Mandelker2014MNRAS.443.3675M,
       author = {{Mandelker}, Nir and {Dekel}, Avishai and {Ceverino}, Daniel and {Tweed}, Dylan and {Moody}, Christopher E. and {Primack}, Joel},
        title = "{The population of giant clumps in simulated high-z galaxies: in situ and ex situ migration and survival}",
      journal = {\mnras},
         year = 2014,
        month = oct,
       volume = {443},
       number = {4},
        pages = {3675-3702},
          doi = {10.1093/mnras/stu1340},
archivePrefix = {arXiv},
       eprint = {1311.0013},
 primaryClass = {astro-ph.CO},
       adsurl = {https://ui.adsabs.harvard.edu/abs/2014MNRAS.443.3675M}
}

@ARTICLE{Renaud2014MNRAS.442L..33R,
       author = {{Renaud}, F. and {Bournaud}, F. and {Kraljic}, K. and {Duc}, P.-A.},
        title = "{Starbursts triggered by intergalactic tides andinterstellar compressive turbulence.}",
      journal = {\mnras},
         year = 2014,
        month = jul,
       volume = {442},
        pages = {L33-L37},
          doi = {10.1093/mnrasl/slu050},
archivePrefix = {arXiv},
       eprint = {1403.7316},
 primaryClass = {astro-ph.GA},
       adsurl = {https://ui.adsabs.harvard.edu/abs/2014MNRAS.442L..33R}
}

@ARTICLE{Jeong2025ApJ...980...10J,
       author = {{Jeong}, Tae Bong and {Jeon}, Myoungwon and {Song}, Hyunmi and {Bromm}, Volker},
        title = "{Simulating High-redshift Galaxies: Enhancing UV Luminosity with Star Formation Efficiency and a Top-heavy IMF}",
      journal = {\apj},
         year = 2025,
        month = feb,
       volume = {980},
       number = {1},
          eid = {10},
        pages = {10},
          doi = {10.3847/1538-4357/ada27d},
archivePrefix = {arXiv},
       eprint = {2411.17007},
 primaryClass = {astro-ph.GA},
       adsurl = {https://ui.adsabs.harvard.edu/abs/2025ApJ...980...10J}
}

@ARTICLE{Pallottini2023A&A...677L...4P,
       author = {{Pallottini}, A. and {Ferrara}, A.},
        title = "{Stochastic star formation in early galaxies: Implications for the James Webb Space Telescope}",
      journal = {\aap},
         year = 2023,
        month = sep,
       volume = {677},
          eid = {L4},
        pages = {L4},
          doi = {10.1051/0004-6361/202347384},
archivePrefix = {arXiv},
       eprint = {2307.03219},
 primaryClass = {astro-ph.GA},
       adsurl = {https://ui.adsabs.harvard.edu/abs/2023A&A...677L...4P}
}

@ARTICLE{Mason2023MNRAS.521..497M,
       author = {{Mason}, Charlotte A. and {Trenti}, Michele and {Treu}, Tommaso},
        title = "{The brightest galaxies at cosmic dawn}",
      journal = {\mnras},
         year = 2023,
        month = may,
       volume = {521},
       number = {1},
        pages = {497-503},
          doi = {10.1093/mnras/stad035},
archivePrefix = {arXiv},
       eprint = {2207.14808},
 primaryClass = {astro-ph.GA},
       adsurl = {https://ui.adsabs.harvard.edu/abs/2023MNRAS.521..497M}
}

@ARTICLE{Carvajal-Bohorquez2025A&A...704A.290C,
       author = {{Carvajal-Bohorquez}, C. and {Ciesla}, L. and {Laporte}, N. and {Boquien}, M. and {Buat}, V. and {Ilbert}, O. and {Aufort}, G. and {Shuntov}, M. and {Witten}, C. and {Oesch}, P.~A. and {Covelo-Paz}, A.},
        title = "{Stochastic star formation activity of galaxies within the first billion years probed by JWST}",
      journal = {\aap},
         year = 2025,
        month = dec,
       volume = {704},
          eid = {A290},
        pages = {A290},
          doi = {10.1051/0004-6361/202556471},
archivePrefix = {arXiv},
       eprint = {2507.13160},
 primaryClass = {astro-ph.GA},
       adsurl = {https://ui.adsabs.harvard.edu/abs/2025A&A...704A.290C}
}

@ARTICLE{Asada2024MNRAS.52711372A,
       author = {{Asada}, Yoshihisa and {Sawicki}, Marcin and {Abraham}, Roberto and {Brada{\v{c}}}, Maru{\v{s}}a and {Brammer}, Gabriel and {Desprez}, Guillaume and {Estrada-Carpenter}, Vince and {Iyer}, Kartheik and {Martis}, Nicholas and {Matharu}, Jasleen and {Mowla}, Lamiya and {Muzzin}, Adam and {Noirot}, Ga{\"e}l and {Sarrouh}, Ghassan T.~E. and {Strait}, Victoria and {Willott}, Chris J. and {Harshan}, Anishya},
        title = "{Bursty star formation and galaxy-galaxy interactions in low-mass galaxies 1 Gyr after the Big Bang}",
      journal = {\mnras},
         year = 2024,
        month = feb,
       volume = {527},
       number = {4},
        pages = {11372-11392},
          doi = {10.1093/mnras/stad3902},
archivePrefix = {arXiv},
       eprint = {2310.02314},
 primaryClass = {astro-ph.GA},
       adsurl = {https://ui.adsabs.harvard.edu/abs/2024MNRAS.52711372A}
}

@ARTICLE{Inoue2011MNRAS.415.2920I,
       author = {{Inoue}, Akio K.},
        title = "{Rest-frame ultraviolet-to-optical spectral characteristics of extremely metal-poor and metal-free galaxies}",
      journal = {\mnras},
         year = 2011,
        month = aug,
       volume = {415},
       number = {3},
        pages = {2920-2931},
          doi = {10.1111/j.1365-2966.2011.18906.x},
archivePrefix = {arXiv},
       eprint = {1102.5150},
 primaryClass = {astro-ph.CO},
       adsurl = {https://ui.adsabs.harvard.edu/abs/2011MNRAS.415.2920I}
}

@ARTICLE{Storey2000MNRAS.312..813S,
       author = {{Storey}, P.~J. and {Zeippen}, C.~J.},
        title = "{Theoretical values for the [OIII] 5007/4959 line-intensity ratio and homologous cases}",
      journal = {\mnras},
         year = 2000,
        month = mar,
       volume = {312},
       number = {4},
        pages = {813-816},
          doi = {10.1046/j.1365-8711.2000.03184.x},
       adsurl = {https://ui.adsabs.harvard.edu/abs/2000MNRAS.312..813S}
}

@ARTICLE{Carnall2019MNRAS.490..417C,
       author = {{Carnall}, A.~C. and {McLure}, R.~J. and {Dunlop}, J.~S. and {Cullen}, F. and {McLeod}, D.~J. and {Wild}, V. and {Johnson}, B.~D. and {Appleby}, S. and {Dav{\'e}}, R. and {Amorin}, R. and {Bolzonella}, M. and {Castellano}, M. and {Cimatti}, A. and {Cucciati}, O. and {Gargiulo}, A. and {Garilli}, B. and {Marchi}, F. and {Pentericci}, L. and {Pozzetti}, L. and {Schreiber}, C. and {Talia}, M. and {Zamorani}, G.},
        title = "{The VANDELS survey: the star-formation histories of massive quiescent galaxies at 1.0 < z < 1.3}",
      journal = {\mnras},
         year = 2019,
        month = nov,
       volume = {490},
       number = {1},
        pages = {417-439},
          doi = {10.1093/mnras/stz2544},
archivePrefix = {arXiv},
       eprint = {1903.11082},
 primaryClass = {astro-ph.GA},
       adsurl = {https://ui.adsabs.harvard.edu/abs/2019MNRAS.490..417C}
}

@ARTICLE{Boylan-Kolchin2023,
       author = {{Boylan-Kolchin}, Michael},
        title = "{Stress testing {\ensuremath{\Lambda}}CDM with high-redshift galaxy candidates}",
      journal = {Nature Astronomy},
         year = 2023,
        month = jun,
       volume = {7},
        pages = {731-735},
          doi = {10.1038/s41550-023-01937-7},
archivePrefix = {arXiv},
       eprint = {2208.01611},
 primaryClass = {astro-ph.CO},
       adsurl = {https://ui.adsabs.harvard.edu/abs/2023NatAs...7..731B}
}

@ARTICLE{Carnall2018MNRAS.480.4379C,
       author = {{Carnall}, A.~C. and {McLure}, R.~J. and {Dunlop}, J.~S. and {Dav{\'e}}, R.},
        title = "{Inferring the star formation histories of massive quiescent galaxies with BAGPIPES: evidence for multiple quenching mechanisms}",
      journal = {\mnras},
         year = 2018,
        month = nov,
       volume = {480},
       number = {4},
        pages = {4379-4401},
          doi = {10.1093/mnras/sty2169},
archivePrefix = {arXiv},
       eprint = {1712.04452},
 primaryClass = {astro-ph.GA},
       adsurl = {https://ui.adsabs.harvard.edu/abs/2018MNRAS.480.4379C}
}

@ARTICLE{2013A&A...558A..33A,
       author = {{Astropy Collaboration} and {Robitaille}, Thomas P. and {Tollerud}, Erik J. and {Greenfield}, Perry and {Droettboom}, Michael and {Bray}, Erik and {Aldcroft}, Tom and {Davis}, Matt and {Ginsburg}, Adam and {Price-Whelan}, Adrian M. and {Kerzendorf}, Wolfgang E. and {Conley}, Alexander and {Crighton}, Neil and {Barbary}, Kyle and {Muna}, Demitri and {Ferguson}, Henry and {Grollier}, Fr{\'e}d{\'e}ric and {Parikh}, Madhura M. and {Nair}, Prasanth H. and {Unther}, Hans M. and {Deil}, Christoph and {Woillez}, Julien and {Conseil}, Simon and {Kramer}, Roban and {Turner}, James E.~H. and {Singer}, Leo and {Fox}, Ryan and {Weaver}, Benjamin A. and {Zabalza}, Victor and {Edwards}, Zachary I. and {Azalee Bostroem}, K. and {Burke}, D.~J. and {Casey}, Andrew R. and {Crawford}, Steven M. and {Dencheva}, Nadia and {Ely}, Justin and {Jenness}, Tim and {Labrie}, Kathleen and {Lim}, Pey Lian and {Pierfederici}, Francesco and {Pontzen}, Andrew and {Ptak}, Andy and {Refsdal}, Brian and {Servillat}, Mathieu and {Streicher}, Ole},
        title = "{Astropy: A community Python package for astronomy}",
      journal = {\aap},
         year = 2013,
        month = oct,
       volume = {558},
          eid = {A33},
        pages = {A33},
          doi = {10.1051/0004-6361/201322068},
archivePrefix = {arXiv},
       eprint = {1307.6212},
 primaryClass = {astro-ph.IM},
       adsurl = {https://ui.adsabs.harvard.edu/abs/2013A&A...558A..33A}
}

@ARTICLE{2018AJ....156..123A,
       author = {{Astropy Collaboration} and {Price-Whelan}, A.~M. and {Sip{\H{o}}cz}, B.~M. and {G{\"u}nther}, H.~M. and {Lim}, P.~L. and {Crawford}, S.~M. and {Conseil}, S. and {Shupe}, D.~L. and {Craig}, M.~W. and {Dencheva}, N. and {Ginsburg}, A. and {VanderPlas}, J.~T. and {Bradley}, L.~D. and {P{\'e}rez-Su{\'a}rez}, D. and {de Val-Borro}, M. and {Aldcroft}, T.~L. and {Cruz}, K.~L. and {Robitaille}, T.~P. and {Tollerud}, E.~J. and {Ardelean}, C. and {Babej}, T. and {Bach}, Y.~P. and {Bachetti}, M. and {Bakanov}, A.~V. and {Bamford}, S.~P. and {Barentsen}, G. and {Barmby}, P. and {Baumbach}, A. and {Berry}, K.~L. and {Biscani}, F. and {Boquien}, M. and {Bostroem}, K.~A. and {Bouma}, L.~G. and {Brammer}, G.~B. and {Bray}, E.~M. and {Breytenbach}, H. and {Buddelmeijer}, H. and {Burke}, D.~J. and {Calderone}, G. and {Cano Rodr{\'\i}guez}, J.~L. and {Cara}, M. and {Cardoso}, J.~V.~M. and {Cheedella}, S. and {Copin}, Y. and {Corrales}, L. and {Crichton}, D. and {D'Avella}, D. and {Deil}, C. and {Depagne}, {\'E}. and {Dietrich}, J.~P. and {Donath}, A. and {Droettboom}, M. and {Earl}, N. and {Erben}, T. and {Fabbro}, S. and {Ferreira}, L.~A. and {Finethy}, T. and {Fox}, R.~T. and {Garrison}, L.~H. and {Gibbons}, S.~L.~J. and {Goldstein}, D.~A. and {Gommers}, R. and {Greco}, J.~P. and {Greenfield}, P. and {Groener}, A.~M. and {Grollier}, F. and {Hagen}, A. and {Hirst}, P. and {Homeier}, D. and {Horton}, A.~J. and {Hosseinzadeh}, G. and {Hu}, L. and {Hunkeler}, J.~S. and {Ivezi{\'c}}, {\v{Z}}. and {Jain}, A. and {Jenness}, T. and {Kanarek}, G. and {Kendrew}, S. and {Kern}, N.~S. and {Kerzendorf}, W.~E. and {Khvalko}, A. and {King}, J. and {Kirkby}, D. and {Kulkarni}, A.~M. and {Kumar}, A. and {Lee}, A. and {Lenz}, D. and {Littlefair}, S.~P. and {Ma}, Z. and {Macleod}, D.~M. and {Mastropietro}, M. and {McCully}, C. and {Montagnac}, S. and {Morris}, B.~M. and {Mueller}, M. and {Mumford}, S.~J. and {Muna}, D. and {Murphy}, N.~A. and {Nelson}, S. and {Nguyen}, G.~H. and {Ninan}, J.~P. and {N{\"o}the}, M. and {Ogaz}, S. and {Oh}, S. and {Parejko}, J.~K. and {Parley}, N. and {Pascual}, S. and {Patil}, R. and {Patil}, A.~A. and {Plunkett}, A.~L. and {Prochaska}, J.~X. and {Rastogi}, T. and {Reddy Janga}, V. and {Sabater}, J. and {Sakurikar}, P. and {Seifert}, M. and {Sherbert}, L.~E. and {Sherwood-Taylor}, H. and {Shih}, A.~Y. and {Sick}, J. and {Silbiger}, M.~T. and {Singanamalla}, S. and {Singer}, L.~P. and {Sladen}, P.~H. and {Sooley}, K.~A. and {Sornarajah}, S. and {Streicher}, O. and {Teuben}, P. and {Thomas}, S.~W. and {Tremblay}, G.~R. and {Turner}, J.~E.~H. and {Terr{\'o}n}, V. and {van Kerkwijk}, M.~H. and {de la Vega}, A. and {Watkins}, L.~L. and {Weaver}, B.~A. and {Whitmore}, J.~B. and {Woillez}, J. and {Zabalza}, V. and {Astropy Contributors}},
        title = "{The Astropy Project: Building an Open-science Project and Status of the v2.0 Core Package}",
      journal = {\aj},
         year = 2018,
        month = sep,
       volume = {156},
       number = {3},
          eid = {123},
        pages = {123},
          doi = {10.3847/1538-3881/aabc4f},
archivePrefix = {arXiv},
       eprint = {1801.02634},
 primaryClass = {astro-ph.IM},
       adsurl = {https://ui.adsabs.harvard.edu/abs/2018AJ....156..123A}
}

@ARTICLE{2022ApJ...935..167A,
       author = {{Astropy Collaboration} and {Price-Whelan}, Adrian M. and {Lim}, Pey Lian and {Earl}, Nicholas and {Starkman}, Nathaniel and {Bradley}, Larry and {Shupe}, David L. and {Patil}, Aarya A. and {Corrales}, Lia and {Brasseur}, C.~E. and {N{\"o}the}, Maximilian and {Donath}, Axel and {Tollerud}, Erik and {Morris}, Brett M. and {Ginsburg}, Adam and {Vaher}, Eero and {Weaver}, Benjamin A. and {Tocknell}, James and {Jamieson}, William and {van Kerkwijk}, Marten H. and {Robitaille}, Thomas P. and {Merry}, Bruce and {Bachetti}, Matteo and {G{\"u}nther}, H. Moritz and {Aldcroft}, Thomas L. and {Alvarado-Montes}, Jaime A. and {Archibald}, Anne M. and {B{\'o}di}, Attila and {Bapat}, Shreyas and {Barentsen}, Geert and {Baz{\'a}n}, Juanjo and {Biswas}, Manish and {Boquien}, M{\'e}d{\'e}ric and {Burke}, D.~J. and {Cara}, Daria and {Cara}, Mihai and {Conroy}, Kyle E. and {Conseil}, Simon and {Craig}, Matthew W. and {Cross}, Robert M. and {Cruz}, Kelle L. and {D'Eugenio}, Francesco and {Dencheva}, Nadia and {Devillepoix}, Hadrien A.~R. and {Dietrich}, J{\"o}rg P. and {Eigenbrot}, Arthur Davis and {Erben}, Thomas and {Ferreira}, Leonardo and {Foreman-Mackey}, Daniel and {Fox}, Ryan and {Freij}, Nabil and {Garg}, Suyog and {Geda}, Robel and {Glattly}, Lauren and {Gondhalekar}, Yash and {Gordon}, Karl D. and {Grant}, David and {Greenfield}, Perry and {Groener}, Austen M. and {Guest}, Steve and {Gurovich}, Sebastian and {Handberg}, Rasmus and {Hart}, Akeem and {Hatfield-Dodds}, Zac and {Homeier}, Derek and {Hosseinzadeh}, Griffin and {Jenness}, Tim and {Jones}, Craig K. and {Joseph}, Prajwel and {Kalmbach}, J. Bryce and {Karamehmetoglu}, Emir and {Ka{\l}uszy{\'n}ski}, Miko{\l}aj and {Kelley}, Michael S.~P. and {Kern}, Nicholas and {Kerzendorf}, Wolfgang E. and {Koch}, Eric W. and {Kulumani}, Shankar and {Lee}, Antony and {Ly}, Chun and {Ma}, Zhiyuan and {MacBride}, Conor and {Maljaars}, Jakob M. and {Muna}, Demitri and {Murphy}, N.~A. and {Norman}, Henrik and {O'Steen}, Richard and {Oman}, Kyle A. and {Pacifici}, Camilla and {Pascual}, Sergio and {Pascual-Granado}, J. and {Patil}, Rohit R. and {Perren}, Gabriel I. and {Pickering}, Timothy E. and {Rastogi}, Tanuj and {Roulston}, Benjamin R. and {Ryan}, Daniel F. and {Rykoff}, Eli S. and {Sabater}, Jose and {Sakurikar}, Parikshit and {Salgado}, Jes{\'u}s and {Sanghi}, Aniket and {Saunders}, Nicholas and {Savchenko}, Volodymyr and {Schwardt}, Ludwig and {Seifert-Eckert}, Michael and {Shih}, Albert Y. and {Jain}, Anany Shrey and {Shukla}, Gyanendra and {Sick}, Jonathan and {Simpson}, Chris and {Singanamalla}, Sudheesh and {Singer}, Leo P. and {Singhal}, Jaladh and {Sinha}, Manodeep and {Sip{\H{o}}cz}, Brigitta M. and {Spitler}, Lee R. and {Stansby}, David and {Streicher}, Ole and {{\v{S}}umak}, Jani and {Swinbank}, John D. and {Taranu}, Dan S. and {Tewary}, Nikita and {Tremblay}, Grant R. and {de Val-Borro}, Miguel and {Van Kooten}, Samuel J. and {Vasovi{\'c}}, Zlatan and {Verma}, Shresth and {de Miranda Cardoso}, Jos{\'e} Vin{\'\i}cius and {Williams}, Peter K.~G. and {Wilson}, Tom J. and {Winkel}, Benjamin and {Wood-Vasey}, W.~M. and {Xue}, Rui and {Yoachim}, Peter and {Zhang}, Chen and {Zonca}, Andrea and {Astropy Project Contributors}},
        title = "{The Astropy Project: Sustaining and Growing a Community-oriented Open-source Project and the Latest Major Release (v5.0) of the Core Package}",
      journal = {\apj},
         year = 2022,
        month = aug,
       volume = {935},
       number = {2},
          eid = {167},
        pages = {167},
          doi = {10.3847/1538-4357/ac7c74},
archivePrefix = {arXiv},
       eprint = {2206.14220},
 primaryClass = {astro-ph.IM},
       adsurl = {https://ui.adsabs.harvard.edu/abs/2022ApJ...935..167A}
}

@software{Brammer2022zndo...7299500B,
       author = {{Brammer}, Gabriel},
        title = "{msaexp: NIRSpec analyis tools}",
         year = 2023,
        month = sep,
          eid = {10.5281/zenodo.7299500},
          doi = {10.5281/zenodo.7299500},
      version = {0.6.17},
    publisher = {Zenodo},
       adsurl = {https://ui.adsabs.harvard.edu/abs/2022zndo...7299500B}
}

@software{bradley_2026_19636730,
  author       = {Bradley, Larry and
                  Sipőcz, Brigitta M. and
                  Robitaille, T. P. and
                  Tollerud, E. J. and
                  Vinícius, Zé and
                  Deil, Christoph and
                  Barbary, Kyle and
                  Wilson, Tom J. and
                  Busko, Ivo and
                  Donath, Axel and
                  Günther, Hans Moritz and
                  Cara, Mihai and
                  Lim, P. L. and
                  Meßlinger, Sebastian and
                  Conseil, Simon and
                  Droettboom, Michael and
                  Bostroem, K. Azalee and
                  Bray, E. M. and
                  Bratholm, Lars Andersen and
                  Burnett, Zach and
                  Jamieson, William and
                  Ginsburg, Adam and
                  Taranu, Dan and
                  Barentsen, Geert and
                  Craig, Matthew W. and
                  Morris, Brett M. and
                  Perrin, Marshall and
                  Rathi, Shivangee},
  title        = {Photutils},
  month        = apr,
  year         = 2026,
  publisher    = {Zenodo},
  version      = {3.0.0},
  doi          = {10.5281/zenodo.19636730},
  url          = {https://doi.org/10.5281/zenodo.19636730},
  swhid        = {swh:1:dir:5ea90432cd987336a26af201c1b50343b12e2daa
                   ;origin=https://doi.org/10.5281/zenodo.596036;visi
                   t=swh:1:snp:25dd84d95353ec3813baa8785c248017da7192
                   33;anchor=swh:1:rel:253cbde27398f9aea9d7368f386661
                   8c00fe16f9;path=astropy-photutils-3322558
                  },
}

@ARTICLE{Burgarella2026arXiv260509829B,
       author = {{Burgarella}, D. and {Buat}, V. and {Inoue}, A.~K. and {Takeuchi}, T.~T. and {Aurin}, C. and {Bouret}, J.-C. and {Dayal}, P. and {Dewachter}, T. and {Dickinson}, M. and {Kobayashi}, C. and {Nikopoulos}, G.~P. and {Somerville}, R.~S.},
        title = "{Stardust Galaxies at z>9: A Dust-Origin Transition Behind the Excess of UV-Bright Galaxies}",
      journal = {arXiv e-prints},
         year = 2026,
        month = may,
          eid = {arXiv:2605.09829},
        pages = {arXiv:2605.09829},
          doi = {10.48550/arXiv.2605.09829},
archivePrefix = {arXiv},
       eprint = {2605.09829},
 primaryClass = {astro-ph.GA},
       adsurl = {https://ui.adsabs.harvard.edu/abs/2026arXiv260509829B}
}

@ARTICLE{Burgarella2025A&A...699A.336B,
       author = {{Burgarella}, Denis and {Buat}, V{\'e}ronique and {Theul{\'e}}, Patrice and {Zavala}, Jorge and {Dickinson}, Mark and {Arrabal Haro}, Pablo and {Bagley}, Micaela B. and {Boquien}, M{\'e}d{\'e}ric and {Cleri}, Nikko and {Dewachter}, Tim and {Ferguson}, Henry C. and {Fern{\`a}ndez}, Vital and {Finkelstein}, Steven L. and {Gawiser}, Eric and {Grazian}, Andrea and {Grogin}, Norman and {Holwerda}, Benne W. and {Kartaltepe}, Jeyhan S. and {Kewley}, Lisa and {Kirkpatrick}, Allison and {Kocevski}, Dale and {Koekemoer}, Anton M. and {Long}, Arianna and {Lotz}, Jennifer and {Lucas}, Ray A. and {Mobasher}, Bahram and {Papovich}, Casey and {P{\'e}rez-Gonz{\`a}lez}, Pablo G. and {Pirzkal}, Nor and {Ravindranath}, Swara and {Rodighiero}, Giulia and {Roehlly}, Yannick and {Rose}, Caitlin and {Seill{\'e}}, Lise-Marie and {Somerville}, Rachel and {Wilkins}, Steve and {Yang}, Guang and {Yung}, L.~Y. Aaron},
        title = "{CEERS: Possibly forging the first dust grains in the universe: A population of galaxies with spectroscopically derived extremely low dust attenuation (GELDA) at 4.0 < z {\ensuremath{\lesssim}} 11.4}",
      journal = {\aap},
         year = 2025,
        month = jul,
       volume = {699},
          eid = {A336},
        pages = {A336},
          doi = {10.1051/0004-6361/202554231},
archivePrefix = {arXiv},
       eprint = {2504.13118},
 primaryClass = {astro-ph.GA},
       adsurl = {https://ui.adsabs.harvard.edu/abs/2025A&A...699A.336B}
}

@ARTICLE{Vika2013MNRAS.435..623V,
       author = {{Vika}, Marina and {Bamford}, Steven P. and {H{\"a}u{\ss}ler}, Boris and {Rojas}, Alex L. and {Borch}, Andrea and {Nichol}, Robert C.},
        title = "{MegaMorph - multiwavelength measurement of galaxy structure. S{\'e}rsic profile fits to galaxies near and far}",
      journal = {\mnras},
         year = 2013,
        month = oct,
       volume = {435},
       number = {1},
        pages = {623-649},
          doi = {10.1093/mnras/stt1320},
archivePrefix = {arXiv},
       eprint = {1307.4996},
 primaryClass = {astro-ph.CO},
       adsurl = {https://ui.adsabs.harvard.edu/abs/2013MNRAS.435..623V}
}

@ARTICLE{Peng2002AJ....124..266P,
       author = {{Peng}, Chien Y. and {Ho}, Luis C. and {Impey}, Chris D. and {Rix}, Hans-Walter},
        title = "{Detailed Structural Decomposition of Galaxy Images}",
      journal = {\aj},
         year = 2002,
        month = jul,
       volume = {124},
       number = {1},
        pages = {266-293},
          doi = {10.1086/340952},
archivePrefix = {arXiv},
       eprint = {astro-ph/0204182},
 primaryClass = {astro-ph},
       adsurl = {https://ui.adsabs.harvard.edu/abs/2002AJ....124..266P}
}

@ARTICLE{Boquien2019A&A...622A.103B,
       author = {{Boquien}, M. and {Burgarella}, D. and {Roehlly}, Y. and {Buat}, V. and {Ciesla}, L. and {Corre}, D. and {Inoue}, A.~K. and {Salas}, H.},
        title = "{CIGALE: a python Code Investigating GALaxy Emission}",
      journal = {\aap},
         year = 2019,
        month = feb,
       volume = {622},
          eid = {A103},
        pages = {A103},
          doi = {10.1051/0004-6361/201834156},
archivePrefix = {arXiv},
       eprint = {1811.03094},
 primaryClass = {astro-ph.GA},
       adsurl = {https://ui.adsabs.harvard.edu/abs/2019A&A...622A.103B}
}

@ARTICLE{Cullen2025MNRAS.540.2176C,
       author = {{Cullen}, F. and {Carnall}, A.~C. and {Scholte}, D. and {McLeod}, D.~J. and {McLure}, R.~J. and {Arellano-C{\'o}rdova}, K.~Z. and {Stanton}, T.~M. and {Donnan}, C.~T. and {Dunlop}, J.~S. and {Shapley}, A.~E. and {Barrufet}, L. and {Begley}, R. and {Bondestam}, C. and {Cirasuolo}, M. and {Leung}, H.-H. and {Pollock}, C.~L. and {Stevenson}, S.},
        title = "{The JWST EXCELS survey: an extremely metal-poor galaxy at z = 8.271 hosting an unusual population of massive stars}",
      journal = {\mnras},
         year = 2025,
        month = jul,
       volume = {540},
       number = {3},
        pages = {2176-2194},
          doi = {10.1093/mnras/staf838},
archivePrefix = {arXiv},
       eprint = {2501.11099},
 primaryClass = {astro-ph.GA},
       adsurl = {https://ui.adsabs.harvard.edu/abs/2025MNRAS.540.2176C}
}

@ARTICLE{Yanagisawa2025ApJ...988...86Y,
       author = {{Yanagisawa}, Hiroto and {Ouchi}, Masami and {Nakajima}, Kimihiko and {Harikane}, Yuichi and {Fujimoto}, Seiji and {Ono}, Yoshiaki and {Umeda}, Hiroya and {Nakane}, Minami and {Yajima}, Hidenobu and {Fukushima}, Hajime and {Xu}, Yi},
        title = "{A Galaxy with an Extremely Blue Ultraviolet Slope {\ensuremath{\beta}} = {\ensuremath{-}}3 at z = 9.25 Identified by JWST Spectroscopy: Evidence for a Weak Nebular Continuum and Efficient Ionizing Photon Escape?}",
      journal = {\apj},
         year = 2025,
        month = jul,
       volume = {988},
       number = {1},
          eid = {86},
        pages = {86},
          doi = {10.3847/1538-4357/adcecd},
archivePrefix = {arXiv},
       eprint = {2411.19893},
 primaryClass = {astro-ph.GA},
       adsurl = {https://ui.adsabs.harvard.edu/abs/2025ApJ...988...86Y}
}

@ARTICLE{Topping2022ApJ...941..153T,
       author = {{Topping}, Michael W. and {Stark}, Daniel P. and {Endsley}, Ryan and {Plat}, Adele and {Whitler}, Lily and {Chen}, Zuyi and {Charlot}, St{\'e}phane},
        title = "{Searching for Extremely Blue UV Continuum Slopes at z = 7-11 in JWST/NIRCam Imaging: Implications for Stellar Metallicity and Ionizing Photon Escape in Early Galaxies}",
      journal = {\apj},
         year = 2022,
        month = dec,
       volume = {941},
       number = {2},
          eid = {153},
        pages = {153},
          doi = {10.3847/1538-4357/aca522},
archivePrefix = {arXiv},
       eprint = {2208.01610},
 primaryClass = {astro-ph.GA},
       adsurl = {https://ui.adsabs.harvard.edu/abs/2022ApJ...941..153T}
}

@ARTICLE{Zackrisson2013ApJ...777...39Z,
       author = {{Zackrisson}, Erik and {Inoue}, Akio K. and {Jensen}, Hannes},
        title = "{The Spectral Evolution of the First Galaxies. II. Spectral Signatures of Lyman Continuum Leakage from Galaxies in the Reionization Epoch}",
      journal = {\apj},
         year = 2013,
        month = nov,
       volume = {777},
       number = {1},
          eid = {39},
        pages = {39},
          doi = {10.1088/0004-637X/777/1/39},
archivePrefix = {arXiv},
       eprint = {1304.6404},
 primaryClass = {astro-ph.CO},
       adsurl = {https://ui.adsabs.harvard.edu/abs/2013ApJ...777...39Z}
}

@ARTICLE{Nakajima2014MNRAS.442..900N,
       author = {{Nakajima}, Kimihiko and {Ouchi}, Masami},
        title = "{Ionization state of inter-stellar medium in galaxies: evolution, SFR-M$_{*}$-Z dependence, and ionizing photon escape}",
      journal = {\mnras},
         year = 2014,
        month = jul,
       volume = {442},
       number = {1},
        pages = {900-916},
          doi = {10.1093/mnras/stu902},
archivePrefix = {arXiv},
       eprint = {1309.0207},
 primaryClass = {astro-ph.CO},
       adsurl = {https://ui.adsabs.harvard.edu/abs/2014MNRAS.442..900N}
}

@ARTICLE{Adamo2024Natur.632..513A,
       author = {{Adamo}, Angela and {Bradley}, Larry D. and {Vanzella}, Eros and {Claeyssens}, Ad{\'e}la{\"\i}de and {Welch}, Brian and {Diego}, Jose M. and {Mahler}, Guillaume and {Oguri}, Masamune and {Sharon}, Keren and {Abdurro'uf} and {Hsiao}, Tiger Yu-Yang and {Xu}, Xinfeng and {Messa}, Matteo and {Lassen}, Augusto E. and {Zackrisson}, Erik and {Brammer}, Gabriel and {Coe}, Dan and {Kokorev}, Vasily and {Ricotti}, Massimo and {Zitrin}, Adi and {Fujimoto}, Seiji and {Inoue}, Akio K. and {Resseguier}, Tom and {Rigby}, Jane R. and {Jim{\'e}nez-Teja}, Yolanda and {Windhorst}, Rogier A. and {Hashimoto}, Takuya and {Tamura}, Yoichi},
        title = "{Bound star clusters observed in a lensed galaxy 460 Myr after the Big Bang}",
      journal = {\nat},
         year = 2024,
        month = aug,
       volume = {632},
       number = {8025},
        pages = {513-516},
          doi = {10.1038/s41586-024-07703-7},
archivePrefix = {arXiv},
       eprint = {2401.03224},
 primaryClass = {astro-ph.GA},
       adsurl = {https://ui.adsabs.harvard.edu/abs/2024Natur.632..513A}
}

@ARTICLE{Nakazato2024ApJ...975..238N,
       author = {{Nakazato}, Yurina and {Ceverino}, Daniel and {Yoshida}, Naoki},
        title = "{A Merger-driven Scenario for Clumpy Galaxy Formation in the Epoch of Reionization: Physical Properties of Clumps in the FirstLight Simulation}",
      journal = {\apj},
         year = 2024,
        month = nov,
       volume = {975},
       number = {2},
          eid = {238},
        pages = {238},
          doi = {10.3847/1538-4357/ad7d0b},
archivePrefix = {arXiv},
       eprint = {2402.08911},
 primaryClass = {astro-ph.GA},
       adsurl = {https://ui.adsabs.harvard.edu/abs/2024ApJ...975..238N}
}

@ARTICLE{de_la_Vega2026ApJ..1005..169D,
       author = {{de la Vega}, Alexander and {Mobasher}, Bahram and {Sattari}, Zahra and {Chartab}, Nima and {Manesh}, Faezeh and {Sharei}, Niloofar},
        title = "{The Fraction of Clumpy Galaxies in JADES Over 2 < z < 9}",
      journal = {\apj},
         year = 2026,
        month = jul,
       volume = {1005},
       number = {2},
          eid = {169},
        pages = {169},
          doi = {10.3847/1538-4357/ae7b26},
archivePrefix = {arXiv},
       eprint = {2508.14972},
 primaryClass = {astro-ph.GA},
       adsurl = {https://ui.adsabs.harvard.edu/abs/2026ApJ..1005..169D}
}

@ARTICLE{Mandelker2017MNRAS.464..635M,
       author = {{Mandelker}, Nir and {Dekel}, Avishai and {Ceverino}, Daniel and {DeGraf}, Colin and {Guo}, Yicheng and {Primack}, Joel},
        title = "{Giant clumps in simulated high- z Galaxies: properties, evolution and dependence on feedback}",
      journal = {\mnras},
         year = 2017,
        month = jan,
       volume = {464},
       number = {1},
        pages = {635-665},
          doi = {10.1093/mnras/stw2358},
archivePrefix = {arXiv},
       eprint = {1512.08791},
 primaryClass = {astro-ph.GA},
       adsurl = {https://ui.adsabs.harvard.edu/abs/2017MNRAS.464..635M}
}

@ARTICLE{Noguchi1999ApJ...514...77N,
       author = {{Noguchi}, Masafumi},
        title = "{Early Evolution of Disk Galaxies: Formation of Bulges in Clumpy Young Galactic Disks}",
      journal = {\apj},
         year = 1999,
        month = mar,
       volume = {514},
       number = {1},
        pages = {77-95},
          doi = {10.1086/306932},
archivePrefix = {arXiv},
       eprint = {astro-ph/9806355},
 primaryClass = {astro-ph},
       adsurl = {https://ui.adsabs.harvard.edu/abs/1999ApJ...514...77N}
}

@ARTICLE{Kaviraj2014MNRAS.440.2944K,
       author = {{Kaviraj}, Sugata},
        title = "{The importance of minor-merger-driven star formation and black hole growth in disc galaxies}",
      journal = {\mnras},
         year = 2014,
        month = jun,
       volume = {440},
       number = {4},
        pages = {2944-2952},
          doi = {10.1093/mnras/stu338},
archivePrefix = {arXiv},
       eprint = {1402.1166},
 primaryClass = {astro-ph.GA},
       adsurl = {https://ui.adsabs.harvard.edu/abs/2014MNRAS.440.2944K}
}

@ARTICLE{Barnes1991ApJ...370L..65B,
       author = {{Barnes}, Joshua E. and {Hernquist}, Lars E.},
        title = "{Fueling Starburst Galaxies with Gas-rich Mergers}",
      journal = {\apjl},
         year = 1991,
        month = apr,
       volume = {370},
        pages = {L65},
          doi = {10.1086/185978},
       adsurl = {https://ui.adsabs.harvard.edu/abs/1991ApJ...370L..65B}
}

@ARTICLE{Calabro2019A&A...632A..98C,
       author = {{Calabr{\`o}}, Antonello and {Daddi}, Emanuele and {Fensch}, J{\'e}r{\'e}my and {Bournaud}, Fr{\'e}d{\'e}ric and {Cibinel}, Anna and {Puglisi}, Annagrazia and {Jin}, Shuowen and {Delvecchio}, Ivan and {D'Eugenio}, Chiara},
        title = "{Merger induced clump formation in distant infrared luminous starburst galaxies}",
      journal = {\aap},
         year = 2019,
        month = dec,
       volume = {632},
          eid = {A98},
        pages = {A98},
          doi = {10.1051/0004-6361/201935778},
archivePrefix = {arXiv},
       eprint = {1910.01661},
 primaryClass = {astro-ph.GA},
       adsurl = {https://ui.adsabs.harvard.edu/abs/2019A&A...632A..98C}
}

@ARTICLE{Ferrara2023MNRAS.522.3986F,
       author = {{Ferrara}, Andrea and {Pallottini}, Andrea and {Dayal}, Pratika},
        title = "{On the stunning abundance of super-early, luminous galaxies revealed by JWST}",
      journal = {\mnras},
         year = 2023,
        month = jul,
       volume = {522},
       number = {3},
        pages = {3986-3991},
          doi = {10.1093/mnras/stad1095},
archivePrefix = {arXiv},
       eprint = {2208.00720},
 primaryClass = {astro-ph.GA},
       adsurl = {https://ui.adsabs.harvard.edu/abs/2023MNRAS.522.3986F}
}

@ARTICLE{Donnan2024MNRAS.533.3222D,
       author = {{Donnan}, C.~T. and {McLure}, R.~J. and {Dunlop}, J.~S. and {McLeod}, D.~J. and {Magee}, D. and {Arellano-C{\'o}rdova}, K.~Z. and {Barrufet}, L. and {Begley}, R. and {Bowler}, R.~A.~A. and {Carnall}, A.~C. and {Cullen}, F. and {Ellis}, R.~S. and {Fontana}, A. and {Illingworth}, G.~D. and {Grogin}, N.~A. and {Hamadouche}, M.~L. and {Koekemoer}, A.~M. and {Liu}, F.-Y. and {Mason}, C. and {Santini}, P. and {Stanton}, T.~M.},
        title = "{JWST PRIMER: a new multifield determination of the evolving galaxy UV luminosity function at redshifts z ≃ 9 - 15}",
      journal = {\mnras},
         year = 2024,
        month = sep,
       volume = {533},
       number = {3},
        pages = {3222-3237},
          doi = {10.1093/mnras/stae2037},
archivePrefix = {arXiv},
       eprint = {2403.03171},
 primaryClass = {astro-ph.GA},
       adsurl = {https://ui.adsabs.harvard.edu/abs/2024MNRAS.533.3222D}
}

@ARTICLE{Izotov2006A&A...448..955I,
       author = {{Izotov}, Y.~I. and {Stasi{\'n}ska}, G. and {Meynet}, G. and {Guseva}, N.~G. and {Thuan}, T.~X.},
        title = "{The chemical composition of metal-poor emission-line galaxies in the Data Release 3 of the Sloan Digital Sky Survey}",
      journal = {\aap},
         year = 2006,
        month = mar,
       volume = {448},
       number = {3},
        pages = {955-970},
          doi = {10.1051/0004-6361:20053763},
archivePrefix = {arXiv},
       eprint = {astro-ph/0511644},
 primaryClass = {astro-ph},
       adsurl = {https://ui.adsabs.harvard.edu/abs/2006A&A...448..955I}
}

@ARTICLE{Roberts-Borsani2024ApJ...976..193R,
       author = {{Roberts-Borsani}, Guido and {Treu}, Tommaso and {Shapley}, Alice and {Fontana}, Adriano and {Pentericci}, Laura and {Castellano}, Marco and {Morishita}, Takahiro and {Bergamini}, Pietro and {Rosati}, Piero},
        title = "{Between the Extremes: A JWST Spectroscopic Benchmark for High-redshift Galaxies Using {\ensuremath{\sim}}500 Confirmed Sources at z {\ensuremath{\geq}} 5}",
      journal = {\apj},
         year = 2024,
        month = dec,
       volume = {976},
       number = {2},
          eid = {193},
        pages = {193},
          doi = {10.3847/1538-4357/ad85d3},
archivePrefix = {arXiv},
       eprint = {2403.07103},
 primaryClass = {astro-ph.GA},
       adsurl = {https://ui.adsabs.harvard.edu/abs/2024ApJ...976..193R}
}

@ARTICLE{Sanders2024ApJ...962...24S,
       author = {{Sanders}, Ryan L. and {Shapley}, Alice E. and {Topping}, Michael W. and {Reddy}, Naveen A. and {Brammer}, Gabriel B.},
        title = "{Direct T $_{e}$-based Metallicities of z = 2─9 Galaxies with JWST/NIRSpec: Empirical Metallicity Calibrations Applicable from Reionization to Cosmic Noon}",
      journal = {\apj},
         year = 2024,
        month = feb,
       volume = {962},
       number = {1},
          eid = {24},
        pages = {24},
          doi = {10.3847/1538-4357/ad15fc},
archivePrefix = {arXiv},
       eprint = {2303.08149},
 primaryClass = {astro-ph.GA},
       adsurl = {https://ui.adsabs.harvard.edu/abs/2024ApJ...962...24S}
}

@ARTICLE{Ji2024MNRAS.535..881J,
       author = {{Ji}, Xihan and {{\"U}bler}, Hannah and {Maiolino}, Roberto and {D'Eugenio}, Francesco and {Arribas}, Santiago and {Bunker}, Andrew J. and {Charlot}, St{\'e}phane and {Perna}, Michele and {Rodr{\'\i}guez Del Pino}, Bruno and {B{\"o}ker}, Torsten and {Cresci}, Giovanni and {Curti}, Mirko and {Kumari}, Nimisha and {Lamperti}, Isabella},
        title = "{GA-NIFS: an extremely nitrogen-loud and chemically stratified galaxy at z   5.55}",
      journal = {\mnras},
         year = 2024,
        month = nov,
       volume = {535},
       number = {1},
        pages = {881-908},
          doi = {10.1093/mnras/stae2375},
archivePrefix = {arXiv},
       eprint = {2404.04148},
 primaryClass = {astro-ph.GA},
       adsurl = {https://ui.adsabs.harvard.edu/abs/2024MNRAS.535..881J}
}

@ARTICLE{Zamora2025arXiv251209022Z,
       author = {{Zamora}, Sandra and {Carniani}, Stefano and {Bertola}, Elena and {Parlanti}, Eleonora and {P{\'e}rez-Gonz{\'a}lez}, Pablo G. and {Arribas}, Santiago and {B{\"o}ker}, Torsten and {Bunker}, Andrew J. and {D'Eugenio}, Francesco and {Maiolino}, Roberto and {Perna}, Michele and {Rodr{\'\i}guez Del Pino}, Bruno and {{\"U}bler}, Hannah and {Cresci}, Giovanni and {Jones}, Gareth C. and {Lamperti}, Isabella and {Scholtz}, Jan and {Trefoloni}, Bartolomeo and {Venturi}, Giacomo},
        title = "{GA-NIFS: Understanding the ionization nature of EGSY8p7/CEERS-1019. Evidence for a star formation-driven outflow at z = 8.6}",
      journal = {arXiv e-prints},
         year = 2025,
        month = dec,
          eid = {arXiv:2512.09022},
        pages = {arXiv:2512.09022},
          doi = {10.48550/arXiv.2512.09022},
archivePrefix = {arXiv},
       eprint = {2512.09022},
 primaryClass = {astro-ph.GA},
       adsurl = {https://ui.adsabs.harvard.edu/abs/2025arXiv251209022Z}
}

@ARTICLE{Larson2023ApJ...953L..29L,
       author = {{Larson}, Rebecca L. and {Finkelstein}, Steven L. and {Kocevski}, Dale D. and {Hutchison}, Taylor A. and {Trump}, Jonathan R. and {Arrabal Haro}, Pablo and {Bromm}, Volker and {Cleri}, Nikko J. and {Dickinson}, Mark and {Fujimoto}, Seiji and {Kartaltepe}, Jeyhan S. and {Koekemoer}, Anton M. and {Papovich}, Casey and {Pirzkal}, Nor and {Tacchella}, Sandro and {Zavala}, Jorge A. and {Bagley}, Micaela and {Behroozi}, Peter and {Champagne}, Jaclyn B. and {Cole}, Justin W. and {Jung}, Intae and {Morales}, Alexa M. and {Yang}, Guang and {Zhang}, Haowen and {Zitrin}, Adi and {Amor{\'\i}n}, Ricardo O. and {Burgarella}, Denis and {Casey}, Caitlin M. and {Ch{\'a}vez Ortiz}, {\'O}scar A. and {Cox}, Isabella G. and {Chworowsky}, Katherine and {Fontana}, Adriano and {Gawiser}, Eric and {Grazian}, Andrea and {Grogin}, Norman A. and {Harish}, Santosh and {Hathi}, Nimish P. and {Hirschmann}, Michaela and {Holwerda}, Benne W. and {Juneau}, St{\'e}phanie and {Leung}, Gene C.~K. and {Lucas}, Ray A. and {McGrath}, Elizabeth J. and {P{\'e}rez-Gonz{\'a}lez}, Pablo G. and {Rigby}, Jane R. and {Seill{\'e}}, Lise-Marie and {Simons}, Raymond C. and {de La Vega}, Alexander and {Weiner}, Benjamin J. and {Wilkins}, Stephen M. and {Yung}, L.~Y. Aaron and {Ceers Team}},
        title = "{A CEERS Discovery of an Accreting Supermassive Black Hole 570 Myr after the Big Bang: Identifying a Progenitor of Massive z > 6 Quasars}",
      journal = {\apjl},
         year = 2023,
        month = aug,
       volume = {953},
       number = {2},
          eid = {L29},
        pages = {L29},
          doi = {10.3847/2041-8213/ace619},
archivePrefix = {arXiv},
       eprint = {2303.08918},
 primaryClass = {astro-ph.GA},
       adsurl = {https://ui.adsabs.harvard.edu/abs/2023ApJ...953L..29L}
}

@ARTICLE{Pollock2026A&A...708A.203P,
       author = {{Pollock}, Clara L. and {Gottumukkala}, Rashmi and {Heintz}, Kasper E. and {Brammer}, Gabriel B. and {Roberts-Borsani}, Guido and {Oesch}, Pascal A. and {Witstok}, Joris and {Arellano-C{\'o}rdova}, Karla Z. and {Cullen}, Fergus and {Scholte}, Dirk and {Terp}, Chamilla and {Rowland}, Lucie and {Sneppen}, Albert and {Ito}, Kei and {Valentino}, Francesco and {Matthee}, Jorryt and {Watson}, Darach and {Toft}, Sune},
        title = "{Novel z {\ensuremath{\sim}} 10 auroral line measurements extend the gradual offset of the fundamental metallicity relation deep into the first gigayear of cosmic time}",
      journal = {\aap},
         year = 2026,
        month = apr,
       volume = {708},
          eid = {A203},
        pages = {A203},
          doi = {10.1051/0004-6361/202556032},
archivePrefix = {arXiv},
       eprint = {2506.15779},
 primaryClass = {astro-ph.GA},
       adsurl = {https://ui.adsabs.harvard.edu/abs/2026A&A...708A.203P}
}

@ARTICLE{Donnan2025ApJ...993..224D,
       author = {{Donnan}, Callum T. and {Dickinson}, Mark and {Taylor}, Anthony J. and {Arrabal Haro}, Pablo and {Finkelstein}, Steven L. and {Stanton}, Thomas M. and {Jung}, Intae and {Papovich}, Casey and {Akins}, Hollis B. and {Koekemoer}, Anton M. and {McLeod}, Derek J. and {Napolitano}, Lorenzo and {Amor{\'\i}n}, Ricardo O. and {Begley}, Ryan and {Burgarella}, Denis and {Carnall}, Adam C. and {Casey}, Caitlin M. and {Calabr{\`o}}, Antonello and {Cullen}, Fergus and {Dunlop}, James S. and {Ellis}, Richard S. and {Fern{\'a}ndez}, Vital and {Giavalisco}, Mauro and {Hirschmann}, Michaela and {Hu}, Weida and {Illingworth}, Garth and {Kartaltepe}, Jeyhan S. and {Kocevski}, Dale D. and {Kokorev}, Vasily and {Leung}, Ho-Hin and {Lucas}, Ray A. and {Morales}, Alexa M. and {McLure}, Ross and {Pentericci}, Laura and {P{\'e}rez-Gonz{\'a}lez}, Pablo G. and {Somerville}, Rachel S. and {Stevenson}, Struan and {Trump}, Jonathan R. and {Yung}, L.~Y. Aaron and {Zavala}, Jorge A.},
        title = "{Very Bright, Very Blue, and Very Red: JWST CAPERS Analysis of Highly Luminous Galaxies with Extreme Ultraviolet Slopes at z = 10}",
      journal = {\apj},
         year = 2025,
        month = nov,
       volume = {993},
       number = {2},
          eid = {224},
        pages = {224},
          doi = {10.3847/1538-4357/ae0a1f},
archivePrefix = {arXiv},
       eprint = {2507.10518},
 primaryClass = {astro-ph.GA},
       adsurl = {https://ui.adsabs.harvard.edu/abs/2025ApJ...993..224D}
}

@ARTICLE{Helton2026ApJ..1004L..30H,
       author = {{Helton}, Jakob M. and {Morrison}, Jane E. and {Hainline}, Kevin N. and {D'Eugenio}, Francesco and {Rieke}, George H. and {Alberts}, Stacey and {Carniani}, Stefano and {Leja}, Joel and {Li}, Yijia and {Rinaldi}, Pierluigi and {Scholtz}, Jan and {Stone}, Meredith and {Willmer}, Christopher N.~A. and {Wu}, Zihao and {Baker}, William M. and {Bunker}, Andrew J. and {Charlot}, Stephane and {Chevallard}, Jacopo and {Cleri}, Nikko J. and {Curti}, Mirko and {Curtis-Lake}, Emma and {Egami}, Eiichi and {Eisenstein}, Daniel J. and {Jakobsen}, Peter and {Ji}, Zhiyuan and {Johnson}, Benjamin D. and {Kumari}, Nimisha and {Lin}, Xiaojing and {Lyu}, Jianwei and {Maiolino}, Roberto and {Maseda}, Michael and {P{\'e}rez-Gonz{\'a}lez}, Pablo G. and {Rieke}, Marcia J. and {Robertson}, Brant and {Saxena}, Aayush and {Sun}, Fengwu and {Tacchella}, Sandro and {{\"U}bler}, Hannah and {Venturi}, Giacomo and {Williams}, Christina C. and {Willott}, Chris and {Witstok}, Joris and {Zhu}, Yongda},
        title = "{Ionizing Photon Production Efficiencies and Chemical Abundances at Cosmic Dawn Revealed by Ultradeep Rest-frame Optical Spectroscopy of JADES-GS-z14-0}",
      journal = {\apjl},
         year = 2026,
        month = jun,
       volume = {1004},
       number = {2},
          eid = {L30},
        pages = {L30},
          doi = {10.3847/2041-8213/ae7446},
archivePrefix = {arXiv},
       eprint = {2512.19695},
 primaryClass = {astro-ph.GA},
       adsurl = {https://ui.adsabs.harvard.edu/abs/2026ApJ..1004L..30H}
}

@ARTICLE{Bunker2023A&A...677A..88B,
       author = {{Bunker}, Andrew J. and {Saxena}, Aayush and {Cameron}, Alex J. and {Willott}, Chris J. and {Curtis-Lake}, Emma and {Jakobsen}, Peter and {Carniani}, Stefano and {Smit}, Renske and {Maiolino}, Roberto and {Witstok}, Joris and {Curti}, Mirko and {D'Eugenio}, Francesco and {Jones}, Gareth C. and {Ferruit}, Pierre and {Arribas}, Santiago and {Charlot}, Stephane and {Chevallard}, Jacopo and {Giardino}, Giovanna and {de Graaff}, Anna and {Looser}, Tobias J. and {L{\"u}tzgendorf}, Nora and {Maseda}, Michael V. and {Rawle}, Tim and {Rix}, Hans-Walter and {Del Pino}, Bruno Rodr{\'\i}guez and {Alberts}, Stacey and {Egami}, Eiichi and {Eisenstein}, Daniel J. and {Endsley}, Ryan and {Hainline}, Kevin and {Hausen}, Ryan and {Johnson}, Benjamin D. and {Rieke}, George and {Rieke}, Marcia and {Robertson}, Brant E. and {Shivaei}, Irene and {Stark}, Daniel P. and {Sun}, Fengwu and {Tacchella}, Sandro and {Tang}, Mengtao and {Williams}, Christina C. and {Willmer}, Christopher N.~A. and {Baker}, William M. and {Baum}, Stefi and {Bhatawdekar}, Rachana and {Bowler}, Rebecca and {Boyett}, Kristan and {Chen}, Zuyi and {Circosta}, Chiara and {Helton}, Jakob M. and {Ji}, Zhiyuan and {Kumari}, Nimisha and {Lyu}, Jianwei and {Nelson}, Erica and {Parlanti}, Eleonora and {Perna}, Michele and {Sandles}, Lester and {Scholtz}, Jan and {Suess}, Katherine A. and {Topping}, Michael W. and {{\"U}bler}, Hannah and {Wallace}, Imaan E.~B. and {Whitler}, Lily},
        title = "{JADES NIRSpec Spectroscopy of GN-z11: Lyman-{\ensuremath{\alpha}} emission and possible enhanced nitrogen abundance in a z = 10.60 luminous galaxy}",
      journal = {\aap},
         year = 2023,
        month = sep,
       volume = {677},
          eid = {A88},
        pages = {A88},
          doi = {10.1051/0004-6361/202346159},
archivePrefix = {arXiv},
       eprint = {2302.07256},
 primaryClass = {astro-ph.GA},
       adsurl = {https://ui.adsabs.harvard.edu/abs/2023A&A...677A..88B}
}

@ARTICLE{Curti2025A&A...697A..89C,
       author = {{Curti}, Mirko and {Witstok}, Joris and {Jakobsen}, Peter and {Kobayashi}, Chiaki and {Curtis-Lake}, Emma and {Hainline}, Kevin and {Ji}, Xihan and {D'Eugenio}, Francesco and {Chevallard}, Jacopo and {Maiolino}, Roberto and {Scholtz}, Jan and {Carniani}, Stefano and {Arribas}, Santiago and {Baker}, William M. and {Bhatawdekar}, Rachana and {Boyett}, Kristan and {Bunker}, Andrew J. and {Cameron}, Alex and {Cargile}, Phillip A. and {Charlot}, St{\'e}phane and {Eisenstein}, Daniel J. and {Ji}, Zhiyuan and {Johnson}, Benjamin D. and {Kumari}, Nimisha and {Maseda}, Michael V. and {Robertson}, Brant and {Silcock}, Maddie S. and {Tacchella}, Sandro and {{\"U}bler}, Hannah and {Venturi}, Giacomo and {Williams}, Christina C. and {Willmer}, Christopher N.~A. and {Willott}, Chris},
        title = "{JADES: The star formation and chemical enrichment history of a luminous galaxy at z {\ensuremath{\sim}} 9.43 probed by ultra-deep JWST/NIRSpec spectroscopy}",
      journal = {\aap},
         year = 2025,
        month = may,
       volume = {697},
          eid = {A89},
        pages = {A89},
          doi = {10.1051/0004-6361/202451410},
archivePrefix = {arXiv},
       eprint = {2407.02575},
 primaryClass = {astro-ph.GA},
       adsurl = {https://ui.adsabs.harvard.edu/abs/2025A&A...697A..89C}
}

@ARTICLE{Schaerer2024A&A...687L..11S,
       author = {{Schaerer}, D. and {Marques-Chaves}, R. and {Xiao}, M. and {Korber}, D.},
        title = "{Discovery of a new N-emitter in the epoch of reionization}",
      journal = {\aap},
         year = 2024,
        month = jul,
       volume = {687},
          eid = {L11},
        pages = {L11},
          doi = {10.1051/0004-6361/202450721},
archivePrefix = {arXiv},
       eprint = {2406.08408},
 primaryClass = {astro-ph.GA},
       adsurl = {https://ui.adsabs.harvard.edu/abs/2024A&A...687L..11S}
}

@ARTICLE{Boyett2024NatAs...8..657B,
       author = {{Boyett}, Kristan and {Trenti}, Michele and {Leethochawalit}, Nicha and {Calabr{\'o}}, Antonello and {Metha}, Benjamin and {Roberts-Borsani}, Guido and {Dalmasso}, Nicol{\'o} and {Yang}, Lilan and {Santini}, Paola and {Treu}, Tommaso and {Jones}, Tucker and {Henry}, Alaina and {Mason}, Charlotte A. and {Morishita}, Takahiro and {Nanayakkara}, Themiya and {Roy}, Namrata and {Wang}, Xin and {Fontana}, Adriano and {Merlin}, Emiliano and {Castellano}, Marco and {Paris}, Diego and {Brada{\v{c}}}, Maru{\v{s}}a and {Malkan}, Matt and {Marchesini}, Danilo and {Mascia}, Sara and {Glazebrook}, Karl and {Pentericci}, Laura and {Vanzella}, Eros and {Vulcani}, Benedetta},
        title = "{A massive interacting galaxy 510 million years after the Big Bang}",
      journal = {Nature Astronomy},
         year = 2024,
        month = may,
       volume = {8},
        pages = {657-672},
          doi = {10.1038/s41550-024-02218-7},
archivePrefix = {arXiv},
       eprint = {2303.00306},
 primaryClass = {astro-ph.GA},
       adsurl = {https://ui.adsabs.harvard.edu/abs/2024NatAs...8..657B}
}

@ARTICLE{Hsiao2023ApJ...949L..34H,
       author = {{Hsiao}, Tiger Yu-Yang and {Coe}, Dan and {Abdurro'uf} and {Whitler}, Lily and {Jung}, Intae and {Khullar}, Gourav and {Meena}, Ashish Kumar and {Dayal}, Pratika and {Barrow}, Kirk S.~S. and {Santos-Olmsted}, Lillian and {Casselman}, Adam and {Vanzella}, Eros and {Nonino}, Mario and {Jim{\'e}nez-Teja}, Yolanda and {Oguri}, Masamune and {Stark}, Daniel P. and {Furtak}, Lukas J. and {Zitrin}, Adi and {Adamo}, Angela and {Brammer}, Gabriel and {Bradley}, Larry and {Diego}, Jose M. and {Zackrisson}, Erik and {Finkelstein}, Steven L. and {Windhorst}, Rogier A. and {Bhatawdekar}, Rachana and {Hutchison}, Taylor A. and {Broadhurst}, Tom and {Dimauro}, Paola and {Andrade-Santos}, Felipe and {Eldridge}, Jan J. and {Acebron}, Ana and {Avila}, Roberto J. and {Bayliss}, Matthew B. and {Ben{\'\i}tez}, Alex and {Binggeli}, Christian and {Bolan}, Patricia and {Brada{\v{c}}}, Maru{\v{s}}a and {Carnall}, Adam C. and {Conselice}, Christopher J. and {Donahue}, Megan and {Frye}, Brenda and {Fujimoto}, Seiji and {Henry}, Alaina and {James}, Bethan L. and {Kassin}, Susan A. and {Kewley}, Lisa and {Larson}, Rebecca L. and {Lauer}, Tod and {Law}, David and {Mahler}, Guillaume and {Mainali}, Ramesh and {McCandliss}, Stephan and {Nicholls}, David and {Pirzkal}, Norbert and {Postman}, Marc and {Rigby}, Jane R. and {Ryan}, Russell and {Senchyna}, Peter and {Sharon}, Keren and {Shimizu}, Ikko and {Strait}, Victoria and {Tang}, Mengtao and {Trenti}, Michele and {Vikaeus}, Anton and {Welch}, Brian},
        title = "{JWST Reveals a Possible z {\ensuremath{\sim}} 11 Galaxy Merger in Triply Lensed MACS0647─JD}",
      journal = {\apjl},
         year = 2023,
        month = jun,
       volume = {949},
       number = {2},
          eid = {L34},
        pages = {L34},
          doi = {10.3847/2041-8213/acc94b},
archivePrefix = {arXiv},
       eprint = {2210.14123},
 primaryClass = {astro-ph.GA},
       adsurl = {https://ui.adsabs.harvard.edu/abs/2023ApJ...949L..34H}
}

@ARTICLE{Williams2023Sci...380..416W,
       author = {{Williams}, Hayley and {Kelly}, Patrick L. and {Chen}, Wenlei and {Brammer}, Gabriel and {Zitrin}, Adi and {Treu}, Tommaso and {Scarlata}, Claudia and {Koekemoer}, Anton M. and {Oguri}, Masamune and {Lin}, Yu-Heng and {Diego}, Jose M. and {Nonino}, Mario and {Hjorth}, Jens and {Langeroodi}, Danial and {Broadhurst}, Tom and {Rogers}, Noah and {Perez-Fournon}, Ismael and {Foley}, Ryan J. and {Jha}, Saurabh and {Filippenko}, Alexei V. and {Strolger}, Lou and {Pierel}, Justin and {Poidevin}, Frederick and {Yang}, Lilan},
        title = "{A magnified compact galaxy at redshift 9.51 with strong nebular emission lines}",
      journal = {Science},
         year = 2023,
        month = apr,
       volume = {380},
       number = {6643},
        pages = {416-420},
          doi = {10.1126/science.adf5307},
archivePrefix = {arXiv},
       eprint = {2210.15699},
 primaryClass = {astro-ph.GA},
       adsurl = {https://ui.adsabs.harvard.edu/abs/2023Sci...380..416W}
}

@ARTICLE{Stiavelli2023ApJ...957L..18S,
       author = {{Stiavelli}, Massimo and {Morishita}, Takahiro and {Chiaberge}, Marco and {Grillo}, Claudio and {Leethochawalit}, Nicha and {Rosati}, Piero and {Schuldt}, Stefan and {Trenti}, Michele and {Treu}, Tommaso},
        title = "{The Puzzling Properties of the MACS1149-JD1 Galaxy at z = 9.11}",
      journal = {\apjl},
         year = 2023,
        month = nov,
       volume = {957},
       number = {2},
          eid = {L18},
        pages = {L18},
          doi = {10.3847/2041-8213/ad0159},
archivePrefix = {arXiv},
       eprint = {2308.14696},
 primaryClass = {astro-ph.GA},
       adsurl = {https://ui.adsabs.harvard.edu/abs/2023ApJ...957L..18S}
}

@ARTICLE{Curti2024A&A...684A..75C,
       author = {{Curti}, Mirko and {Maiolino}, Roberto and {Curtis-Lake}, Emma and {Chevallard}, Jacopo and {Carniani}, Stefano and {D'Eugenio}, Francesco and {Looser}, Tobias J. and {Scholtz}, Jan and {Charlot}, Stephane and {Cameron}, Alex and {{\"U}bler}, Hannah and {Witstok}, Joris and {Boyett}, Kristian and {Laseter}, Isaac and {Sandles}, Lester and {Arribas}, Santiago and {Bunker}, Andrew and {Giardino}, Giovanna and {Maseda}, Michael V. and {Rawle}, Tim and {Rodr{\'\i}guez Del Pino}, Bruno and {Smit}, Renske and {Willott}, Chris J. and {Eisenstein}, Daniel J. and {Hausen}, Ryan and {Johnson}, Benjamin and {Rieke}, Marcia and {Robertson}, Brant and {Tacchella}, Sandro and {Williams}, Christina C. and {Willmer}, Christopher and {Baker}, William M. and {Bhatawdekar}, Rachana and {Egami}, Eiichi and {Helton}, Jakob M. and {Ji}, Zhiyuan and {Kumari}, Nimisha and {Perna}, Michele and {Shivaei}, Irene and {Sun}, Fengwu},
        title = "{JADES: Insights into the low-mass end of the mass-metallicity-SFR relation at 3 < z < 10 from deep JWST/NIRSpec spectroscopy}",
      journal = {\aap},
         year = 2024,
        month = apr,
       volume = {684},
          eid = {A75},
        pages = {A75},
          doi = {10.1051/0004-6361/202346698},
archivePrefix = {arXiv},
       eprint = {2304.08516},
 primaryClass = {astro-ph.GA},
       adsurl = {https://ui.adsabs.harvard.edu/abs/2024A&A...684A..75C}
}

@ARTICLE{Bunker2024A&A...690A.288B,
       author = {{Bunker}, Andrew J. and {Cameron}, Alex J. and {Curtis-Lake}, Emma and {Jakobsen}, Peter and {Carniani}, Stefano and {Curti}, Mirko and {Witstok}, Joris and {Maiolino}, Roberto and {D'Eugenio}, Francesco and {Looser}, Tobias J. and {Willott}, Chris and {Bonaventura}, Nina and {Hainline}, Kevin and {{\"U}bler}, Hannah and {Willmer}, Christopher N.~A. and {Saxena}, Aayush and {Smit}, Renske and {Alberts}, Stacey and {Arribas}, Santiago and {Baker}, William M. and {Baum}, Stefi and {Bhatawdekar}, Rachana and {Bowler}, Rebecca A.~A. and {Boyett}, Kristan and {Charlot}, Stephane and {Chen}, Zuyi and {Chevallard}, Jacopo and {Circosta}, Chiara and {DeCoursey}, Christa and {de Graaff}, Anna and {Egami}, Eiichi and {Eisenstein}, Daniel J. and {Endsley}, Ryan and {Ferruit}, Pierre and {Giardino}, Giovanna and {Hausen}, Ryan and {Helton}, Jakob M. and {Hviding}, Raphael E. and {Ji}, Zhiyuan and {Johnson}, Benjamin D. and {Jones}, Gareth C. and {Kumari}, Nimisha and {Laseter}, Isaac and {L{\"u}tzgendorf}, Nora and {Maseda}, Michael V. and {Nelson}, Erica and {Parlanti}, Eleonora and {Perna}, Michele and {Rauscher}, Bernard J. and {Rawle}, Tim and {Rix}, Hans-Walter and {Rieke}, Marcia and {Robertson}, Brant and {Rodr{\'\i}guez Del Pino}, Bruno and {Sandles}, Lester and {Scholtz}, Jan and {Sharpe}, Katherine and {Skarbinski}, Maya and {Stark}, Daniel P. and {Sun}, Fengwu and {Tacchella}, Sandro and {Topping}, Michael W. and {Villanueva}, Natalia C. and {Wallace}, Imaan E.~B. and {Williams}, Christina C. and {Woodrum}, Charity},
        title = "{JADES NIRSpec initial data release for the Hubble Ultra Deep Field: Redshifts and line fluxes of distant galaxies from the deepest JWST Cycle 1 NIRSpec multi-object spectroscopy}",
      journal = {\aap},
         year = 2024,
        month = oct,
       volume = {690},
          eid = {A288},
        pages = {A288},
          doi = {10.1051/0004-6361/202347094},
archivePrefix = {arXiv},
       eprint = {2306.02467},
 primaryClass = {astro-ph.GA},
       adsurl = {https://ui.adsabs.harvard.edu/abs/2024A&A...690A.288B}
}

@ARTICLE{Cameron2023A&A...677A.115C,
       author = {{Cameron}, Alex J. and {Saxena}, Aayush and {Bunker}, Andrew J. and {D'Eugenio}, Francesco and {Carniani}, Stefano and {Maiolino}, Roberto and {Curtis-Lake}, Emma and {Ferruit}, Pierre and {Jakobsen}, Peter and {Arribas}, Santiago and {Bonaventura}, Nina and {Charlot}, Stephane and {Chevallard}, Jacopo and {Curti}, Mirko and {Looser}, Tobias J. and {Maseda}, Michael V. and {Rawle}, Tim and {Rodr{\'\i}guez Del Pino}, Bruno and {Smit}, Renske and {{\"U}bler}, Hannah and {Willott}, Chris and {Witstok}, Joris and {Egami}, Eiichi and {Eisenstein}, Daniel J. and {Johnson}, Benjamin D. and {Hainline}, Kevin and {Rieke}, Marcia and {Robertson}, Brant E. and {Stark}, Daniel P. and {Tacchella}, Sandro and {Williams}, Christina C. and {Willmer}, Christopher N.~A. and {Bhatawdekar}, Rachana and {Bowler}, Rebecca and {Boyett}, Kristan and {Circosta}, Chiara and {Helton}, Jakob M. and {Jones}, Gareth C. and {Kumari}, Nimisha and {Ji}, Zhiyuan and {Nelson}, Erica and {Parlanti}, Eleonora and {Sandles}, Lester and {Scholtz}, Jan and {Sun}, Fengwu},
        title = "{JADES: Probing interstellar medium conditions at z {\ensuremath{\sim}} 5.5-9.5 with ultra-deep JWST/NIRSpec spectroscopy}",
      journal = {\aap},
         year = 2023,
        month = sep,
       volume = {677},
          eid = {A115},
        pages = {A115},
          doi = {10.1051/0004-6361/202346107},
archivePrefix = {arXiv},
       eprint = {2302.04298},
 primaryClass = {astro-ph.GA},
       adsurl = {https://ui.adsabs.harvard.edu/abs/2023A&A...677A.115C}
}

@ARTICLE{Calabro2023A&A...679A..80C,
       author = {{Calabr{\`o}}, Antonello and {Pentericci}, Laura and {Feltre}, Anna and {Arrabal Haro}, Pablo and {Radovich}, Mario and {Seill{\'e}}, Lise-Marie and {Oliva}, Ernesto and {Daddi}, Emanuele and {Amor{\'\i}n}, Ricardo and {Bagley}, Micaela B. and {Bisigello}, Laura and {Buat}, V{\'e}ronique and {Castellano}, Marco and {Cleri}, Nikko J. and {Dickinson}, Mark and {Fern{\'a}ndez}, Vital and {Finkelstein}, Steven L. and {Giavalisco}, Mauro and {Grazian}, Andrea and {Hathi}, Nimish P. and {Hirschmann}, Michaela and {Juneau}, St{\'e}phanie and {Kartaltepe}, Jeyhan S. and {Koekemoer}, Anton M. and {Lucas}, Ray A. and {Papovich}, Casey and {P{\'e}rez-Gonz{\'a}lez}, Pablo G. and {Pirzkal}, Nor and {Santini}, Paola and {Trump}, Jonathan and {de la Vega}, Alexander and {Wilkins}, Stephen M. and {Yung}, L.~Y. Aaron and {Cassata}, Paolo and {Gobat}, Raphael A.~S. and {Mascia}, Sara and {Napolitano}, Lorenzo and {Vulcani}, Benedetta},
        title = "{Near-infrared emission line diagnostics for AGN from the local Universe to z {\ensuremath{\sim}} 3}",
      journal = {\aap},
         year = 2023,
        month = nov,
       volume = {679},
          eid = {A80},
        pages = {A80},
          doi = {10.1051/0004-6361/202347190},
archivePrefix = {arXiv},
       eprint = {2306.08605},
 primaryClass = {astro-ph.GA},
       adsurl = {https://ui.adsabs.harvard.edu/abs/2023A&A...679A..80C}
}

@ARTICLE{Nakajima2023ApJS..269...33N,
       author = {{Nakajima}, Kimihiko and {Ouchi}, Masami and {Isobe}, Yuki and {Harikane}, Yuichi and {Zhang}, Yechi and {Ono}, Yoshiaki and {Umeda}, Hiroya and {Oguri}, Masamune},
        title = "{JWST Census for the Mass-Metallicity Star Formation Relations at z = 4-10 with Self-consistent Flux Calibration and Proper Metallicity Calibrators}",
      journal = {\apjs},
         year = 2023,
        month = dec,
       volume = {269},
       number = {2},
          eid = {33},
        pages = {33},
          doi = {10.3847/1538-4365/acd556},
archivePrefix = {arXiv},
       eprint = {2301.12825},
 primaryClass = {astro-ph.GA},
       adsurl = {https://ui.adsabs.harvard.edu/abs/2023ApJS..269...33N}
}

@ARTICLE{Chakraborty2025ApJ...985...24C,
       author = {{Chakraborty}, Priyanka and {Sarkar}, Arnab and {Smith}, Randall and {Ferland}, Gary J. and {McDonald}, Michael and {Forman}, William and {Vogelsberger}, Mark and {Torrey}, Paul and {Garcia}, Alex M. and {Bautz}, Mark and {Foster}, Adam and {Miller}, Eric and {Grant}, Catherine},
        title = "{Unveiling the Cosmic Chemistry. II. ``Direct'' T$_{e}$-based Metallicity of Galaxies at 3 < z < 10 with JWST/NIRSpec}",
      journal = {\apj},
         year = 2025,
        month = may,
       volume = {985},
       number = {1},
          eid = {24},
        pages = {24},
          doi = {10.3847/1538-4357/adc7b5},
archivePrefix = {arXiv},
       eprint = {2412.15435},
 primaryClass = {astro-ph.GA},
       adsurl = {https://ui.adsabs.harvard.edu/abs/2025ApJ...985...24C}
}

@ARTICLE{Sanders2026ApJ..1003..228S,
       author = {{Sanders}, Ryan L. and {Shapley}, Alice E. and {Topping}, Michael W. and {Reddy}, Naveen A. and {Berg}, Danielle A. and {Khostovan}, Ali Ahmad and {Bouwens}, Rychard J. and {Brammer}, Gabriel and {Carnall}, Adam C. and {Cullen}, Fergus and {Dav{\'e}}, Romeel and {Dunlop}, James S. and {Ellis}, Richard S. and {F{\"o}rster Schreiber}, N.~M. and {Furlanetto}, Steven R. and {Glazebrook}, Karl and {Illingworth}, Garth D. and {Jones}, Tucker and {Kriek}, Mariska and {McLeod}, Derek J. and {McLure}, Ross J. and {Narayanan}, Desika and {Oesch}, Pascal A. and {Pahl}, Anthony J. and {Pettini}, Max and {Schaerer}, Daniel and {Stark}, Daniel P. and {Steidel}, Charles C. and {Tang}, Mengtao and {Clarke}, Leonardo and {Donnan}, Callum T. and {Kehoe}, Emily},
        title = "{The AURORA Survey: High-redshift Empirical Metallicity Calibrations from Electron Temperature Measurements at z = 2─10}",
      journal = {\apj},
         year = 2026,
        month = jun,
       volume = {1003},
       number = {2},
          eid = {228},
        pages = {228},
          doi = {10.3847/1538-4357/ae66e2},
archivePrefix = {arXiv},
       eprint = {2508.10099},
 primaryClass = {astro-ph.GA},
       adsurl = {https://ui.adsabs.harvard.edu/abs/2026ApJ..1003..228S}
}

@ARTICLE{Chemerynska2024MNRAS.531.2615C,
       author = {{Chemerynska}, Iryna and {Atek}, Hakim and {Furtak}, Lukas J. and {Zitrin}, Adi and {Greene}, Jenny E. and {Dayal}, Pratika and {Weibel}, Andrea and {Fujimoto}, Seiji and {Kokorev}, Vasily and {Goulding}, Andy D. and {Williams}, Christina C. and {Nanayakkara}, Themiya and {Bezanson}, Rachel and {Brammer}, Gabriel and {Cutler}, Sam E. and {Labbe}, Ivo and {Leja}, Joel and {Pan}, Richard and {Price}, Sedona H. and {van Dokkum}, Pieter and {Wang}, Bingjie and {Weaver}, John R. and {Whitaker}, Katherine E.},
        title = "{JWST UNCOVER: the overabundance of ultraviolet-luminous galaxies at z > 9}",
      journal = {\mnras},
         year = 2024,
        month = jun,
       volume = {531},
       number = {2},
        pages = {2615-2625},
          doi = {10.1093/mnras/stae1260},
archivePrefix = {arXiv},
       eprint = {2312.05030},
 primaryClass = {astro-ph.GA},
       adsurl = {https://ui.adsabs.harvard.edu/abs/2024MNRAS.531.2615C}
}

@ARTICLE{Hutter2025A&A...694A.254H,
       author = {{Hutter}, Anne and {Cueto}, Elie R. and {Dayal}, Pratika and {Gottl{\"o}ber}, Stefan and {Trebitsch}, Maxime and {Yepes}, Gustavo},
        title = "{ASTRAEUS: X. Indications of a top-heavy initial mass function in highly star-forming galaxies from JWST observations at z > 10}",
      journal = {\aap},
         year = 2025,
        month = feb,
       volume = {694},
          eid = {A254},
        pages = {A254},
          doi = {10.1051/0004-6361/202452460},
archivePrefix = {arXiv},
       eprint = {2410.00730},
 primaryClass = {astro-ph.GA},
       adsurl = {https://ui.adsabs.harvard.edu/abs/2025A&A...694A.254H}
}
\bibliographystyle{aasjournalv7}



\end{CJK*}
\end{document}